\documentclass[journal,final,twoside]{IEEEtran}
\usepackage{cite}

\usepackage{amssymb}
\usepackage{blindtext}

\usepackage{graphicx}
\graphicspath{{images/}}
\usepackage[font=small]{caption}
\usepackage[labelformat=simple,font=footnotesize]{subcaption}

\usepackage{amsfonts,bm}
\usepackage[cmex10]{amsmath}
\usepackage[colorlinks=true, allcolors=blue]{hyperref} 
\usepackage{bbm}

\usepackage[dvipsnames]{xcolor}
\usepackage[normalem]{ulem} %provides strike-through
\usepackage{algorithm, algorithmic}
\usepackage{cleveref}

\usepackage{paralist}
\usepackage{multirow}
\usepackage{rotating}

\newcommand{\norm}[1]{\lVert #1 \rVert}

\DeclareFontFamily{U}{mathx}{\hyphenchar\font45}
\DeclareFontShape{U}{mathx}{m}{n}{
      <5> <6> <7> <8> <9> <10>
      <10.95> <12> <14.4> <17.28> <20.74> <24.88>
      mathx10
      }{}
\DeclareSymbolFont{mathx}{U}{mathx}{m}{n}
\DeclareFontSubstitution{U}{mathx}{m}{n}
\DeclareMathAccent{\widecheck}{0}{mathx}{"71}

\newcommand{\Frac}[2]{{{#1}/{#2}}}  % an "inert" form of \frac
\newcommand{\beq}{\begin{equation}}
\newcommand{\eeq}{\end{equation}}
\newcommand{\beqan}{\begin{eqnarray*}}
\newcommand{\eeqan}{\end{eqnarray*}}

\newcommand{\openCase}  {\left\{ \begin{array}{@{\,}ll}}
\newcommand{\openCasell}{\left\{ \begin{array}{@{\,}ll}}
\newcommand{\openCasecl}{\left\{ \begin{array}{@{\,}cl}}
\newcommand{\openCaserl}{\left\{ \begin{array}{@{\,}rl}}
\newcommand{\openCaseTablell}{\left\{ \begin{array}{@{}ll}}
\newcommand{\openCaseTablecl}{\left\{ \begin{array}{@{}cl}}
\newcommand{\openCaseTablerl}{\left\{ \begin{array}{@{}rl}}
\newcommand{\closeCase} {\end{array} \right.}

\DeclareMathOperator*{\argmin}{arg\,min}
\DeclareMathOperator*{\argmax}{arg\,max}

\def\smid{\,|\,}  % Like \mid, but with less space; use for conditioning
\def\sMid{\,;\,}  % Version of \smid for non-Bayesian case
\def\etaVec{\bm{\eta}}
\def\etaVecPrev{\bm{\eta}^{\rm prev}}
\def\yVec{\mathbf{y}}
\def\yconv{\mathbf{y}^{\rm C}}

\def\foracle{f_{\rm oracle}}
\def\fconv{f_{\rm conv}}
\def\ftrml{f_{\rm TRML}}
\def\fqm{f_{\rm QM}}
\def\flqm{f_{\rm LQM}}
\def\mVec{\bm{M}}
\def\lVec{\bm{L}}
\def\etaOracle{\widehat{\etaVec}^{\rm oracle}}
\def\etaQM{\widehat{\etaVec}^{\rm QM}}
\def\etaLQM{\widehat{\etaVec}^{\rm LQM}}
\def\etaG{\widehat{\etaVec}^{\rm naive}}
\def\etaTRML{\widehat{\etaVec}^{\rm TRML}}
\def\IndicatorK{\mathbbm{1}_{\{ \yVec_k^{(i)} > 0\}}}

\def\etaConv{\widehat{\etaVec}^{\rm conv}}

\def\vVec{\bm{v}}
\def\uVec{\bm{u}}
\def\etaHatVec{\widehat{\bm{\eta}}}
\def\vHatVec{\widehat{\bm{v}}}

\def\yVecConv{\bm{y}^{\rm C}}

\DeclareMathOperator{\Poisson}{Poisson}

\newlength{\threewidewidth}
\begin{document}
\title{Denoising Particle Beam Micrographs with
Plug-and-Play Methods}

\author{Minxu Peng,~\IEEEmembership{Student Member,~IEEE,}
        Ruangrawee Kitichotkul,
        Sheila~W.~Seidel,~\IEEEmembership{Student Member,~IEEE,} \\
        Christopher Yu,~\IEEEmembership{Member,~IEEE,}
        and~Vivek K Goyal,~\IEEEmembership{Fellow,~IEEE}
\thanks{This work was supported in part by
a Draper Fellowship
and by the US National Science Foundation under Grant No.~1815896
and Grant No.~2039762.}
\thanks{M. Peng, R. Kitichotkul, S. W. Seidel, and V. K. Goyal are with the Department of Electrical and Computer Engineering, Boston University, Boston, MA 02215 USA
(mxpeng@bu.edu; rkitich@bu.edu; sseidel@bu.edu; v.goyal@ieee.org).}
\thanks{C. Yu is with Charles Stark Draper Laboratory, Cambridge, MA 02139 USA (cyu@draper.com).}
}

\markboth{Peng, Kitichotkul, Seidel, Yu, \& Goyal}{Denoising Particle Beam Micrographs with Plug-and-Play Methods}

\maketitle

\begin{abstract}
In a particle beam microscope,
a raster-scanned focused beam of particles interacts with a sample
to generate a secondary electron (SE) signal pixel by pixel.
Conventionally formed micrographs are noisy because of limitations on acquisition time and dose.
Recent work has shown that estimation methods applicable to a time-resolved measurement paradigm can greatly reduce noise, but these methods apply pixel by pixel without exploiting image structure.
Raw SE count data can be modeled with a compound Poisson (Neyman Type~A) likelihood,
which implies data variance that is signal-dependent and greater than the
variation in the underlying particle--sample interaction.
These statistical properties make methods that assume additive white Gaussian noise ineffective.
This paper introduces methods for
particle beam
micrograph denoising that
use the plug-and-play framework to exploit image structure while being applicable to the unusual data likelihoods of this modality.
Approximations of the data likelihood that vary in accuracy and computational complexity are combined with denoising by total variation regularization, BM3D, and DnCNN\@.
Methods are provided for both conventional
and time-resolved measurements, assuming SE counts are available.
In simulations representative of helium ion microscopy
and scanning electron microscopy,
significant improvements in root mean-squared error (RMSE),
structural similarity index measure (SSIM),
and qualitative appearance are obtained.
Average reductions in RMSE are by factors ranging from 2.24 to 4.11\@.
\end{abstract}


%{\scriptsize \tableofcontents}

\section{Introduction}
\IEEEPARstart{P}{article} beam microscope (PBM) measurements contain
information about topological and chemical composition of a sample. 
The scanning electron microscope (SEM), used widely in materials and life sciences, is a type of PBM that
uses a focused electron beam 
to form micrographs at micro- and nanometer scales~\cite{mcmullan1995scanning}. 
The less common helium ion microscope (HIM)~\cite{WardNE:06} 
employs a beam of helium ions and offers
larger depth of field and sub-nanometer 
resolution~\cite{Notte2016, morgan2006introduction}. 
The HIM has gained increasing popularity in semiconductor and biological 
imaging~\cite{emmrich2016nanopore, joens2013helium, bazargan2012electronic, wirtz2019imaging}.  
Focused beams of heavier ions, such as neon, gallium, and xenon, are also used for imaging.

Conventionally,
acquiring accurate micrographs requires a high dose from a
long dwell time or high beam current.%
\footnote{Here, \textit{dose} is defined as the mean number of incident particles per pixel, since the absolute spatial scale is not considered.
More commonly, dose is the mean number of incident particles per unit area.
} 
Desired acquisition speed is one impediment to high dose.
Furthermore, high dose is especially problematic for HIM or microscopy with heavier ions,
where sputtering and radiation damage are high and
increase with dose~\cite{orloff1996fundamental, livengood2009subsurface, CastaldoHKVM:09}. 
Thus, SEM and HIM micrographs are inherently noisy.
Computational noise reduction through post-processing that
exploits assumed micrograph structure has received relatively little attention,
perhaps because methods that neglect the unusual statistical properties of the measurements
perform poorly.

Though not currently common in commercial PBMs,
direct detection of secondary electrons (SEs)
avoids the noise introduced by scintillators and photomultiplier tubes~\cite{Jiang2018electron}.
Thus, we assume direct SE detection here to characterize favorable performance.
Another concept for improving the accuracy--dose trade-off is time-resolved (TR) measurement~\cite{PengMBBG:20,PengMBG:21}
(see \Cref{sec:TR});
we consider both TR and 
non-TR measurements here.
A concurrent line of research develops models and methods for PBM data without direct SE detection~\cite{Agarwal:2023-U,Agarwal:2023arXiv}.
Advantages from TR measurement persist when direct SE detection is not available~\cite{Agarwal:2023arXiv}.
Measurements in~\cite{Agarwal:2023arXiv} also provide preliminary evidence to support the \emph{Poisson--Poisson--Gaussian} model from~\cite{PengMBBG:20},
which builds directly upon the model used in this paper.
The amount of data that can be collected with the oscilloscope-based setup described in~\cite{Agarwal:2023arXiv}
is limited by the on-board oscilloscope memory;
it is not practical for collecting datasets for images with many pixels.
The standard interface for a PBM produces 8-bit pixel values on an arbitrary scale that cannot be unambiguously mapped to SE counts.
This data is thus not amenable to the goal of developing physics-based, quantitative image formation methods.
For these reasons, the results in this paper are limited to an idealization of PBM data that emphasizes the main characteristic that distinguishes a PBM from an optical
microscope or digital camera: the compound Poisson nature of the SEs.

In a PBM,
the numbers of incident particles and 
the numbers of detected SEs per incident particle
are both random. 
Each of these random numbers may be modeled as Poisson distributed, 
resulting in a Neyman Type~A distribution for an SE measurement. 
This distribution results in signal-dependent noise variance
that is greater than the variance under a Poisson model for the SE count.
Using the distribution explicitly in a model-based reconstruction method
is difficult:
to write the measurement likelihood with elementary functions requires an infinite series
(see \Cref{sec:abstract-model}),
the negative log likelihood is not convex,
and minimization problems involving the negative log likelihood do not generally have closed-form solutions.

The central goal of this paper is to provide methods to include PBM measurement distributions in model-based reconstruction.
In particular,
plug-and-play (PnP) methods have achieved state-of-the-art performance in many applications~\cite{ahmad2020plug, zhang2017learning, he2018optimizing, chan2016plug} 
and are gaining in popularity~\cite{Kamilov:2023-SPM}.
In the PnP alternating direction method of multipliers (ADMM) framework~\cite{venkatakrishnan2013plug},
one desires an efficient computation of the proximal operator of the data fidelity term. 
While no such algorithm is known for PBM data,
we introduce approximate data fidelities based on Poisson distributions, which admit simple proximal operators similar to those used in PnP methods for Poisson models~\cite{figueiredo2010restoration, rond2016poisson}.
Similarly, for a PnP fast iterative shrinkage-thresholding algorithm (FISTA),
one desires an efficient computation of the gradient of the data fidelity term,
and we provide this for TR measurements.
These proximal operators and gradients may have other uses as well.

TR estimation techniques introduced in~\cite{PengMBBG:20, PengMBG:21} operate pixelwise.
The only previous regularized reconstructions from TR measurements are restricted to
total variation (TV) regularization~\cite{rudin1992nonlinear},
and they do not emphasize efficient implementations~\cite{PengCDIKKG:21,Seidel2022TCI}.
In this work, we use ADMM or FISTA for different data fidelity terms depending on whether the proximal operator of the data fidelity term can be computed efficiently.
In addition to denoising by TV regularization,
we plug in the
BM3D image denoiser~\cite{dabov2007image2} 
and a pre-trained deep neural network denoiser DnCNN~\cite{zhang2017beyond}.

\subsection{Main Contributions}
\begin{itemize}
    \item \textit{Introduction of data fidelity terms
    to make PnP methods efficiently applicable to denoising of particle beam micrographs}.
    Five data fidelity terms are presented:
    one for conventional (non-TR) measurements,
    one to study performance assuming an oracle provides the number of incident particles, and
    three for TR measurements.
    
    \item \textit{Experimental evaluations of PnP methods in emulations of HIM and SEM}\@.
    We combined the five data fidelity terms with three denoisers: TV, BM3D, and DnCNN\@.
    The improvement from regularization is greater for HIM than for SEM
    and for conventional data than for TR data.
    RMSE reduction factors range from 2.24 for HIM with TR data
    to 4.11 for SEM with conventional data.
    
\end{itemize}

\subsection{Outline}
In \Cref{sec:meas_model}, we summarize the abstraction, measurement models and pixelwise estimators in PBM,
and in \Cref{sec:pnp_admm}, we review the PnP ADMM and PnP FISTA frameworks.
The key novelties of this paper are in \Cref{sec:data_fidelity_terms},
where we introduce data fidelity terms that allow the application of PnP methods to PBM denoising.
The data fidelity terms  have varying levels of complexity and
accuracy to the negative log-likelihood function of the physical data generation process.
\Cref{sec:proposed_method} describes the collection of PnP algorithms 
we obtain by combining the data fidelity terms with three different denoisers.
We present experimental results in both 
simulated HIM and SEM settings in \Cref{sec:experiment_result}. 
These show that time-resolved measurements and PnP methods provide significant improvements in estimation accuracy.

\section{Measurement Models and Pixelwise Estimators}
\label{sec:meas_model}
In this section, we introduce our
measurement model for direct SE detection in PBM
and several pixelwise estimators;
see~\cite{PengMBBG:20, PengMBG:21} for additional details.
This paper develops regularized estimators analogous to these pixelwise estimators.
While the incident particles may be electrons or ions, we refer to them as ions for simplicity.

\subsection{Abstract Model}
\label{sec:abstract-model}
A sample is raster scanned with a focused beam of ions.
During a fixed dwell time $t$ at any one raster scan location, 
the number of incident ions $M$ is well modeled as a Poisson random variable
with mean $\lambda = \Lambda t$, 
where $\Lambda$ represents the known
rate of incident ions per unit time.\footnote{Although certain operating conditions may cause the beam current to stray from the desired setting due to contamination~\cite{rahman2013observation},
we have shown that the unknown beam current can be estimated at each pixel using time-resolved measurements~\cite{Seidel2022TCI, Seidel2022_MM}.
Estimates from TR measurements also have an inherent insensitivity to knowledge of $\lambda$~\cite{Watkins2021a,Watkins2021b}.}
Incident ion $j$ interacts with the sample, 
generating $X_j$ number of detected SEs.
Each $X_j$ can be described as a Poisson random variable 
with mean $\eta$, where $\eta$ is called the SE yield. 
This is the physical quantity that we wish to estimate at each raster scan location to produce a pixel of the micrograph.
Over the duration of the acquisition, the total detected SEs $Y = \sum^M_{j=1}X_j$ is 
a Neyman Type~A random variable with probability mass function (PMF)
\begin{equation}
\label{eqn:neyman_pmf}
    \mathrm{P}_{Y}(y;\eta,\lambda) 
        = \frac{e^{-\lambda}\eta^y}{y!}
          \sum^\infty_{m=0}\frac{(\lambda e^{-\eta})^m m^y}{m!},
          \qquad
          y = 0,\,1,\,\ldots,
\end{equation}
mean
\begin{equation}
\label{eq:neyman_mean}
    \mathbbm{E}[Y] = \lambda\eta,
\end{equation}
and variance
\begin{equation}
\label{eqn:neyman_var}
    \text{Var}(Y) = \lambda\eta(\eta + 1).
\end{equation}
Notice the dependence of both the mean and variance
on $\eta$ and that this
differs substantially from Poisson-distributed data, assuming $\eta$ is not too small.

\subsection{Conventional Measurement}
The conventional measurement $\yconv\in\mathbb{R}^d$ gathers measurements across all $d$ pixels,
with each entry drawn from the distribution in \eqref{eqn:neyman_pmf}.
Because the entries of $\yconv$ are independent, its joint PMF is given by
\begin{equation}
    \label{eq:Conv-distribution}
    \mathrm{P}_{\yconv}(y^{\rm C} \sMid \etaVec,\lambda)
    = \prod_{k=1}^d \mathrm{P}_Y\big( y^{\rm C}_k \sMid \etaVec_k,\lambda \big),
\end{equation}
where $\etaVec_k$ is the SE yield at the $k$th pixel and $\lambda$ is the per-pixel dose.

\subsection{Time-Resolved Measurement}
\label{sec:TR}
With TR measurement, the per-pixel dwell time $t$ is split
into $n$ sub-acquisitions, each of length $t/n$, yielding an $n$-length vector of measurements at each pixel. The vector $\yVec\in\mathbb{R}^{dn}$ gathers these TR measurements across all pixels with the measurement vector at the $k$th pixel given by $\yVec_k = [\yVec^{(1)}_k,\yVec^{(2)}_k,\dots,\yVec^{(n)}_k]$. Each observation is sampled from the distribution in~\eqref{eqn:neyman_pmf} with $\lambda$ being replaced by $\lambda/n$. We note that at a given pixel, the conventional measurement may be obtained by summing the vector of time-resolved measurements:
\begin{equation}
   \yconv_k =  \sum_{i = 1}^n \yVec_k^{(i)}.
\end{equation}
The entries in $\yVec$ are independent so its joint PMF is given by
\begin{equation}
    \label{eq:TR-distribution}
    \mathrm{P}_{\yVec}(\yVec \sMid \etaVec,\lambda)
    = \prod_{k=1}^d\prod_{i=1}^n \mathrm{P}_Y\big( y^{(i)}_k \sMid \etaVec_k,\lambda/n \big).
\end{equation}

\subsection{Pixelwise Estimators}
\label{ssec:estimator_no_reg}
We now review  estimation methods that operate at a single pixel without regularization. 
We include an oracle estimator that relies on knowledge of the number of incident ions $M$
(which is \emph{not} available in practice)
and estimators that can be applied with conventional or TR data.

\subsubsection{Conventional Estimator}
From~\eqref{eq:neyman_mean}, scaling the observed SE counts at the $k$th pixel by $\lambda$ yields an unbiased estimator:
\begin{equation}
\label{eqn:conv_form}
\etaConv_k = \yconv_k/\lambda.
\end{equation}
From \eqref{eqn:neyman_var},
its mean-squared error (MSE) is 
\begin{equation}
\label{eqn:conv_mse}
\mathrm{MSE}(\etaConv_k)  = \frac{\etaVec_k(\etaVec_k + 1)}{\lambda}. 
\end{equation}
The $\etaVec_k + 1$ factor in the MSE stems from the 
excess variance in~\eqref{eqn:neyman_var} compared with 
the variance of a Poisson random variable with the same mean. 
This excess variance 
can be attributed to \emph{source shot noise}---the randomness of the number of incident ions.

\subsubsection{Oracle Estimator}
If one were able to know the number of incident ions $\mVec_k$ at the $k$th pixel, dividing $\yconv_k$ by $\mVec_k$ will produce a superior estimator
\begin{equation}
\label{eqn:oracle_form}
\etaOracle_k = \yconv_k/\mVec_k,
\end{equation}
which has the MSE
\begin{equation}
\mathrm{MSE}(\etaOracle_k) = e^{-\lambda}(1 - e^{-\lambda})(\etaVec_k-\eta_0)^2 + \sum_{m=1}^\infty \frac{\etaVec_k}{m} \frac{\lambda^m}{m!} e^{-\lambda},
\end{equation}
where $\eta_0$ is the estimate assigned when $\mVec_k = 0$~\cite{PengMBBG:20}.
For large enough $\lambda$, the choice of $\eta_0$ 
has little effect on the MSE\@.
One can show~\cite[App.~A]{PengMBG:21} that for large $\lambda$, the MSE satisfies
\begin{equation}
\label{eq:oracle-mse}
\mathrm{MSE}(\etaOracle_k) \approx \frac{\etaVec_k}{\lambda},
\end{equation}
which eliminates the excess MSE due to the randomness of incident ions.
We emphasize that such an estimator is unimplementable since $\mVec_k$
cannot be known exactly from only observing $\yconv_k$.

\subsubsection{Quotient Mode Estimator}
To approach the performance of the oracle estimator, we may naturally seek a proxy for $\mVec_k$ that is computable from observed quantities.
When $n$ is large enough, the dose for each subacquisition $\lambda/n$ becomes so small that
the probability that more than one ion will arrive during one subacquisition is negligible.
Assuming that $\eta$ is large enough, most ions will produce at least one SE\@.
In this case, the number of subacquisitions at the $k$th pixel measuring a positive number of SEs, 
\begin{equation}
\label{eq:indicator}
    \lVec_k = \sum^n_{i=1}\IndicatorK,
\end{equation}
is a good approximation for the number of incident ions $\mVec_k$,
where $\IndicatorK$ is equal to 1 when $\{\yVec_k^{(i)}>0\}$ and is equal to 0 otherwise.
Analogous to the oracle estimator in~\eqref{eqn:oracle_form},
the \emph{quotient mode} (QM) estimator is defined as
\begin{equation}
\label{eqn:QM_form}
\etaQM_k
= \frac{\yconv_k}{\lVec_k}
=
\frac{\sum_{i = 1}^n \yVec^{(i)}_k}{\sum_{i=1}^{n}\IndicatorK}.
\end{equation}
The QM name is taken from a similar concept proposed by John Notte in~\cite{Notte2013},
where counting of the analog-domain pulses produced by SE bursts is adopted as the denominator of an estimator similar to~\eqref{eqn:QM_form}.
A closed-form expression for the MSE of $\etaQM_k$ is given in~\cite{PengMBG:21}. 
The MSE of $\etaQM_k$ is significantly lower than that of $\etaConv_k$ except when $\eta$ is small.

\subsubsection{Lambert Quotient Mode Estimator}
When $\eta$ is small, $\lVec_k = \sum_{i=1}^n \IndicatorK$ 
significantly underestimates $\mVec_k$
because the probability of an ion generating zero detected SEs cannot be neglected. 
In this case, the bias in $\etaQM$ caused by the underestimation 
can be reduced by replacing $\lVec_k$ with
$(1-e^{-\etaVec_k})^{-1} \lVec_k$. 
Since the probability of an incident ion 
resulting in at least 1 detected SE is $1-e^{-\etaVec_k}$, 
the adjusted $(1-e^{-\etaVec_k})^{-1}\sum^n_{k=1}\IndicatorK$ 
is a more accurate estimate of $\mVec_k$.
Since
$\etaVec_k$ is unknown, 
the substitution
results in a transcendental equation, 
which has the solution
\begin{equation}
\label{eqn:lqm_form}
\etaLQM_k = W\!\left(-\etaQM_k e^{-\etaQM_k}\right) + \etaQM_k,
\end{equation}
where $W(\cdot)$ represents the Lambert W function~\cite{CorlessGHJK:96}. 
Hence, we name $\etaLQM_k$ the \emph{Lambert quotient mode} (LQM) estimator. 
In MSE, $\etaLQM_k$ significantly improves upon $\etaQM$ when $\eta$ is small and is nearly indistinguishable from $\etaQM$ otherwise.

\subsubsection{Time-Resolved Maximum Likelihood Estimator}
With TR data at the $k$th pixel, $ [\yVec^{(1)}_k,\yVec^{(2)}_k,\dots,\yVec^{(n)}_k]$,
the value of $\eta$
that maximizes the joint likelihood
is the \emph{time-resolved maximum likelihood} (TRML) estimator:
\begin{equation}
\label{eq:etaTRML}
    \etaTRML_k
    = \argmax_{\eta \in [0,\infty)} \, 
      \prod_{i=1}^{n} \mathrm{P}_{Y}(\yVec^{(i)}_k; \eta, \lambda/n),
\end{equation}
where $\mathrm{P}_Y(\cdot \sMid \cdot, \cdot)$ is given by~\eqref{eqn:neyman_pmf}.
According to~\cite[Fig.~5(a)]{PengMBG:21}, $\etaTRML_k$ has lower MSE than $\etaConv_k$, $\etaQM_k$, and $\etaLQM_k$ across all $\eta$ values.
However, since that domination does not hold for continuous-time observations (see~\cite[Fig.~3(a)]{PengMBG:21}),
in the discrete-time setting of the present work, the relative performances of estimators may depend on the choices of $\lambda$ and $n$.

\subsection{Comments on Pixelwise Estimator Performances}
\label{ssec:pixelwise-performance}
The estimators in \Cref{ssec:estimator_no_reg} are analyzed and simulated in~\cite{PengMBG:21}.
The key takeaway from Fisher information analyses is to expect%
\footnote{This is the reciprocal of the Fisher information for a \emph{continuous-time} measurement~\cite[Eq.~(26)]{PengMBG:21}
so it assumes an efficient estimator and $\lambda/n \ll 1$.}
\begin{equation}
    \label{eq:theoretical-mse}
\mathrm{MSE}(\etaTRML_k) \approx \frac{\etaVec_k (1-\eta e^{-\etaVec_k})^{-1}}{\lambda}.
\end{equation}
This is always better than the conventional estimator MSE \eqref{eqn:conv_mse}.
It is worse than the oracle estimator MSE \eqref{eq:oracle-mse} by a factor $(1-\etaVec_k e^{-\etaVec_k})^{-1}$,
which is upper-bounded by $e/(e-1) \approx 1.58$ and approaches 1 for $\etaVec_k \rightarrow 0$ and $\etaVec_k \rightarrow \infty$.
Thus, roughly speaking, TR measurement provides an MSE reduction by approximately a factor of $\etaVec_k + 1$.

The gains from TR measurement in \emph{unregularized} estimation have not previously been combined effectively with regularization.

\section{Plug-and-Play Methods}
\label{sec:pnp_admm}
In this section, we review
ADMM, FISTA, and their uses in plug-and-play methods
for image reconstruction incorporating image priors.

\subsection{Problem Formulation}
Given a vector of conventional measurements $\yconv \in \mathbb{R}^d$ or TR measurements $\yVec\in\mathbb{R}^{dn}$,
we seek to reconstruct the underlying SE yield image $\etaVec \in \mathbbm{R}^d$.
This image reconstruction task 
may be written as an optimization problem of the form 
\begin{equation}
\label{eqn:opt_func}
\etaHatVec = \argmin\limits_{\etaVec} f(\etaVec) + \beta g(\etaVec),
\end{equation}
where $f$ is a data fidelity term that encourages consistency with $\yconv$ or $\yVec$, 
$g$ is a regularizer that promotes solutions with desirable properties,
and $\beta$ is a tuning parameter that controls the regularization strength.
The regularizer $g$ is generally non-smooth so solving the optimization problem in~\eqref{eqn:opt_func} is a non-trivial task.

From a Bayesian perspective, \eqref{eqn:opt_func} can arise as the maximum a posteriori (MAP) estimator when
\begin{equation}
  \label{eq:f-bayesian}
    f(\etaVec) = -\log p(\yVec \smid \etaVec),
\end{equation}
the negative log-likelihood of the observation $\yVec$
(or similarly with $\yconv$); and
\begin{equation}
  \label{eq:g-bayesian}
    \beta g(\etaVec) = -\log p(\etaVec),
\end{equation}
the negative log prior of $\etaVec$.

\subsection{ADMM}
\label{sec:admm}
ADMM~\cite{boyd2011distributed} converts the unconstrained problem in~\eqref{eqn:opt_func}
into a constrained one:
\begin{equation}
(\etaHatVec, \vHatVec) 
= \argmin\limits_{\etaVec, \vVec} f(\etaVec) + \beta g(\vVec),
\quad \text{subject to $\etaVec = \vVec$}. \label{eqn:problem}
\end{equation}
This constrained problem can be solved by minimizing its augmented Lagrangian function
\begin{equation}
\mathcal{L}(\etaVec, \vVec) 
= f(\etaVec) + \beta g(\vVec) + \uVec^T(\etaVec - \vVec) + 
\frac{\rho}{2}\|\etaVec - \vVec\|^2,
\end{equation}
where $\rho$ is a penalty parameter 
and $\uVec$ is the Lagrange multiplier.
The solution can be obtained by
iterating the following steps:
\begin{subequations}
\label{eqn:admm}
\begin{align}
\etaVec^{(t + 1)} &= \argmin\limits_{\etaVec \in \mathbbm{R}^d} f(\etaVec) 
                    + \frac{\rho}{2} \|\etaVec - (\vVec^{(t)} - \uVec^{(t)})\|^2, \label{eqn:admm_inversion} \\
\vVec^{(t + 1)} &= \argmin\limits_{\vVec \in \mathbbm{R}^d} g(\vVec) 
                    + \frac{1}{2\sigma^2} \|\vVec - (\etaVec^{(t+1)} + \uVec^{(t)})\|^2, \label{eqn:admm_denoise}\\
\uVec^{(t+1)} &= \uVec^{(t)} + (\etaVec^{(t+1)} - \vVec^{(t+1)}), \label{eqn:admm_update}
\end{align}
\end{subequations}
where $\sigma = \sqrt{\beta/\rho}$.

\subsection{FISTA}
\label{sec:fista}
The fast iterative shrinkage-thresholding algorithm
(FISTA)~\cite{beck2009fast} is another popular method to solve the optimization problem 
\eqref{eqn:opt_func}. 
FISTA
iterates the following steps:
\begin{subequations}
\label{eqn:fista}
\begin{align}
\uVec^{(t + 1)} 
    &= \vVec^{(t)} - \gamma \nabla f(\vVec^{(t)}), \label{eqn:fista_inversion} \\
\etaVec^{(t + 1)}
    &= \argmin\limits_{\etaVec \in \mathbbm{R}^d} g(\etaVec) 
                    + \frac{1}{2\sigma^2} \|\etaVec - \uVec^{(t + 1)} \|^2, \label{eqn:fista_denoise}\\
q_{t + 1} 
    &= \frac{1}{2}\Big(1 + \sqrt{1 + 4q_{t}^2}\Big), \nonumber\\
\vVec^{(t+1)}
    &= \etaVec^{(t + 1)} + \frac{q_{t} - 1}{q_{t + 1}} (\etaVec^{(t+1)} - \etaVec^{(t)}),
\end{align} 
\end{subequations}
where $\gamma$ is the step size, $\sigma = \sqrt{\gamma \beta}$,
and
$q_0 = 1$.
In contrast to ADMM, FISTA only needs to evaluate the gradient $\nabla f(\etaVec)$,
without needing
to solve \eqref{eqn:admm_inversion}.
This is especially important when \eqref{eqn:admm_inversion} does not have a closed-form solution.

\subsection{Plug-in Denoiser}
\label{sec:PnP_admm_subsection}
ADMM and FISTA have similar decouplings into pairs of subproblems.
Specifically, \eqref{eqn:admm_inversion} and \eqref{eqn:fista_inversion}
can be viewed as \emph{inversion} steps
since $f(\etaVec)$ is determined by the forward image measurement model; 
and~\eqref{eqn:admm_denoise} and \eqref{eqn:fista_denoise}
can be viewed as \emph{denoising} steps
since $g(\etaVec)$ represents an image prior. 
In particular, \eqref{eqn:admm_denoise}
is equivalent to Gaussian denoising on $\etaVec^{(t+1)} + \uVec^{(t)}$ 
with noise level $\sigma = \sqrt{\beta/\rho}$.
Based on the intuition that any other Gaussian denoiser could be used instead,
Venkatakrishnan et~al.~\cite{venkatakrishnan2013plug}
proposed the \emph{PnP ADMM} algorithm,
which does not specify $g$ explicitly. 
Instead,~\eqref{eqn:admm_denoise} is replaced with an off-the-shelf denoiser, 
denoted as $\mathcal{D}_{\sigma}$, to yield
\begin{equation}
\label{eqn:pnp_admm_denoiser}
\vVec^{(t + 1)} = \mathcal{D}_{\sigma}(\etaVec^{(t+1)} + \uVec^{(t)}).
\end{equation}
Similarly, replacing \eqref{eqn:fista_denoise} with
\begin{equation}
\label{eqn:pnp_fista_denoiser}
\etaVec^{(t + 1)} = \mathcal{D}_{\sigma}(\uVec^{(t+1)}),
\end{equation}
where $\sigma = \sqrt{\gamma \beta}$, results in a \emph{PnP FISTA} \cite{kamilov17pnpfista}.

\section{Data Fidelity Terms}
\label{sec:data_fidelity_terms}

In many applications of PnP methods,
the data fidelity term $f(\etaVec)$ is derived from a measurement process that involves linear mixing and signal-independent additive white Gaussian noise (AWGN);
\eqref{eq:f-bayesian} then results in $f(\etaVec) \propto \| \yVec - A \etaVec \|_2^2$ for some matrix $A$,
which is quite convenient for computations.
In this work,
we wish to apply PnP methods to particle beam micrograph denoising,
where the challenge is not rooted in linear mixing.
Instead, challenges arise from the measurement likelihood function even though it is separable.
To directly apply \eqref{eq:f-bayesian} with the measurement likelihood function \eqref{eq:Conv-distribution} or \eqref{eq:TR-distribution} is
problematic because of the form of the Neyman Type~A PMF \eqref{eqn:neyman_pmf}.
This is an obstacle to regularized estimation of any form, whether or not one employs PnP methods.

This section introduces several data fidelity terms $f(\etaVec)$ that vary in
their closeness to \eqref{eq:f-bayesian} and
their computational complexity.
For conventional measurement data,
we have a data fidelity term based on a spatially adapted Gaussian approximation.
For oracle or TR data, the data fidelity terms have correspondences with
the oracle, QM, LQM, and TRML estimators of \Cref{sec:meas_model}.
In cases in which \eqref{eqn:admm_inversion} has
a closed-form solution,
the data fidelity term becomes the basis for a PnP ADMM algorithm;
for the remaining case,
we provide an approximation to $\nabla f(\etaVec)$
so that the data fidelity term becomes the basis for a PnP FISTA method.
These PnP methods are detailed in \Cref{sec:proposed_method}.

\subsection{Gaussian}
\label{sec:f-conventional}
We may approximate the entries of $\yconv$ as independent Gaussian random variables, each with mean and variance given in \eqref{eq:neyman_mean} and \eqref{eqn:neyman_var}:
\begin{equation}
\label{eq:dataFid_conventional}
\yconv_k\sim\mathcal{N}(\lambda\etaVec_k,\,\lambda\etaVec_k(\etaVec_k+1)).
\end{equation}
Simple point evaluation of the Gaussian probability density function gives a reasonable approximation of the Neyman Type~A probability mass function
provided that $\lambda\etaVec_k$ is not small,
and there is pointwise convergence of the moment generating function for $\lambda \rightarrow \infty$~\cite{MartinK:62}.
Omitting a constant term, the corresponding negative log-likelihood function is
\begin{equation}
    \fconv(\etaVec) = \sum_{k=1}^d \left(
    \frac{1}{2}\log\etaVec_k + \frac{1}{2}\log(\etaVec_k + 1) + \frac{(\yconv_k-\lambda\etaVec_k)^2}{2\lambda\etaVec_k(\etaVec_k+1)}
    \right).
\end{equation}
To obtain a closed-form solution when we compute the inverse step \eqref{eqn:admm_inversion},
we can further approximate \eqref{eq:dataFid_conventional} as
\begin{equation}
    \label{eq:dataFid_conventional_fast}
\yconv_k\sim\mathcal{N}(\lambda\etaVec_k,\,\lambda\etaVecPrev_k(\etaVecPrev_k+1)),
\end{equation}
where the variance is set to be independent of $\etaVec_k$
by using $\etaVecPrev_k$,
an estimated value of $\etaVec_k$ at the previous iteration.  
In this case, the observation $\yVecConv$ becomes a Gaussian random vector with constant, diagonal covariance matrix, and the negative log-likelihood function is
\begin{equation}
    \fconv(\etaVec) = \sum\limits_{k=1}^d
    \frac{(\yconv_k-\lambda\etaVec_k)^2}{2\lambda\etaVecPrev_k(\etaVecPrev_k+1)}
\end{equation}
after dropping terms that do not depend on $\etaVec$.
With this $f$, the ADMM inversion step \eqref{eqn:admm_inversion} is separable over the $d$ components,
with
closed-form solution
\begin{align}
\etaVec_k^{(t+1)} 
&= \left(\frac{\lambda}{\etaVecPrev_k(\etaVecPrev_k + 1)}+\rho\right)^{-1} \nonumber\\
&\quad \cdot \left(\frac{\yconv_k}{\etaVecPrev_k(\etaVecPrev_k + 1)} + \rho(\vVec_k^{(t)} -\uVec_k^{(t)}) \right).
\label{eq:f-conv-closed-form}
\end{align}

In the model \eqref{eq:dataFid_conventional_fast}, we are positing a variance for pixel $k$
that gives spatially varying strength to the regularizer $g$.
When the detected $\yconv_k$
is unluckily small relative to a moderate or larger underlying true value $\eta_k$, 
estimation will perform poorly if too much confidence is ascribed to the observed $\yconv_k$. 
To avoid this, performance is improved by allowing the local variance estimate
$\lambda\etaVecPrev_k(\etaVecPrev_k+1)$
to depend on a neighborhood of pixel $k$ rather than pixel $k$ alone.
Specifically,
when computing $\etaVec_k^{(t+1)}$,
we choose $\etaVecPrev_k$
to be the average of
$\etaVec_k^{(t)}$ across a neighborhood of pixel $k$:
\begin{equation}
    \label{eq:etaVecPrev}
    \etaVecPrev_k = \frac{1}{9} \sum_{\ell \in \mathcal{N}(k)} \etaVec_\ell^{(t)},
\end{equation}
where $\mathcal{N}(k)$ is a $3 \times 3$ patch centered at pixel $k$.
We chose the patch size to be $3 \times 3$ as using a $5 \times 5$ patch was empirically inferior.

\subsection{Oracle}
When the number of incident ions is exactly known, the number of SEs observed at the $k$th pixel may be modeled as a Poisson random variable with parameter $\mVec_k\etaVec_k$:
\begin{equation}
    \label{eq:dataFid_oracle}
    \yconv_k\sim\Poisson(\mVec_k\etaVec_k).
\end{equation}
The negative log-likelihood function is
\begin{equation}
    \foracle(\etaVec) = \sum\limits_{k=1}^{d} 
    \left(\mVec_k\etaVec_k - \yconv_k\log\etaVec_k\right)
\end{equation}
after dropping terms that do not depend on $\etaVec$.
Since $\foracle$ is convex, the minimum in \eqref{eqn:admm_inversion} is achieved when the gradient of the objective is zero. Hence, we obtain a closed-form update:
\begin{equation}
\label{eq:f-oracle-closed-form}
    \etaVec_k^{(t + 1)} = -\frac{1}{2} \varphi + 
    \frac{1}{2}
    \sqrt{\varphi^2 + \frac{4\yconv_k}{\rho}}
    \quad
    \mbox{where $\varphi = \displaystyle\frac{\mVec_k}{\rho} - \vVec_k + \uVec_k$}.
\end{equation}
Although instruments are not capable of measuring $\mVec_k$, the oracle data fidelity term serves as an interesting benchmark.
We note that the forward model is reduced to sampling from a Poisson distribution. In fact, similar closed-form updates have been used in PnP ADMM methods for Poisson models, such as equation (16) in~\cite{rond2016poisson}.
QM and LQM data fidelity terms, which we will discuss next, also follow Poisson distributions and thus have similar closed-form updates.

\subsection{Quotient Mode}
As in \eqref{eqn:QM_form}, our QM data fidelity term uses the number of subacquisitions where more that one SE was observed as a proxy for $\mVec_k$.
Here we have
\begin{equation}
\label{eq:dataFid_QM}
    \yconv_k \sim \Poisson(\lVec_k\etaVec_k),
\end{equation}
with $\lVec_k$ defined in \eqref{eq:indicator}.
The negative log-likelihood function is
\begin{equation}
    \fqm(\etaVec) = \sum\limits_{k=1}^{d} 
    \left(\lVec_k\etaVec_k - \yconv_k\log\etaVec_k\right)
\end{equation}
after dropping terms that do not depend on $\etaVec$.
Similar to the oracle case,
since $\fqm$ is convex, the ADMM inversion step \eqref{eqn:admm_inversion} again has a closed-form solution:
\begin{equation}
\label{eq:f-qm-closed-form}
    \etaVec_k^{(t + 1)} = -\frac{1}{2} \varphi + 
    \frac{1}{2}
    \sqrt{ \varphi^2 + \frac{4\yconv_k}{\rho}}
    \quad
    \mbox{where $\varphi = \displaystyle\frac{\lVec_k}{\rho} - \vVec_k + \uVec_k$}.
\end{equation}

\subsection{Lambert Quotient Mode}
Using the same adjustment as in \eqref{eqn:lqm_form} to compensate for the underestimate of $\mVec_k$ in $\lVec_k$, our LQM data fidelity term is based upon the model
\begin{equation}
\label{eq:dataFid_LQM}
    \yconv_k \sim \Poisson\!\left((1-e^{-\etaVec_k})^{-1}\lVec_k\etaVec_k\right).
\end{equation}
The negative log-likelihood function after dropping terms that do not depend on $\etaVec$ is
\begin{align}
\flqm(\etaVec) 
&= \sum\limits_{k=1}^{d} 
\bigg(
    \frac{\etaVec_k}{1 - e^{-\etaVec_k}}\lVec_k + \yconv_k\log(1 - e^{-\etaVec_k}) \nonumber\\
& \qquad\quad -\yconv_k \log\etaVec_k
\bigg).
\end{align}
For faster computation, \eqref{eq:dataFid_LQM} may be approximated as
\begin{equation}
\label{eq:dataFid_LQM_approx}
    \yconv_k \sim \Poisson((1-e^{-\etaVecPrev_k})^{-1}\lVec_k\etaVec_k),
\end{equation}
where $\etaVecPrev$ is based on the previous iteration as in \Cref{sec:f-conventional}.
Then
\begin{equation}
    \flqm(\etaVec) \approx  \frac{\lVec_k\etaVec_k}{1-e^{-\etaVecPrev_k}} -\yconv_k\log\etaVec_k.
\end{equation}
With this approximation, the ADMM inversion step \eqref{eqn:admm_inversion} has closed-form solution
\begin{subequations}
\label{eq:f-lqm-closed-form}
\begin{equation}
  \etaVec_k^{(t + 1)} = -\frac{1}{2} \varphi + \frac{1}{2}\sqrt{ \varphi^2 + \frac{4\yconv_k}{\rho}}
\end{equation}
where
\begin{equation}
  \varphi = \frac{\lVec_k}{\rho (1 - e^{-\etaVecPrev_k})} - \vVec_k + \uVec_k.
\end{equation}
\end{subequations}

\subsection{Time-Resolved Maximum Likelihood}
In our TRML data fidelity term, we use the full likelihood in \eqref{eq:TR-distribution}. Here we have
\begin{align}
\label{eq:dataFid_trml}
    \ftrml(\etaVec) = -\log  \mathrm{P}_{\yVec}(\yVec \sMid \etaVec,\lambda).
\end{align}
To use derivatives of \eqref{eq:dataFid_trml} directly with the substitution of
\eqref{eqn:neyman_pmf} and \eqref{eq:TR-distribution}
is computationally expensive and delicate.
However, as derived in \cite[App.~D]{Seidel2022TCI} with the aid of Touchard polynomials, 
the derivatives of \eqref{eq:dataFid_trml} with respect to the entries in $\etaVec$ are approximately
\begin{align}
    \frac{\partial \ftrml(\etaVec)}{\partial \etaVec_k}
    \approx& \ (n - \lVec_k)\frac{\lambda}{n} e^{-\etaVec_k} - \frac{\yconv_k}{\etaVec_k}\nonumber \\
    & \ + \sum_{i \in \mathcal{S}}
            \frac{n+(2^{\yVec_k^{(i)}}  -1)\lambda e^{-\etaVec_k}}
                 {n+(2^{\yVec_k^{(i)}-1}-1)\lambda e^{-\etaVec_k}},
\label{eq:trml_gradient}
\end{align}
where $\mathcal{S} = \{ i \ : \ \yVec_k^{(i)} > 0 \}$.
The approximation is accurate when $\lambda / n$ is small.
The experimental results in \Cref{sec:experiment_result} are for $\lambda / n = 0.1$,
which is small enough for accurate approximation and for most of the gains from TR measurement to be realized~\cite{PengMBG:21}.

\section{Proposed Methods}
\label{sec:proposed_method}

In this work, we test the five data fidelity terms proposed in \Cref{sec:data_fidelity_terms} within the PnP framework using three different denoisers:
TV-regularized least squares, BM3D~\cite{dabov2007image2}, and a deep neural network.
This section explains the overall algorithm design and different types of denoising.

\begin{algorithm}
\caption{Plug-and-Play ADMM}
\label{alg:plugNplay}
\begin{algorithmic}[1]
\renewcommand{\algorithmicrequire}{\textbf{Input:}}
\renewcommand{\algorithmicensure}{\textbf{Initialize:}}
\renewcommand{\algorithmicfor}{\textbf{while}}
\REQUIRE $\alpha$, $\beta$, $\rho$, $\yconv$ or $\yVec = [\yVec^{(1)},\yVec^{(2)},\dots,\yVec^{(n)}]$
\ENSURE $t = 0$, $\etaVec^{(0)} = \vVec^{(0)} = \etaConv$,\! $\uVec^{(0)} = \bm{0}$,\! $\sigma=\sqrt{\beta/\rho}$
    \FOR{not converged}
    \STATE $\etaVec^{(t + 1)} = \argmin\limits_{\etaVec \in \mathbbm{R}^d} f(\etaVec) + \frac{1}{2}\rho \|\etaVec - (\vVec^{(t)}- \uVec^{(t)})\|^2$
    \STATE $\vVec^{(t + 1)} = \mathcal{D}_{\sigma}(\etaVec^{(t+1)} + \uVec^{(t)})$
    \STATE $\uVec^{(t+1)} = \uVec^{(t)} + (\etaVec^{(t+1)} - \vVec^{(t+1)})$
    \STATE $t = t + 1$
    \ENDFOR
    \RETURN $\etaVec^{(t+1)}$
\end{algorithmic}
\end{algorithm}

\begin{algorithm}
\caption{Plug-and-Play FISTA}
\label{alg:plugNplay_FISTA}
\begin{algorithmic}[1]
\renewcommand{\algorithmicrequire}{\textbf{Input:}}
\renewcommand{\algorithmicensure}{\textbf{Initialize:}}
\renewcommand{\algorithmicfor}{\textbf{while}}
\REQUIRE $\alpha$, $\beta$, $\gamma$, $\rho$, $\yconv$ or $\yVec = [\yVec^{(1)},\yVec^{(2)},\dots,\yVec^{(n)}]$
\ENSURE $t = 0$, $\etaVec^{(0)} = \vVec^{(0)} = \etaConv$, $q_0 = 1$, $\sigma=\sqrt{\gamma \beta}$
    \FOR{not converged}
    \STATE $\uVec^{(t + 1)} = \vVec^{(t)} - \gamma \nabla f(\vVec^{(t)}) $
    \STATE $\etaVec^{(t + 1)} = \mathcal{D}_{\sigma}(\uVec^{(t + 1)})$
    \STATE $q_{t + 1} = \frac{1}{2}(1 + \sqrt{1 + 4q_{t}^2})$
    \STATE $\vVec^{(t+1)} = \etaVec^{(t + 1)} + (\Frac{(q_{t} - 1)}{q_{t + 1}}) (\etaVec^{(t+1)} - \etaVec^{(t)})$
    % \STATE $\vVec^{(t+1)} = \etaVec^{(t + 1)} + \frac{q_{t + 1} - 1}{q_t} (\etaVec^{(t+1)} - \etaVec^{(t)})$
    \STATE $t = t + 1$
    \ENDFOR
    \RETURN $\etaVec^{(t+1)}$
\end{algorithmic}
\end{algorithm}

\subsection{Algorithm Overview}
\label{sec:PnP}
Algorithms~\ref{alg:plugNplay} and~\ref{alg:plugNplay_FISTA} outline the key steps of PnP ADMM and PnP FISTA,
as detailed in \Cref{sec:pnp_admm}, adapted to our setting. 
We apply Algorithm~\ref{alg:plugNplay_FISTA} with the TRML data fidelity term since
\eqref{eqn:admm_inversion}
does not have a closed-form solution in that case,
and we apply Algorithm~\ref{alg:plugNplay} to the other cases.

In Algorithm~\ref{alg:plugNplay},
Line~2,
the \textit{inversion step}, incorporates the data fidelity term.
For the Gaussian, oracle, QM or LQM data fidelity term),
it is computed with \eqref{eq:f-conv-closed-form}, \eqref{eq:f-oracle-closed-form}, \eqref{eq:f-qm-closed-form} or \eqref{eq:f-lqm-closed-form}, respectively.
Line~3 is the \textit{denoising} step of \eqref{eqn:pnp_admm_denoiser}.
Line~4 updates the Lagrange multiplier.

In Algorithm~\ref{alg:plugNplay_FISTA},
Line~2 makes $\uVec$ a step from $\vVec$ in the direction of the negative gradient of the data fidelity $f$. 
This is computed with \eqref{eq:trml_gradient} for the TRML data fidelity term.
Line~3 applies a denoiser to remove noise in $\uVec$. 
Line~4 updates $q_t$, which controls the convergence rate.
Line~5 uses $q_t$ to update $\vVec$.

Both algorithms terminate when convergence is achieved.
Here, convergence is declared when
$\Delta_{t+1} \leq \alpha$,
where $\alpha$ is a threshold parameter. For PnP ADMM,
\begin{align}
\label{eq:delta_t}
    \Delta_{t+1} :=& \frac{1}{\sqrt{d}}
                     \big(\norm{\etaVec^{(t+1)}-\etaVec^{(t)}}_2
                    + \norm{\vVec^{(t+1)}-\vVec^{(t)}}_2
                      \nonumber\\
                   & 
                    \qquad + \norm{\uVec^{(t+1)}-\uVec^{(t)}}_2 \big).
\end{align}
For PnP FISTA,
\begin{equation}
    \Delta_{t+1} := \frac{1}{\sqrt{d}} \norm{\etaVec^{(t+1)}-\etaVec^{(t)}}_2.
\end{equation}

If one wishes to constrain $\etaVec$ to a convex set, it is natural to include a projection to that set within each iteration~\cite{figueiredo2010restoration, Bouman:22, ziabari2019physics, ono2017primal}.
Though the physical meaning of $\etaVec$ implies nonnegativity of each entry,
we have not imposed this explicitly because our experimental conditions keep all entries of the estimates nonnegative at all iterations.

\subsection{Denoisers}
\label{ssec:regularization}
We employ three denoising methods.

\subsubsection{Total Variation} Under this formulation,
the denoising step is regularized least-squares estimation \eqref{eqn:admm_denoise} with regularizer
$g(\etaVec) = \norm{\etaVec}_{\rm TV}$.
The isotropic TV cost $\norm{\etaVec}_{\rm TV}$ is given by
\begin{equation}
   \|\etaVec\|_{\rm TV} = \norm{\boldsymbol{D}\etaVec}_2,
\end{equation}
where $\boldsymbol{D}$ denotes the discrete image gradient. This cost term is designed to promote estimates that are piecewise smooth while maintaining edge features. The corresponding denoising step is solved iteratively as in \cite{beck2009fast}.

\subsubsection{BM3D}
One branch of denoising algorithms exploits non-local similarity 
of image patches to recover a clean image from the noisy observation.
BM3D~\cite{dabov2007image2} is one of the most widely used
among these methods.
It groups image patches based on 
similarity, then applies collaborative filtering 
and recombines to yield a reconstructed image.

\subsubsection{Deep Neural Network} 
\label{sec:DnCNN}
In recent years, deep learning-based methods have achieved great success 
in image denoising tasks. 
Most existing deep neural networks use a large number of clean--noisy image pairs
as training samples, which is one key factor contributing to performance improvements. 
However, collecting a large dataset of clean--noisy image pairs can be expensive and challenging, especially in PBM\@.
Without access to such a dataset,
\cite{PengCDIKKG:21} used synthetic data,
generated using accurate knowledge of the PBM forward model,
to train a deep neural network.
The network's weights were optimized to 
minimize the $\mathcal{L}_2$ difference
between reconstructed and ground truth images.
This method did not
explicitly leverage the model information 
during inversion
but instead relied on the PBM model to generate 
a noisy counterpart to each clean training image.
In contrast, we use the PnP framework to combine
knowledge of the PBM model 
with the assistance of a deep neural network 
to extract meaningful features, resulting in improved performance.
In the denoising step (Line 3 of \Cref{alg:plugNplay}),
we adopt a deep neural network called DnCNN~\cite{zhang2017beyond}.
It combines residual learning~\cite{he2016deep} and batch normalization~\cite{ioffe2015batch},
and it has been demonstrated to be effective in removing AWGN\@. 

In the formulation of ADMM in \Cref{sec:admm},
the Gaussian denoising problem \eqref{eqn:admm_denoise} has a noise standard deviation
$\sigma = \sqrt{\beta/\rho}$
derived directly from the regularization parameter $\beta$
and variable-splitting penalty $\rho$.
Similarly, in PnP FISTA, the noise standard deviation $\sigma = \sqrt{\gamma \beta}$ depends on $\beta$ and the step size $\gamma$.
This dependence on $\sigma$ can be troublesome in PnP methods
because the denoiser
may require training that depends on $\sigma$
or it may have no analogous parameter.

Denoiser scaling~\cite{xu2020boosting} is a method for introducing
tunable regularization to a denoiser trained for a specific noise standard deviation $\sigma$.
Suppose only that some denoiser $\mathcal{D}$ is given.
Then a scaled denoiser is defined as
\begin{equation}
\label{eq:scaled_denoiser}
    \mathcal{D}_{\mu}(\etaVec) \equiv (1/\mu)\mathcal{D}(\mu \etaVec),
\end{equation}
where $\mu > 0$
is the denoiser scaling parameter.
This approach makes it possible to use a single pre-trained 
network across a variety of noise levels.

\subsection{Convergence} \label{sec:convergence}

When $f$ and $g$ are convex functions, both ADMM \cite{boyd2011distributed} and FISTA \cite{beck2009fast} will converge to the global minimum of the convex optimization problem \eqref{eqn:problem} under some additional technical conditions.
While a denoiser in PnP methods may not be the proximal operator of any convex function $g$,
they can still have fixed point convergence. Specifically, PnP ADMM and PnP iterative shrinkage-thresholding algorithm (ISTA),
i.e., FISTA without Nesterov acceleration,
are shown to have fixed point convergence under appropriate choices of the parameters $\rho$ and $\gamma$ when $f$ is strongly convex and the residue
$\mathcal{D}_{\sigma} - \mathbf{I}$ has a sufficiently small Lipschitz constant, where $\mathbf{I}$ is the identity operator \cite{ryu19pnp}. 

In our case, not all of the proposed data fidelity terms are convex with respect to $\etaVec$. While $\foracle$ and $\fqm$ are convex, $\fconv$, $\flqm$, and $\ftrml$ are not.
We find that $\flqm$ is locally strongly convex for typical values of $\yVec$ and $\etaVec$ in our experiments.
Consequently, we can establish fixed point convergence of PnP ADMM and PnP ISTA for $\foracle$, $\fqm$, and $\flqm$ under appropriate choices of $\rho$, $\gamma$, and $\beta$, and the denoiser $\mathcal{D}_{\sigma}$.
We note that the TV denoiser is a proximal operator and that DnCNN typically has contractive residue even without further modification \cite{ryu19pnp}, so using these denoisers will not hinder fixed point convergence of PnP methods.
However, BM3D has been shown to violate the residue contractiveness criterion \cite{ryu19pnp}, so the PnP methods that use BM3D may not converge, even if $f$ is convex.
We find that the theory outlined here predicts the empirical convergence of PnP ADMM and PnP FISTA well in our experiments.

\section{Experimental Results}
\label{sec:experiment_result}
In this section,
we compare the performances of the proposed methods using simulated 
HIM and SEM micrograph images. 
While previous works~\cite{PengMBBG:20,PengMBG:21} have characterized
the QM, LQM, and TRML estimators of \Cref{ssec:estimator_no_reg},
here we aim to compare how these and the benchmark conventional and oracle estimators
perform in combination with the three denoisers in \Cref{ssec:regularization},
within the PnP ADMM and PnP FISTA frameworks.
The five estimators are included in the PnP methods through the data fidelity terms detailed in \Cref{sec:data_fidelity_terms}.

\subsection{Datasets and Experiment Details}

We use five crops of images from the ``porous sponge'' NFFA-EUROPE SEM dataset \cite{aversa18semDataset}
as the ground truth images in our experiments.
We scale the images to $\eta \in [2, 8]$ to emulate HIM~\cite{notte2007introduction} and to $\eta \in [1, 2]$ to emulate SEM\@.
Given a ground truth image, we apply~\eqref{eq:TR-distribution} to generate noisy TR measurements pseudorandomly using total dose $\lambda = 20$ split over $n = 200$ sub-acquisitions for each pixel for the HIM sample and total dose $\lambda = 50$ split over $n = 500$ sub-acquisitions per pixel for the SEM sample.
The conventional measurement is obtained by summing over the $n$ subacquisitions at each pixel.
All PnP experiments use a stopping threshold of $\alpha =  5 \times 10^{-4}$ as defined in \Cref{sec:PnP}.
For PnP ADMM methods, we fix $\rho = 2.5$ for HIM and $\rho = 10$ for SEM\@.
For PnP FISTA methods, we fix $\gamma = 0.1$ for HIM and $\gamma = 0.01$ for SEM\@.
We tune $\sigma$ (for TV and BM3D denoisers) and $\mu$ (for the DnCNN denoiser), which implies the $\beta$ that controls the regularization strength, on a hold-out validation image.
For HIM experiments, we use $\sigma \in \{0.24, 0.32, 0.72, 0.96, 1.2\}$ and $\mu \in \{0.5, 0.75, 1, 1.25, 1.5\}$.
For SEM experiments, we use $\sigma \in \{0.1, 0.3, 0.5, 0.7, 0.9\}$ and $\mu \in \{0.25, 0.5, 0.75, 1, 1.25, 1.5\}$.
We use $\sigma$ and $\mu$ values that yield the least root mean-squared error (RMSE) on the validation image for each combination of $f$ and the denoiser in the corresponding experiments with the remaining 4 test images.

We train the DnCNN denoiser to minimize $\ell_2$ loss using 500 natural images from the Berkeley Segmentation Dataset (BSD 500)~\cite{BSDS500}.
The pixel values are scaled to $[0, 1]$, and the standard deviation of the additive white Gaussian noise is $25 / 255$.
When applying DnCNN to an image with a different dynamic range, such as $\eta \in [2, 8]$ in the case of HIM experiments, we shift and scale the image so that the dynamic range of the input to DnCNN is $[0, 1]$.
Then, we scale and shift the output back to the original dynamic range.
Note that denoiser scaling as in \eqref{eq:scaled_denoiser} is performed on the image that is already transformed to the $[0, 1]$ dynamic range.
In addition, DnCNN-related methods are GPU-accelerated for faster computation.

\newlength{\figHeight}
\setlength{\figHeight}{2.97cm}
\newlength{\firstcolwidth}
\setlength{\firstcolwidth}{0.1728\textwidth} % 0.18
\newlength{\restcolwidth}
\setlength{\restcolwidth}{0.1536\textwidth} % 0.16

\begin{figure*}
  \centering
  \begin{tabular}{@{}c@{\,}c@{\,\,}c@{\,}c@{\,}c@{\,}c@{\,}c@{}}
    & \multicolumn{1}{c|}{{\small \bf Ground truth}}
    & {\small Oracle}
    & {\small Conventional}
    & {\small QM}
    & {\small LQM}
    & {\small TRML}
    \\
    \rotatebox[origin=l]{90}{\quad \small No regularization}
    &
    \multicolumn{1}{l|}{
    \begin{subfigure}[t]{\firstcolwidth}
        \includegraphics[height=0.88\figHeight]{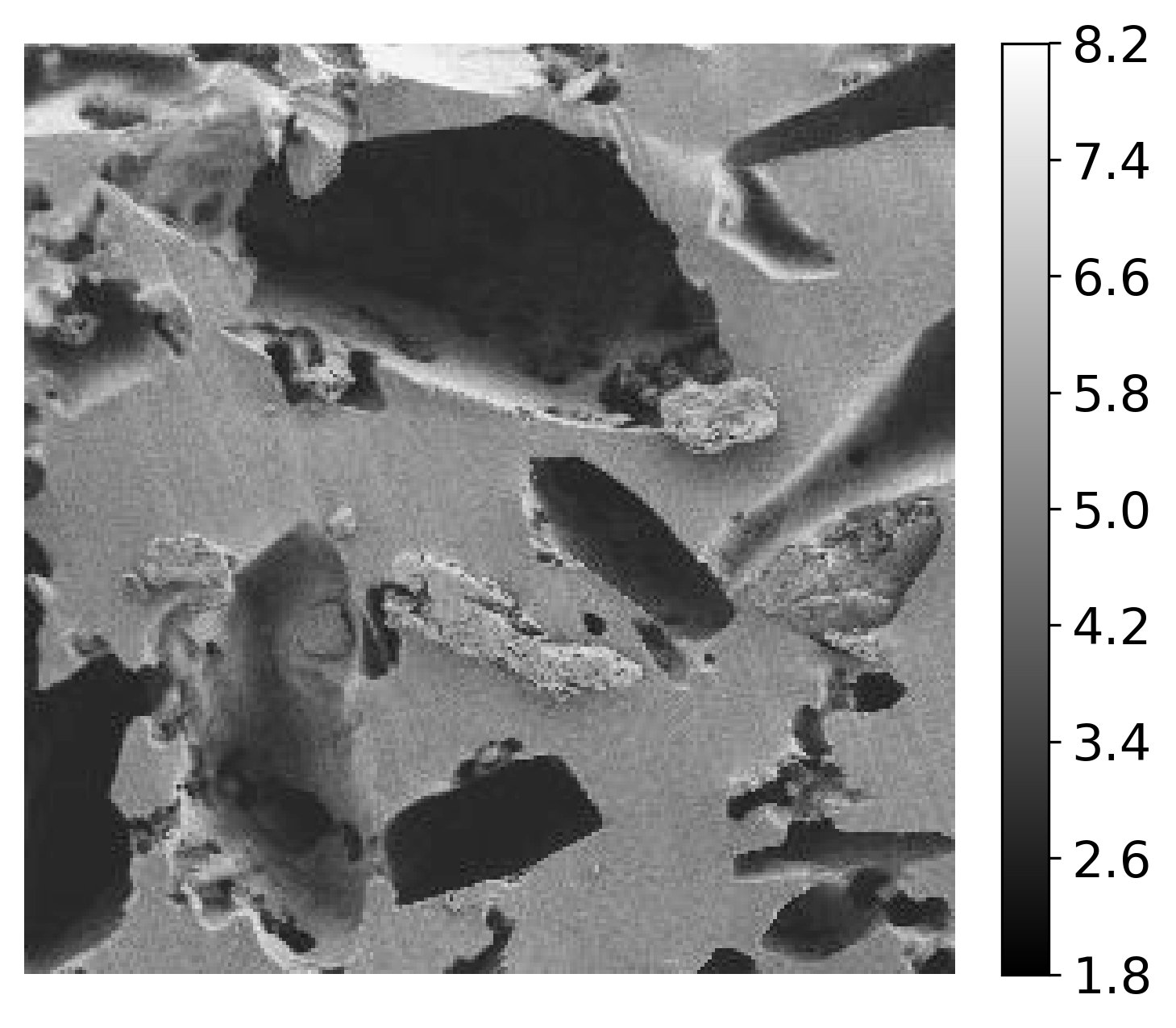}
    \end{subfigure}} &
    \,\begin{subfigure}[t]{\restcolwidth}
        \includegraphics[height=1.01\figHeight]{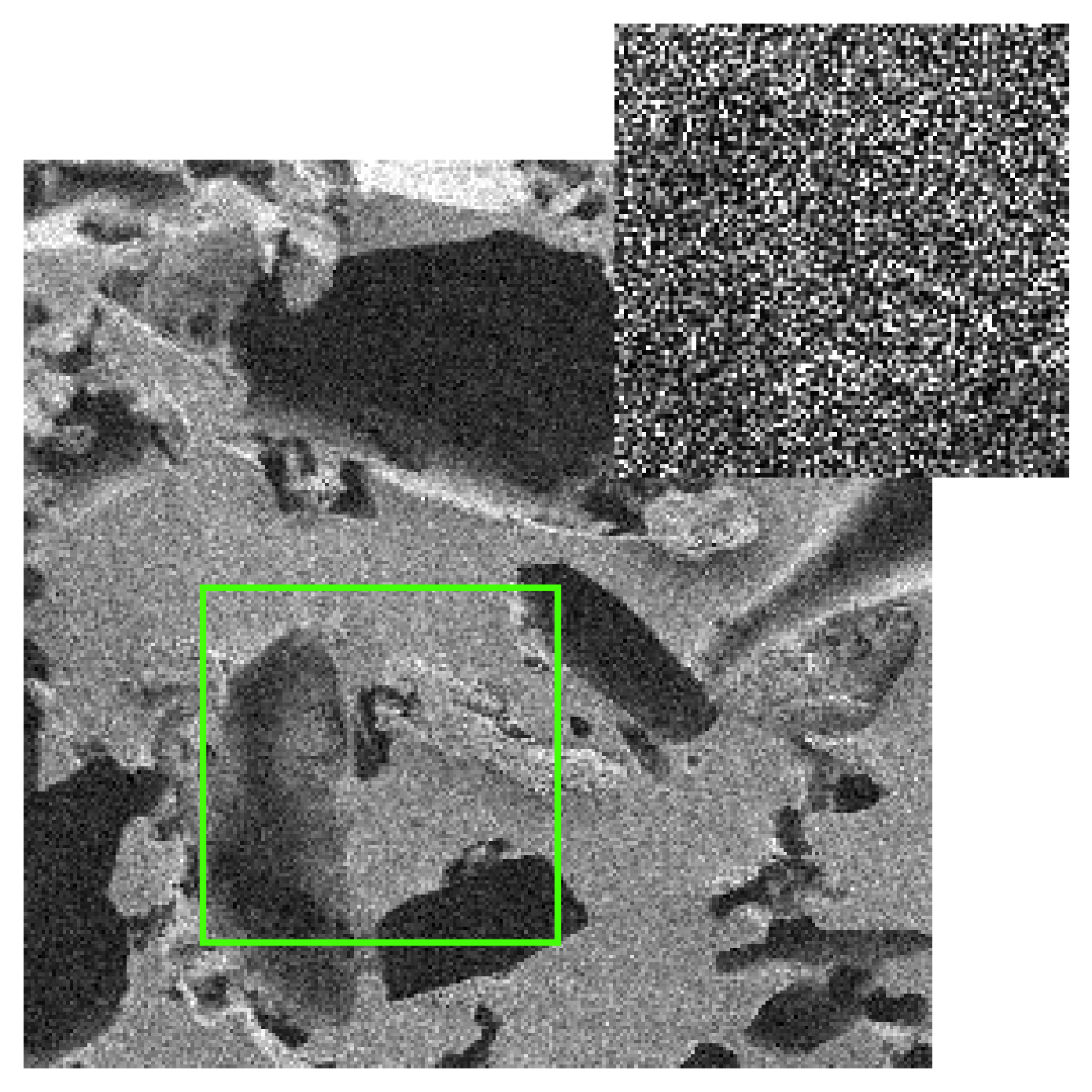}
        \captionsetup{justification=centering}
        \caption*{$\mathrm{RMSE} = 0.495$\\ $\mathrm{SSIM} = 0.536$}
    \end{subfigure} &
    \begin{subfigure}[t]{1.01\restcolwidth}
        \includegraphics[height=\figHeight]{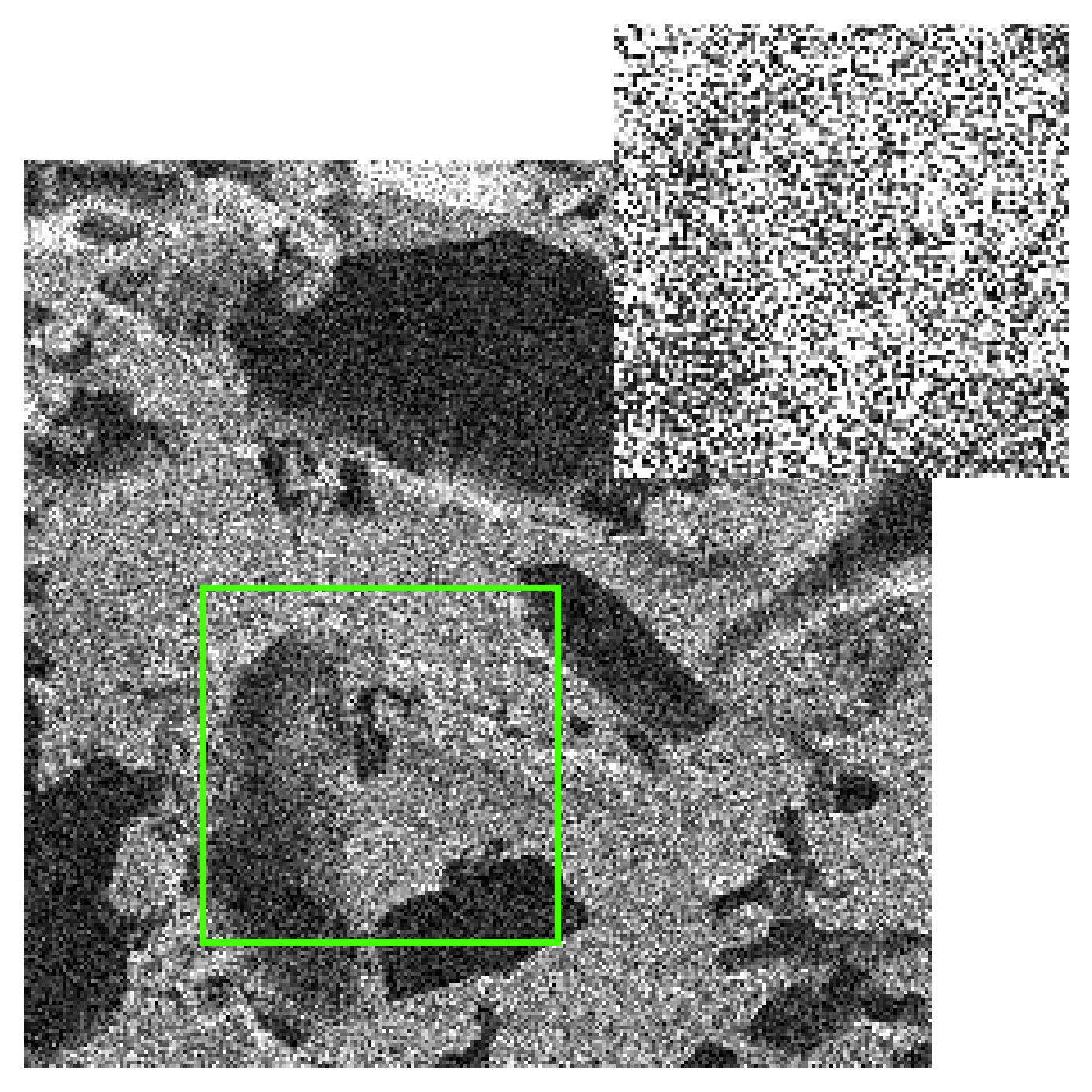}
        \captionsetup{justification=centering}
        \caption*{$\mathrm{RMSE} = 1.171$\\ $\mathrm{SSIM} = 0.261$}
    \end{subfigure} &
    \begin{subfigure}[t]{\restcolwidth}
        \includegraphics[height=\figHeight]{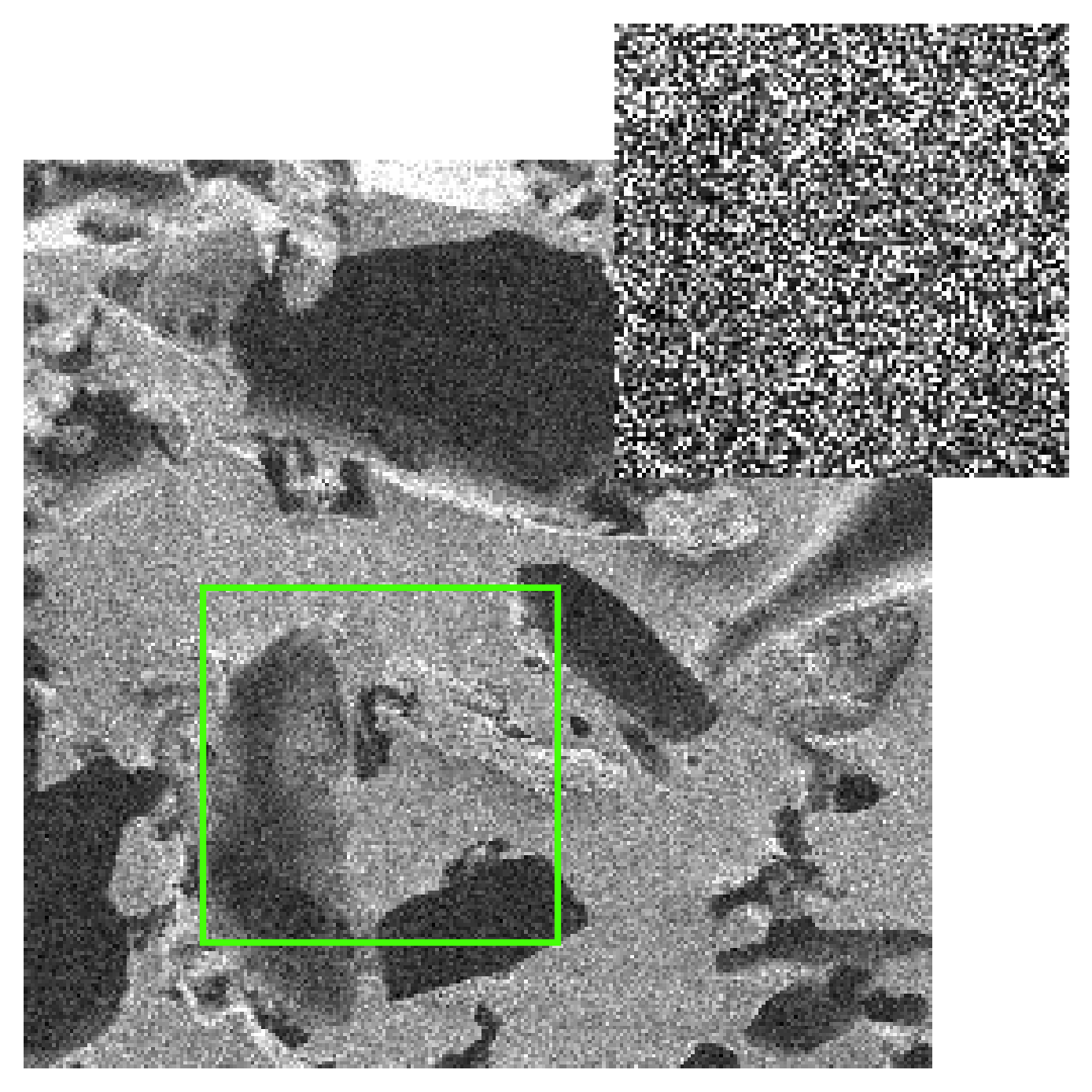}
        \captionsetup{justification=centering}
        \caption*{$\mathrm{RMSE} = 0.649$\\ $\mathrm{SSIM} = 0.487$}
    \end{subfigure} &
    \begin{subfigure}[t]{\restcolwidth}
        \includegraphics[height=\figHeight]{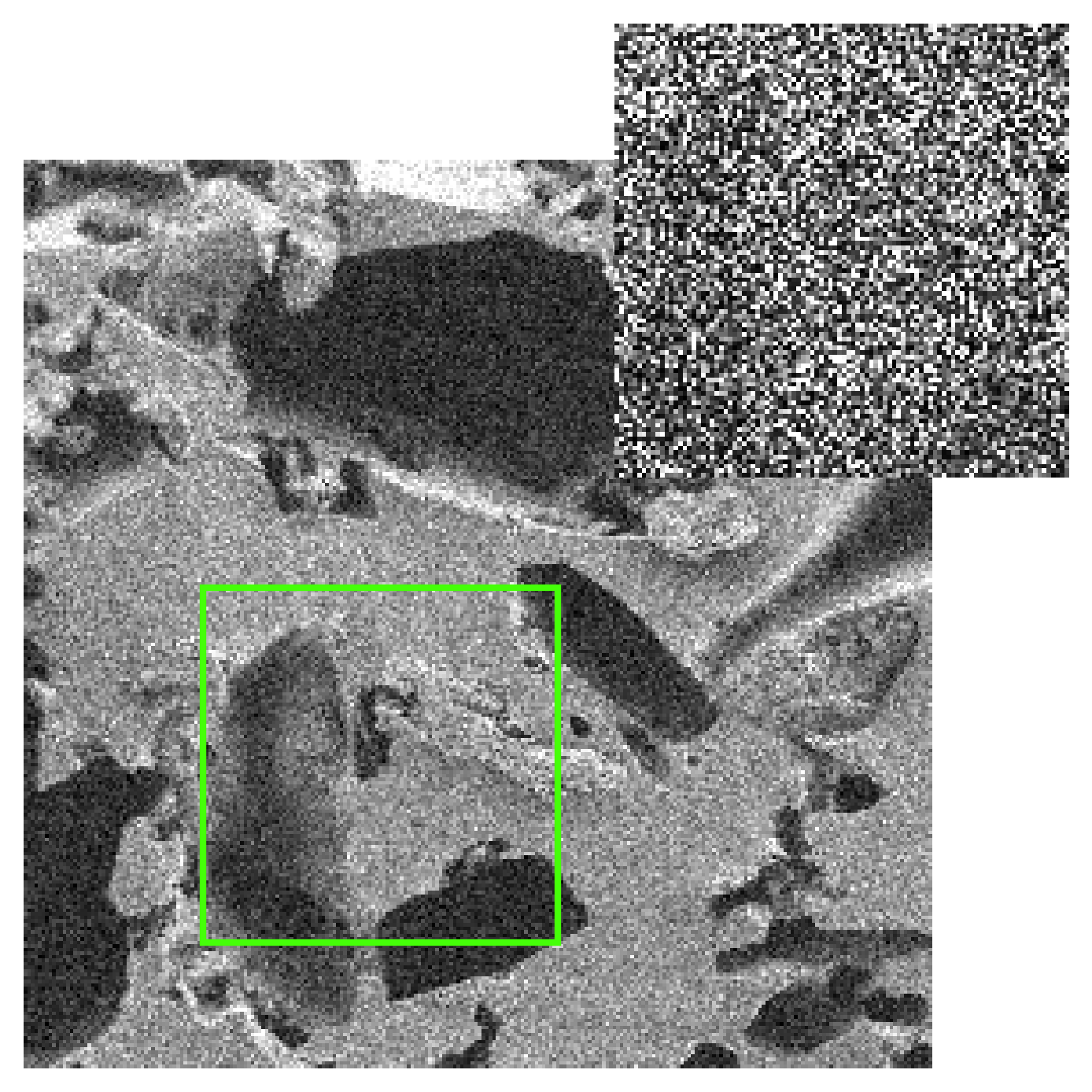}
        \captionsetup{justification=centering}
        \caption*{$\mathrm{RMSE} = 0.644$\\ $\mathrm{SSIM} = 0.477$}
    \end{subfigure} &
    \begin{subfigure}[t]{\restcolwidth}
        \includegraphics[height=\figHeight]{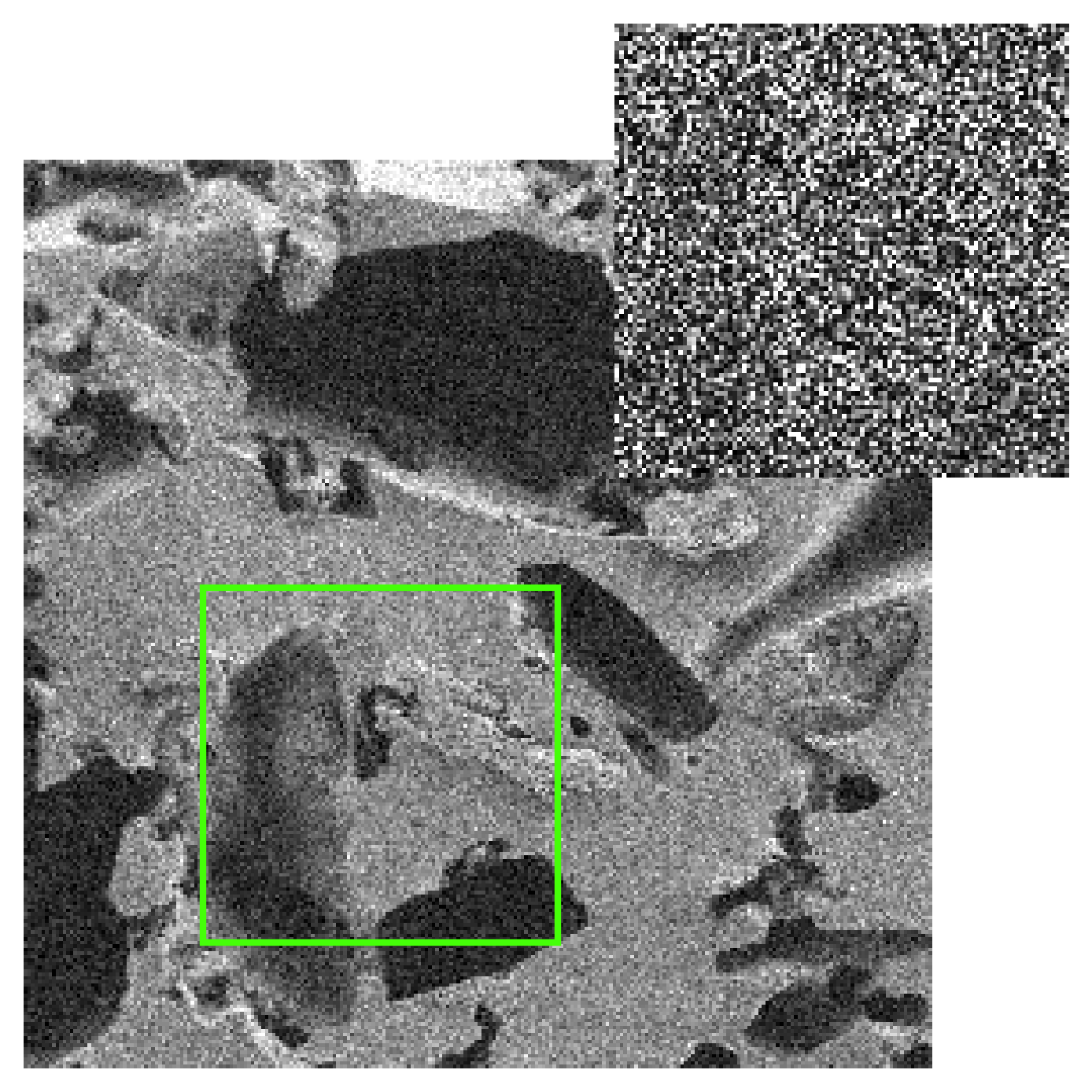}
        \captionsetup{justification=centering}
        \caption*{$\mathrm{RMSE} = 0.561$\\ $\mathrm{SSIM} = 0.493$}
    \end{subfigure}
    \\[5ex]
    \hline
    \\[-1.3ex]
    \multicolumn{3}{l}{{\small \bf PnP estimators}}
    \\
    & {\small Na\"ive}
    & {\small Oracle}
    & {\small Gaussian}
    & {\small QM}
    & {\small LQM}
    & {\small TRML}
    \\
    \rotatebox[origin=l]{90}{\qquad \quad \small TV}
    &
    \begin{subfigure}[t]{\firstcolwidth}
        \includegraphics[height=\figHeight]{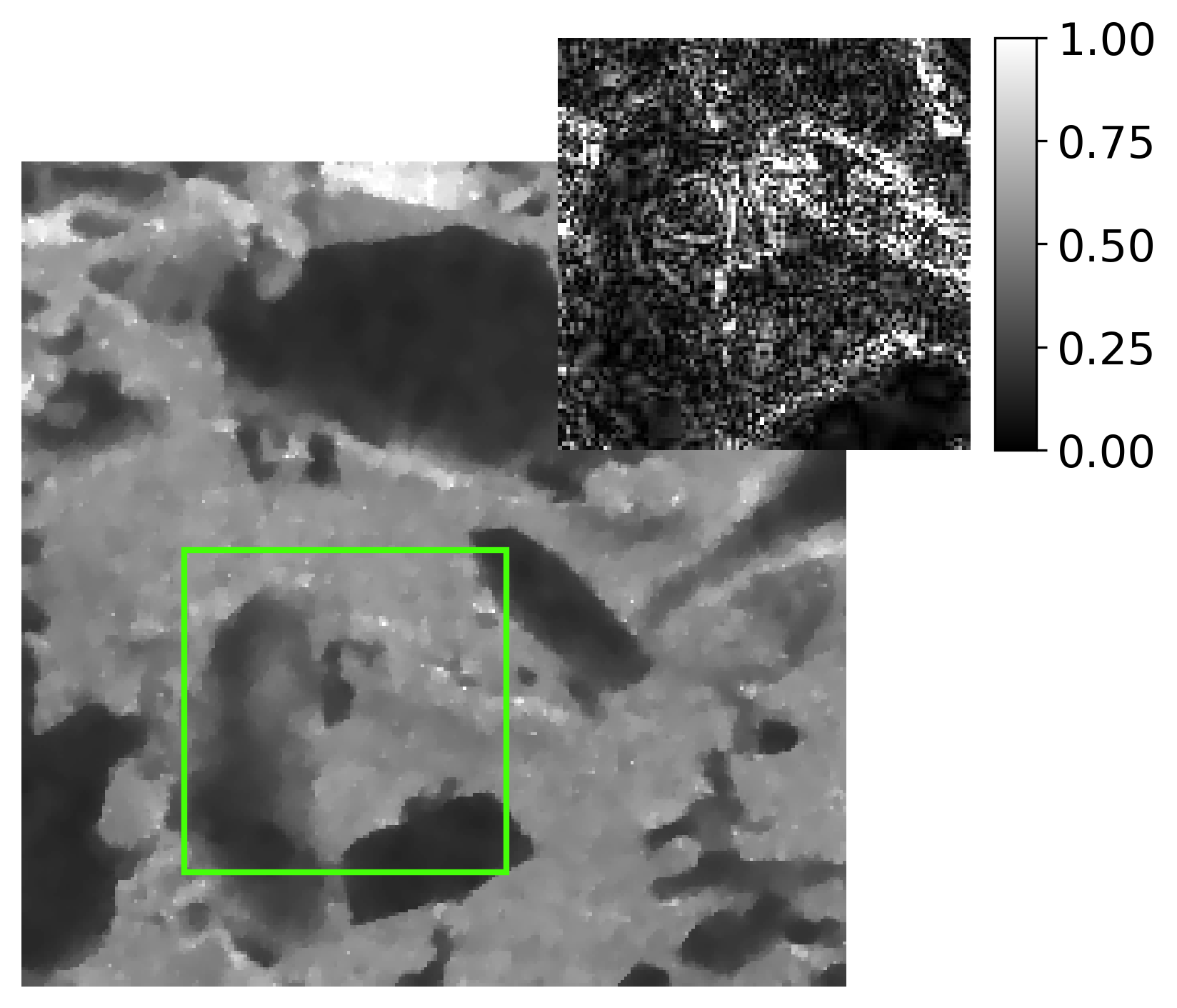}
        \captionsetup{justification=centering}
        \caption*{$\mathrm{RMSE} = 0.391$\\ $\mathrm{SSIM} = 0.650$}
    \end{subfigure} & 
    \begin{subfigure}[t]{\restcolwidth}
        \includegraphics[height=1.01\figHeight]{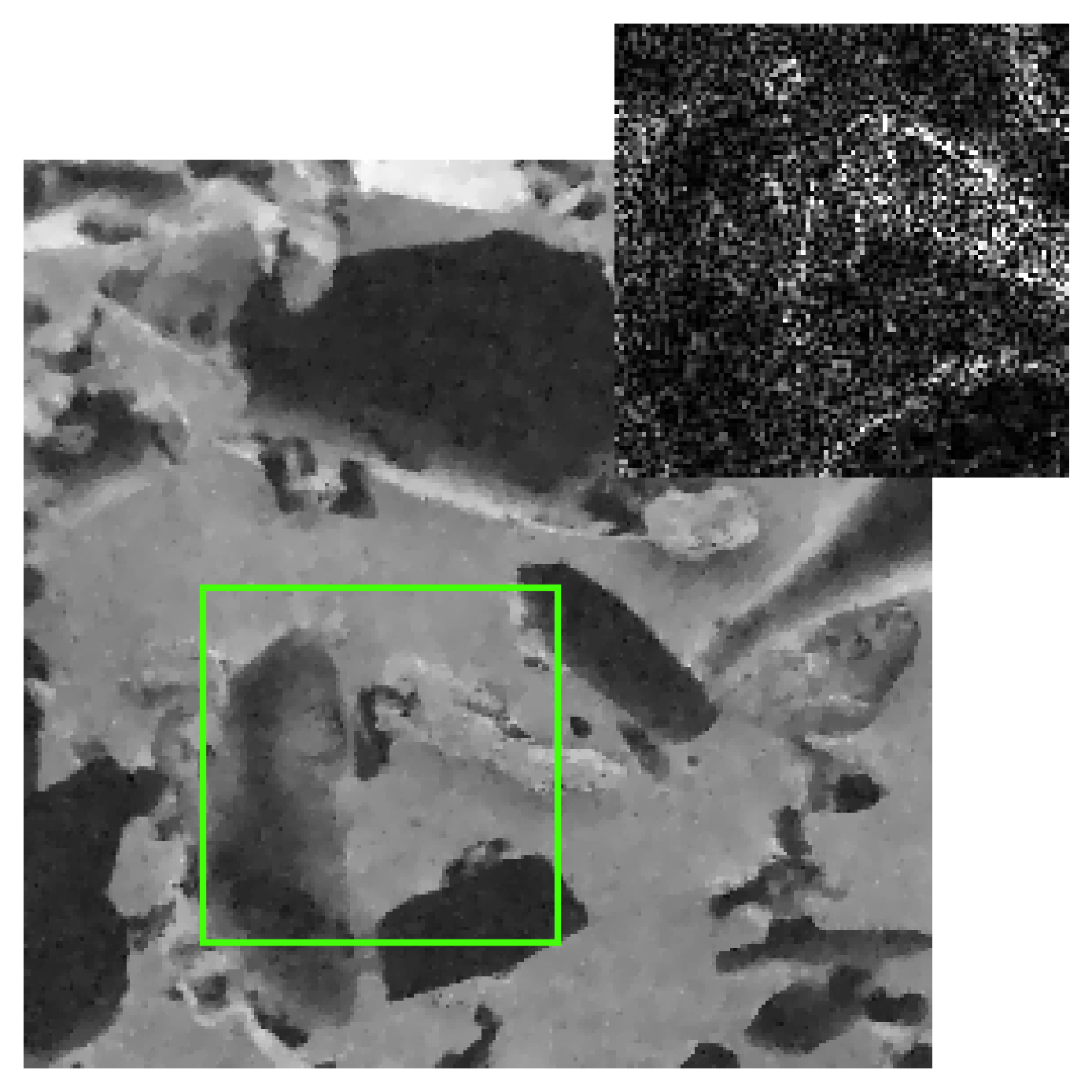}
        \captionsetup{justification=centering}
        \caption*{$\mathrm{RMSE} = 0.263$\\ $\mathrm{SSIM} = 0.765$}
    \end{subfigure} &
    \begin{subfigure}[t]{\restcolwidth}
        \includegraphics[height=\figHeight]{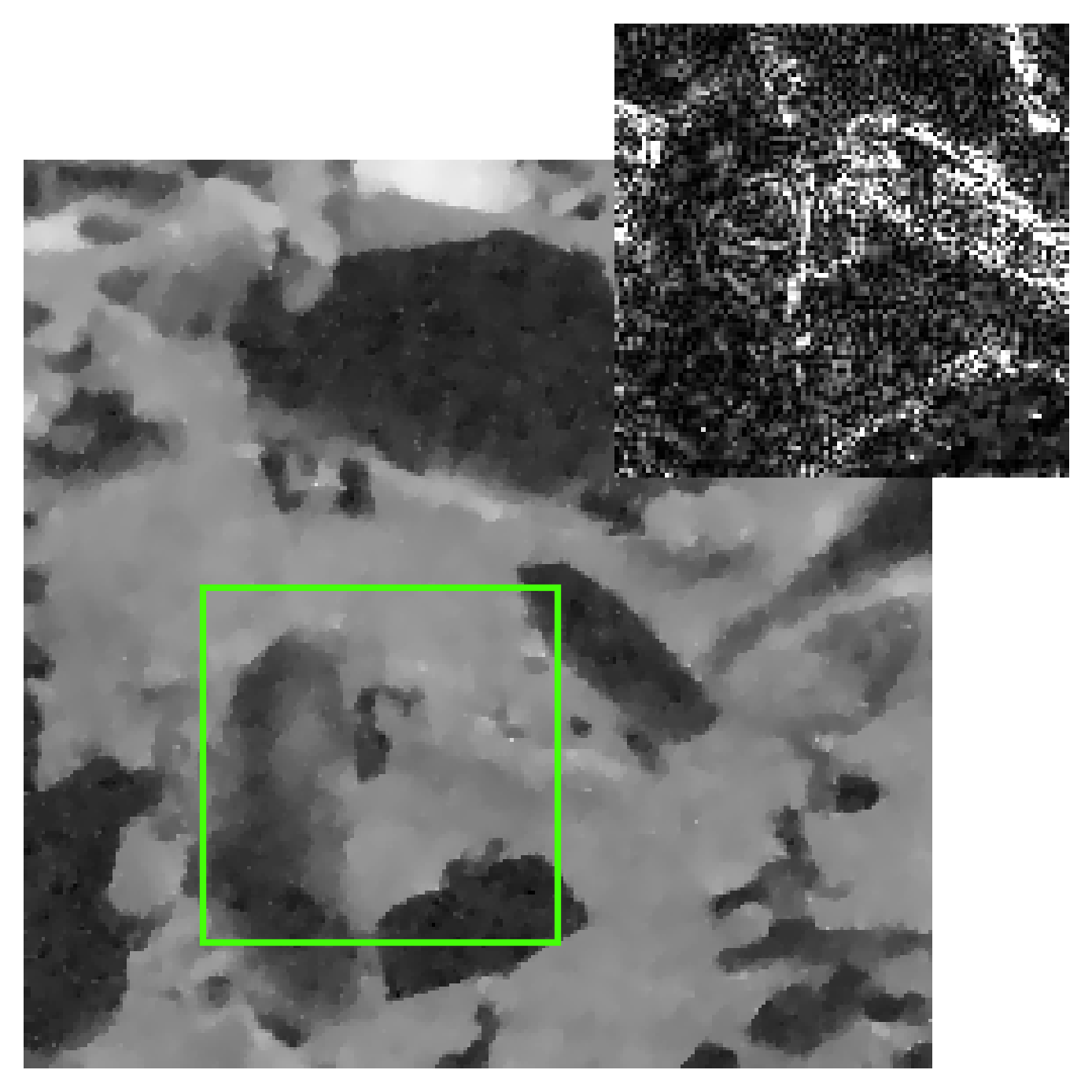}
        \captionsetup{justification=centering}
        \caption*{$\mathrm{RMSE} = 0.394$\\ $\mathrm{SSIM} = 0.619$}
    \end{subfigure} & 
    \begin{subfigure}[t]{\restcolwidth}
        \includegraphics[height=\figHeight]{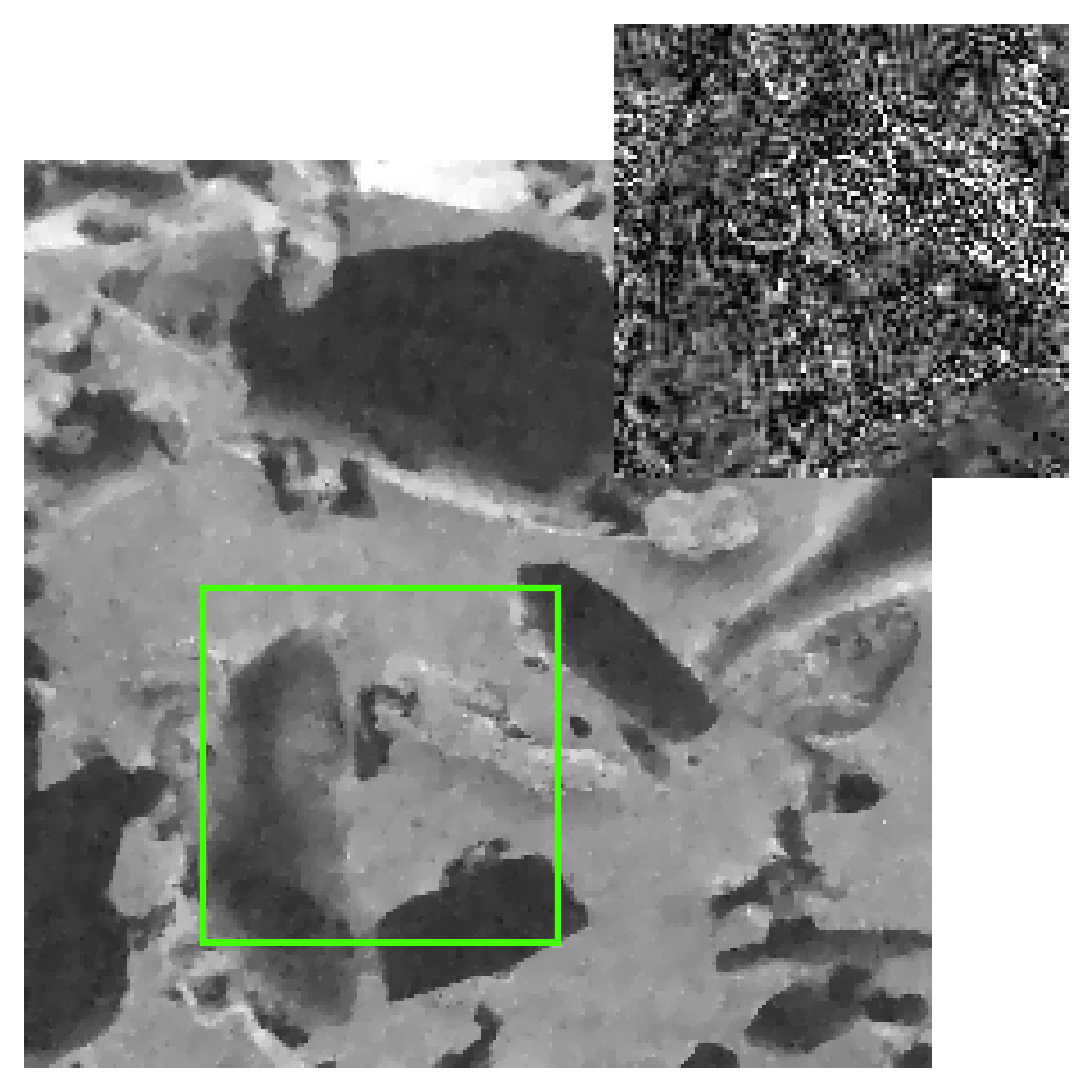}
        \captionsetup{justification=centering}
        \caption*{$\mathrm{RMSE} = 0.396$\\ $\mathrm{SSIM} = 0.755$}
    \end{subfigure} &
    \begin{subfigure}[t]{\restcolwidth}
        \includegraphics[height=\figHeight]{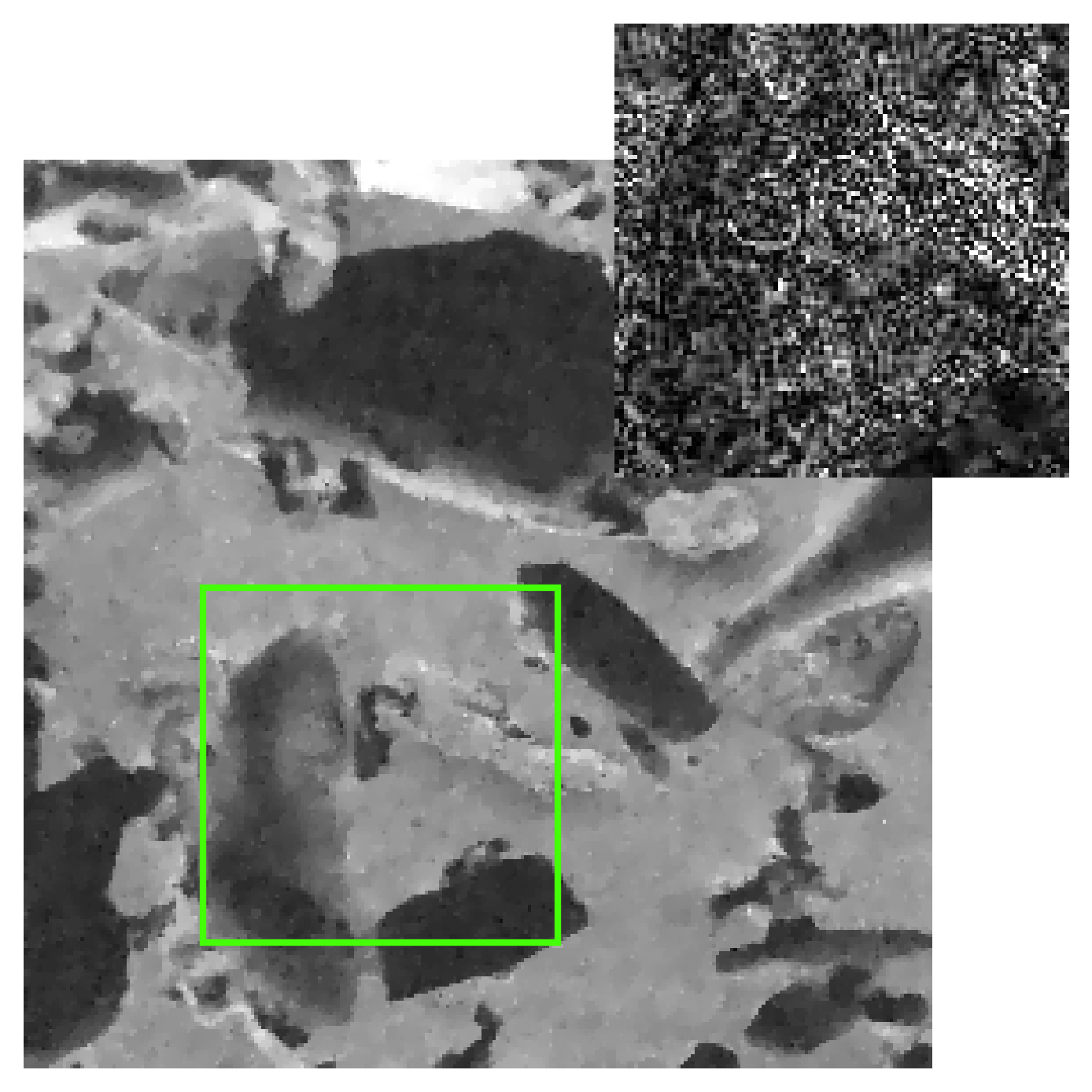}
        \captionsetup{justification=centering}
        \caption*{$\mathrm{RMSE} = 0.358$\\ $\mathrm{SSIM} = 0.751$}
    \end{subfigure} & 
    \begin{subfigure}[t]{\restcolwidth}
        \includegraphics[height=\figHeight]{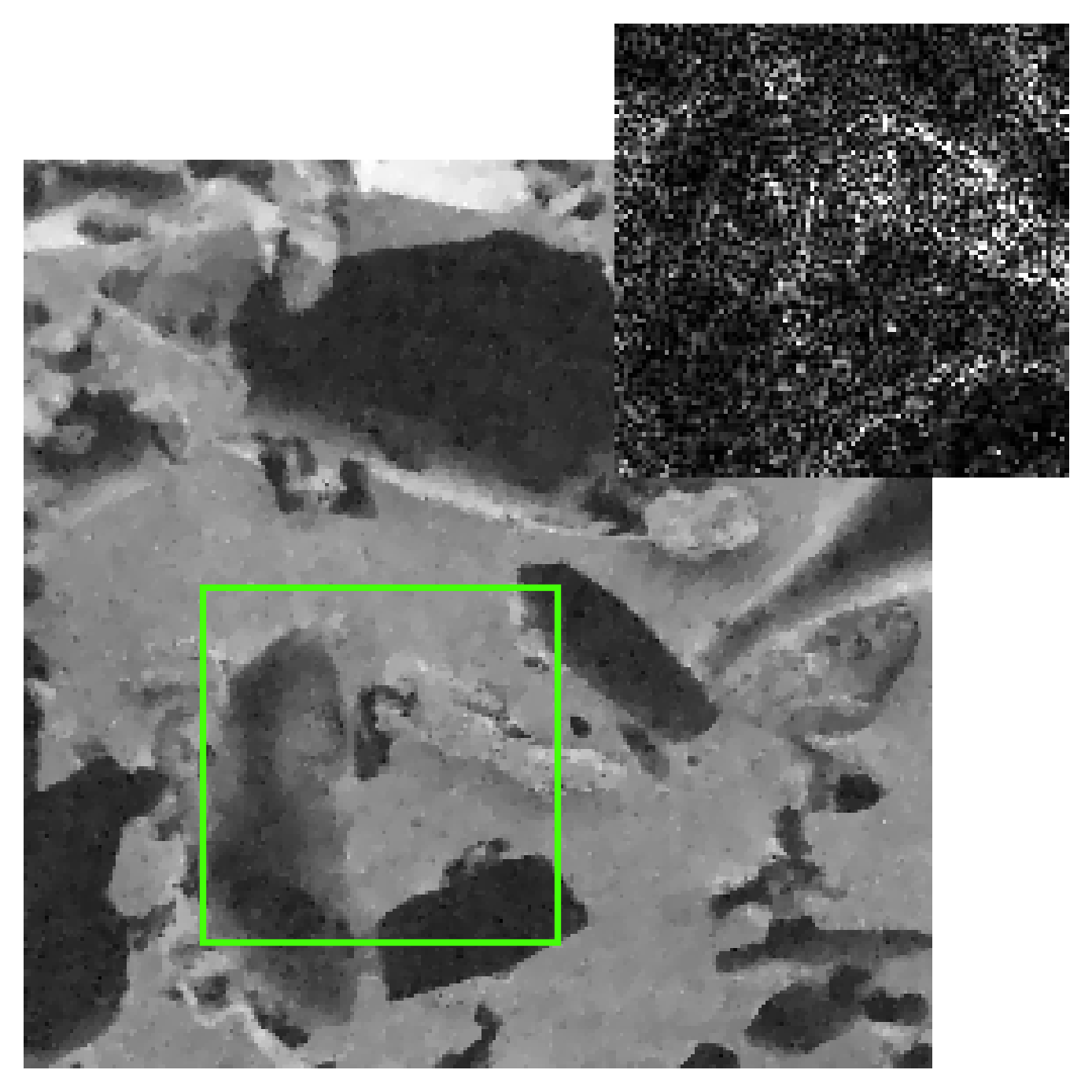}
        \captionsetup{justification=centering}
        \caption*{$\mathrm{RMSE} = 0.275$\\ $\mathrm{SSIM} = 0.746$}
    \end{subfigure}
    \\
    \rotatebox[origin=l]{90}{\qquad \small BM3D}
    &
    \begin{subfigure}[t]{\firstcolwidth}
        \includegraphics[height=\figHeight]{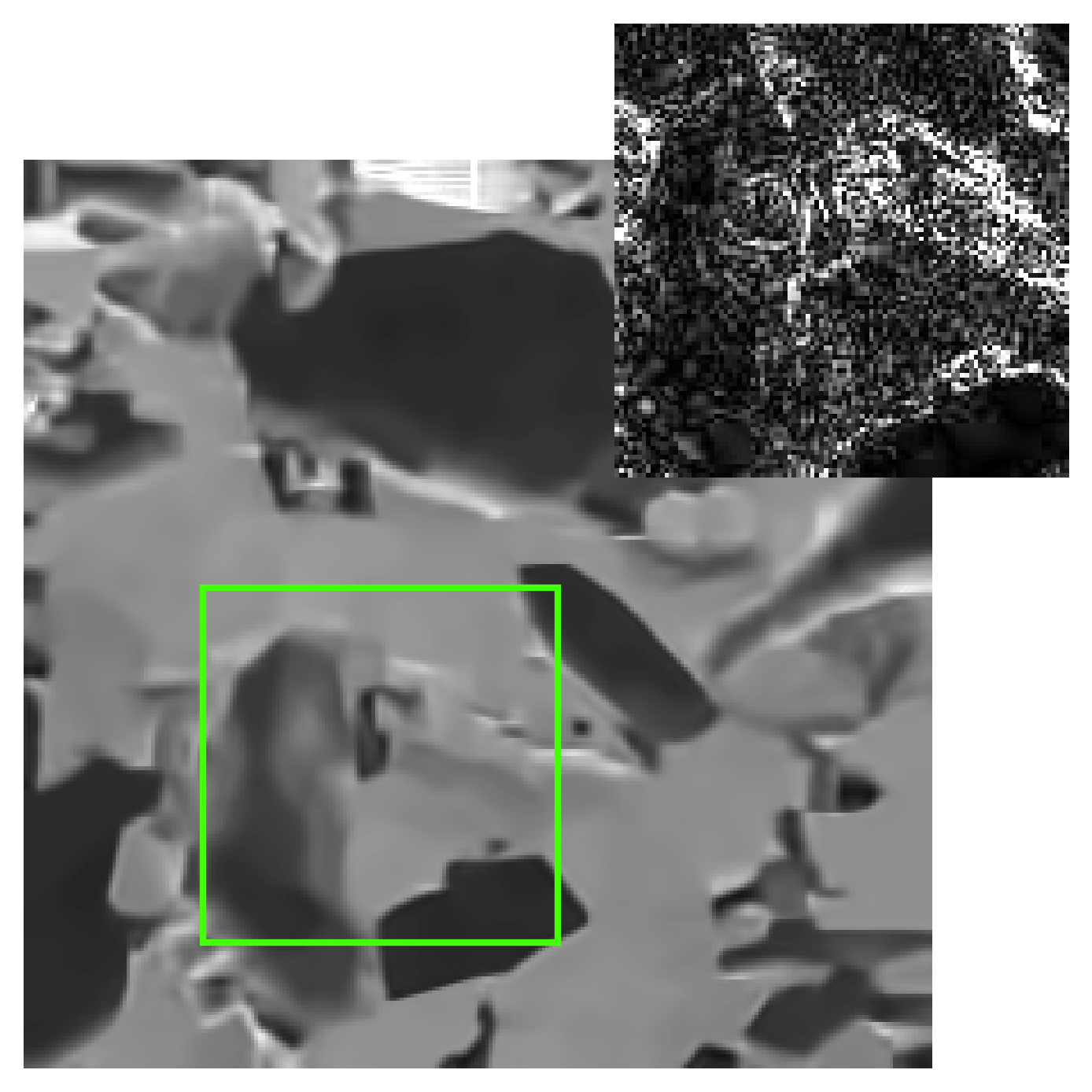}
        \captionsetup{justification=centering}
        \caption*{$\mathrm{RMSE} = 0.371$\\ $\mathrm{SSIM} = 0.667$}
    \end{subfigure} & 
    \begin{subfigure}[t]{\restcolwidth}
        \includegraphics[height=\figHeight]{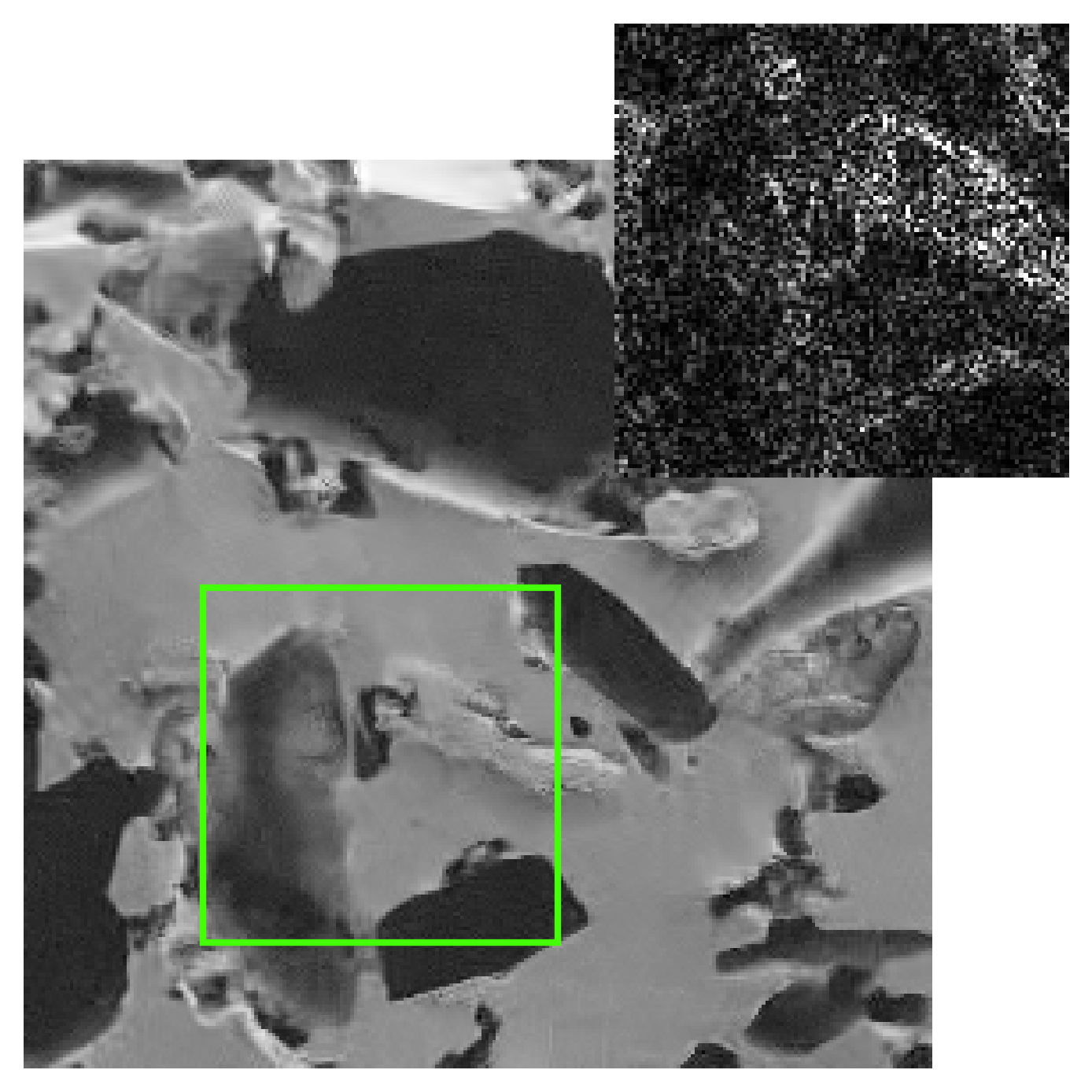}
        \captionsetup{justification=centering}
        \caption*{$\mathrm{RMSE} = 0.248$\\ $\mathrm{SSIM} = 0.775$}
    \end{subfigure} &
    \begin{subfigure}[t]{\restcolwidth}
        \includegraphics[height=\figHeight]{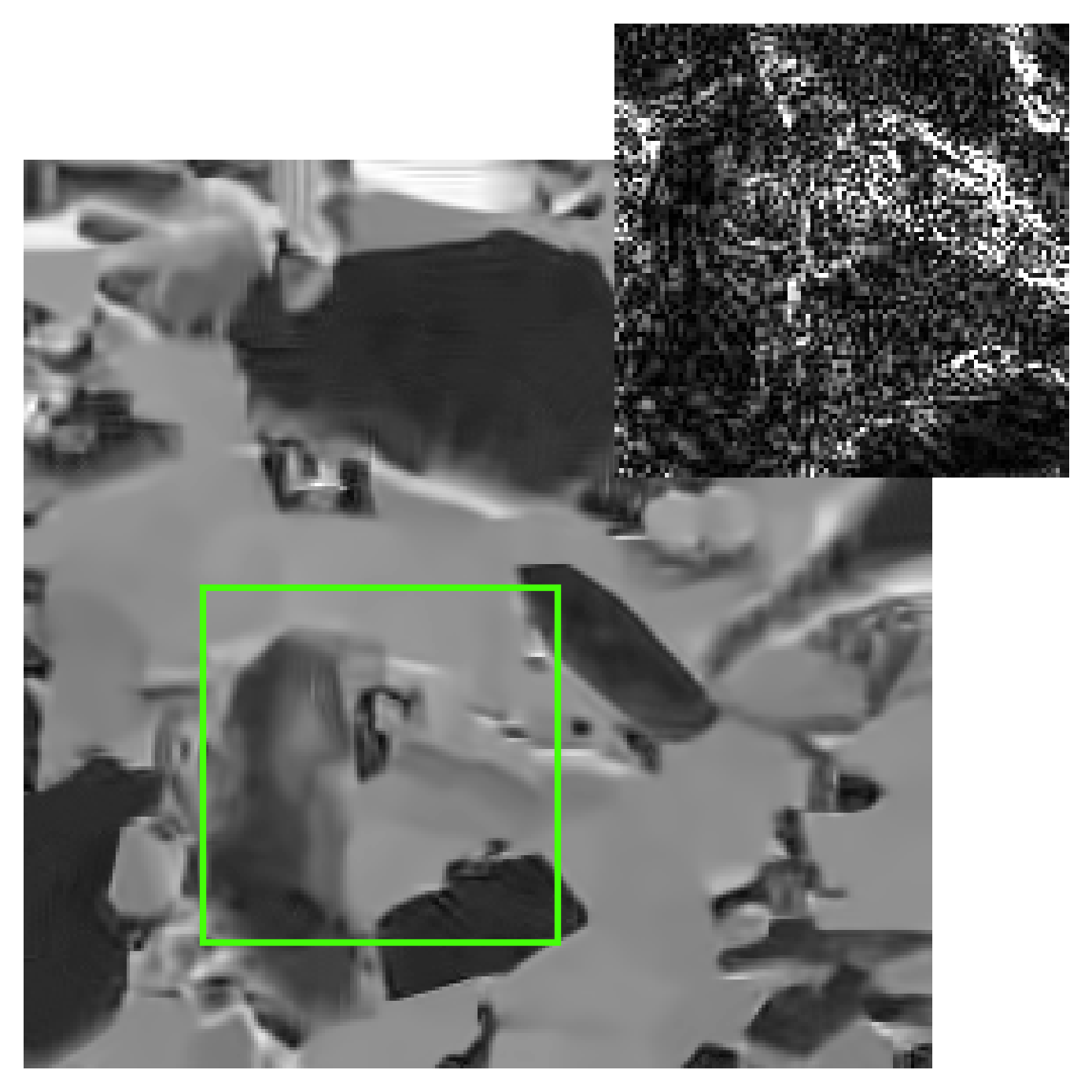}
        \captionsetup{justification=centering}
        \caption*{$\mathrm{RMSE} = 0.363$\\ $\mathrm{SSIM} = 0.675$}
    \end{subfigure} & 
    \begin{subfigure}[t]{\restcolwidth}
        \includegraphics[height=\figHeight]{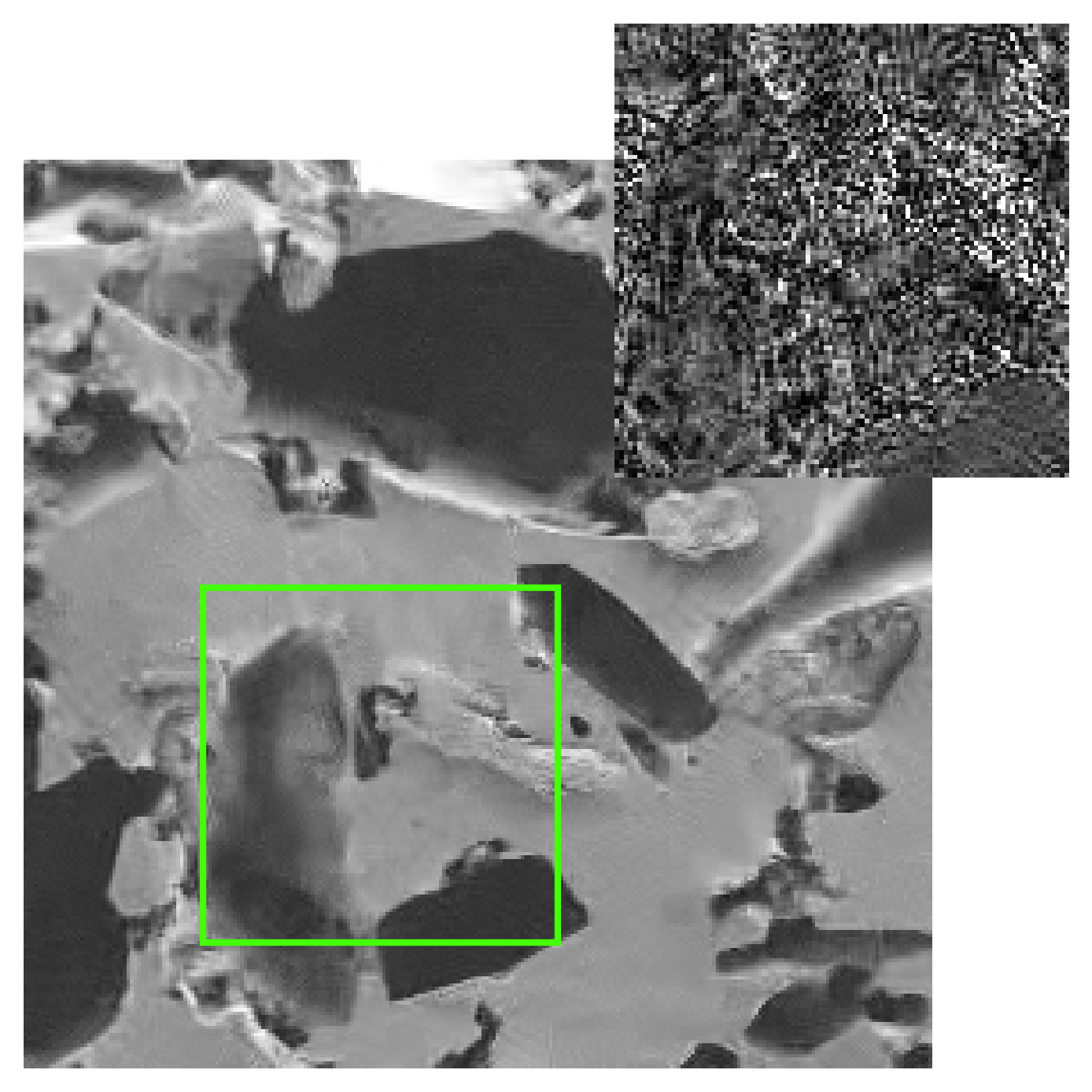}
        \captionsetup{justification=centering}
        \caption*{$\mathrm{RMSE} = 0.386$\\ $\mathrm{SSIM} = 0.776$}
    \end{subfigure} & 
    \begin{subfigure}[t]{\restcolwidth}
        \includegraphics[height=\figHeight]{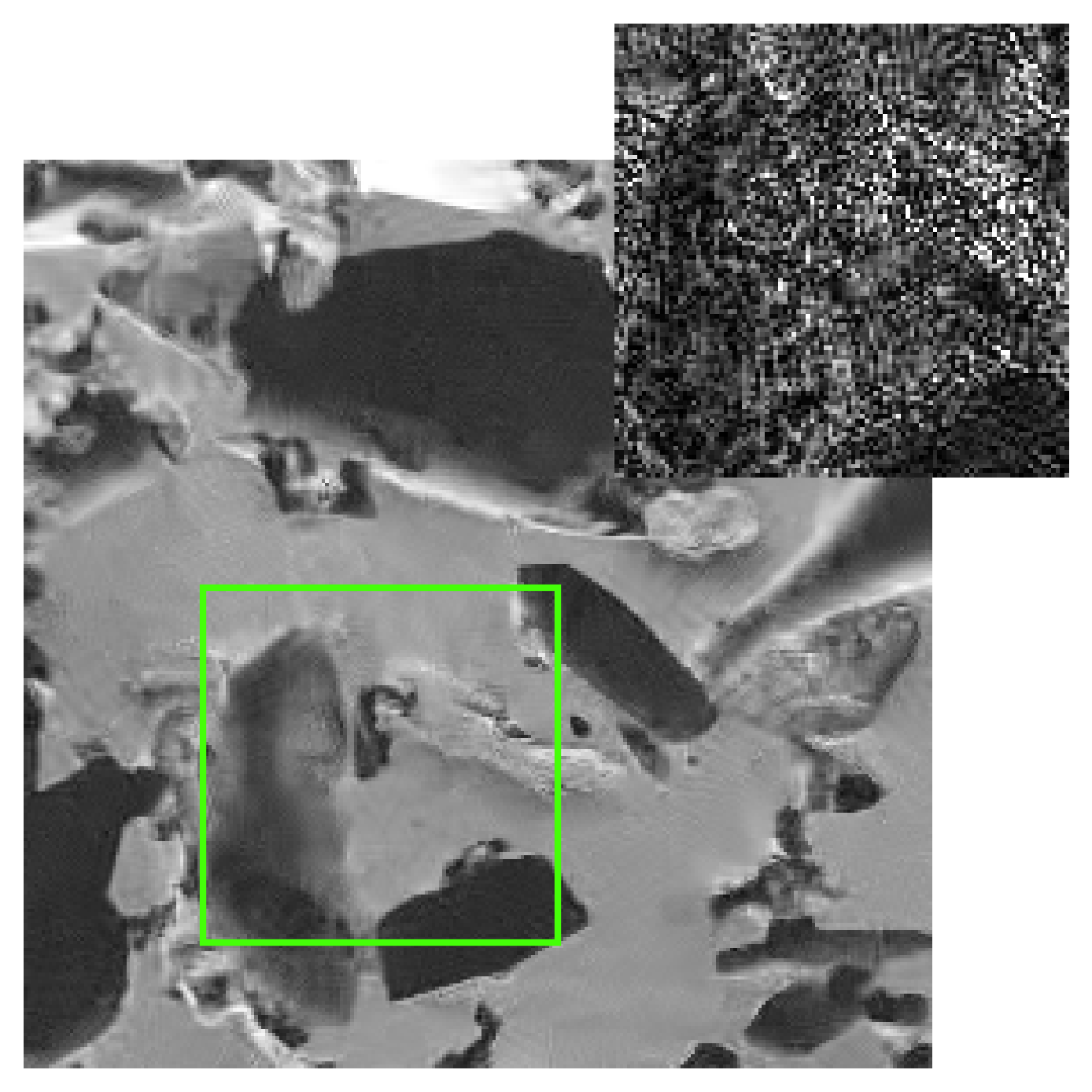}
        \captionsetup{justification=centering}
        \caption*{$\mathrm{RMSE} = 0.350$\\ $\mathrm{SSIM} = 0.770$}
    \end{subfigure} & 
    \begin{subfigure}[t]{\restcolwidth}
        \includegraphics[height=\figHeight]{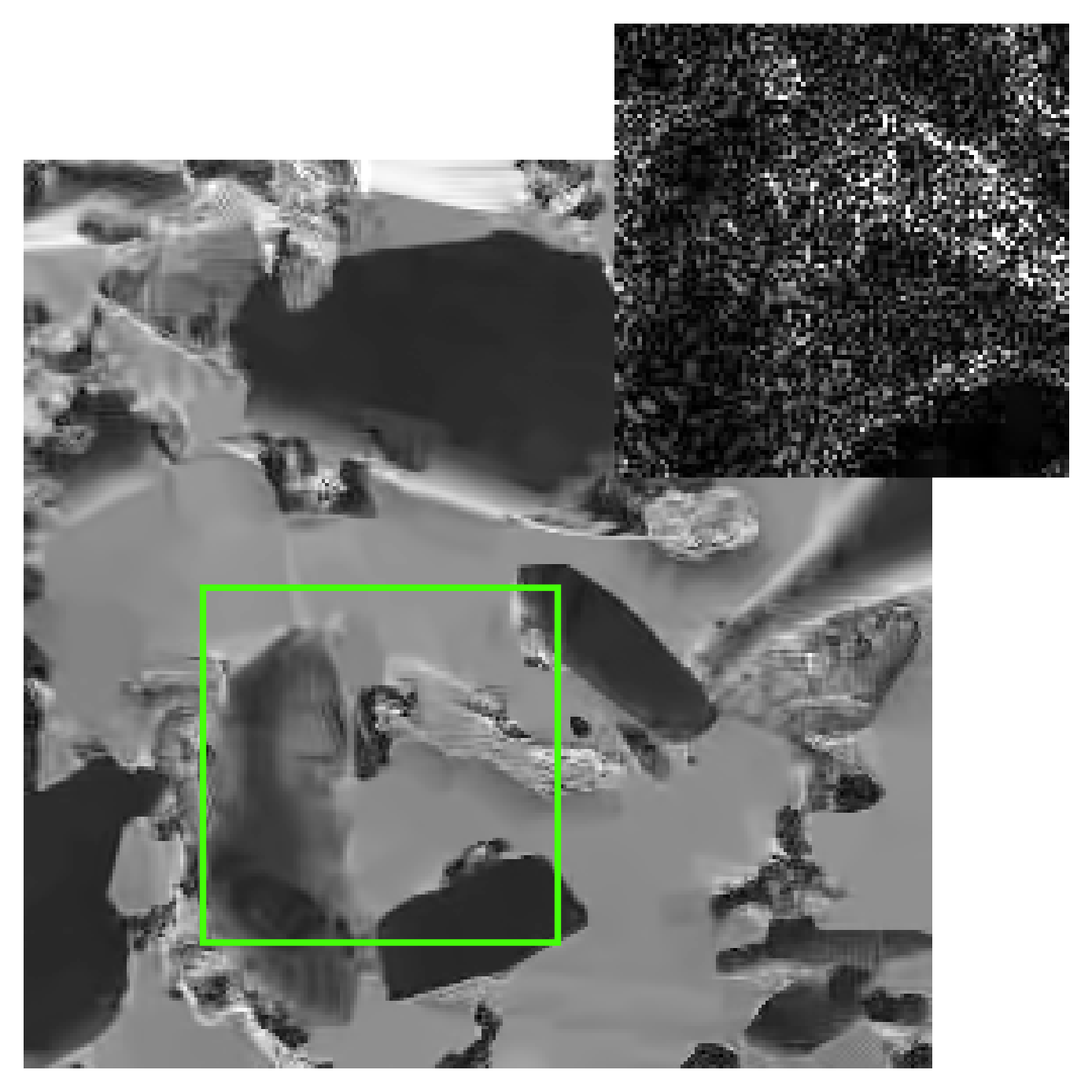}
        \captionsetup{justification=centering}
        \caption*{$\mathrm{RMSE} = 0.265$\\ $\mathrm{SSIM} = 0.776$}
    \end{subfigure}
    \\
    \rotatebox[origin=l]{90}{\qquad \small DnCNN}
    &
    \begin{subfigure}[t]{\firstcolwidth}
        \includegraphics[height=\figHeight]{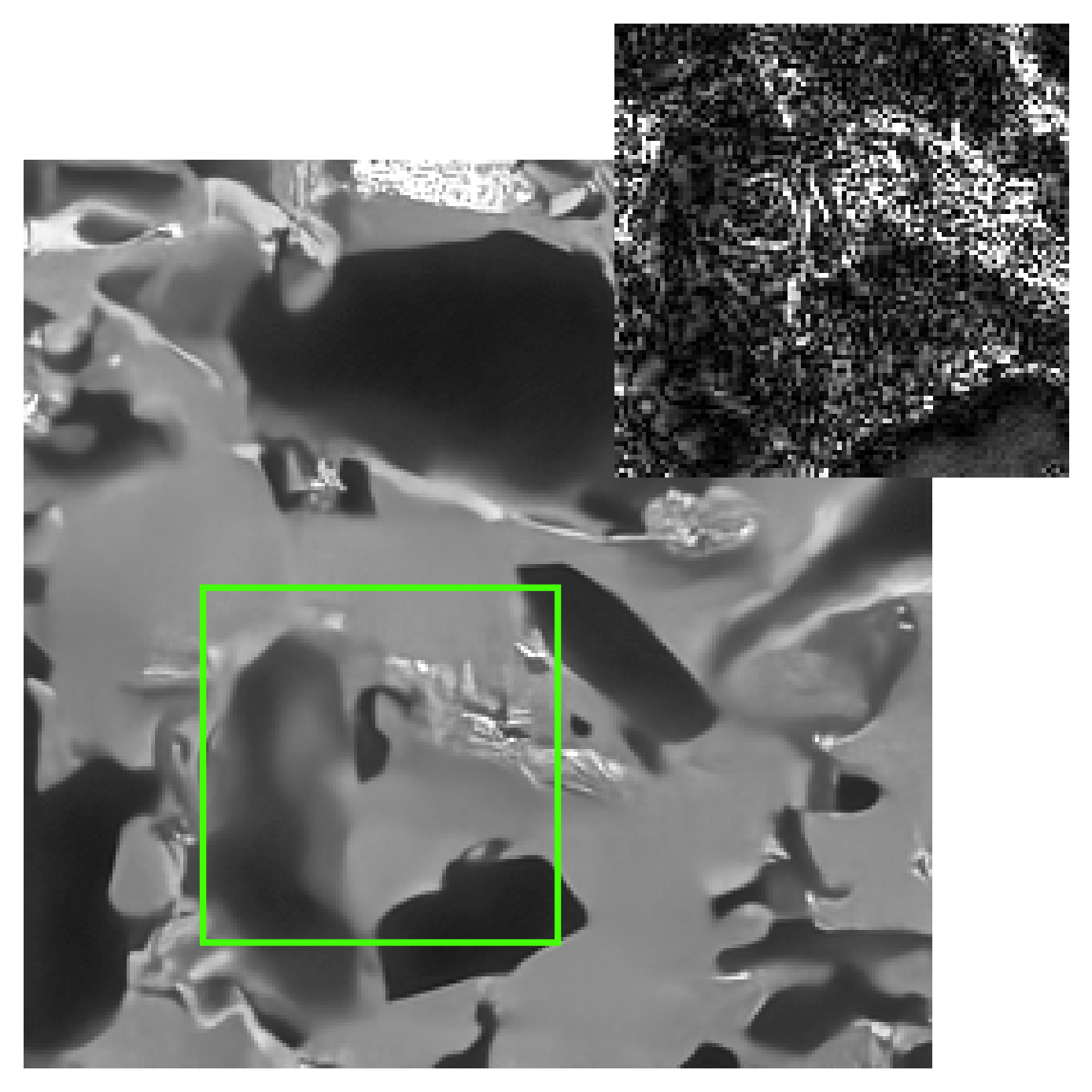}
        \captionsetup{justification=centering}
        \caption*{$\mathrm{RMSE} = 0.382$\\ $\mathrm{SSIM} = 0.678$}
    \end{subfigure} & 
    \begin{subfigure}[t]{\restcolwidth}
        \includegraphics[height=\figHeight]{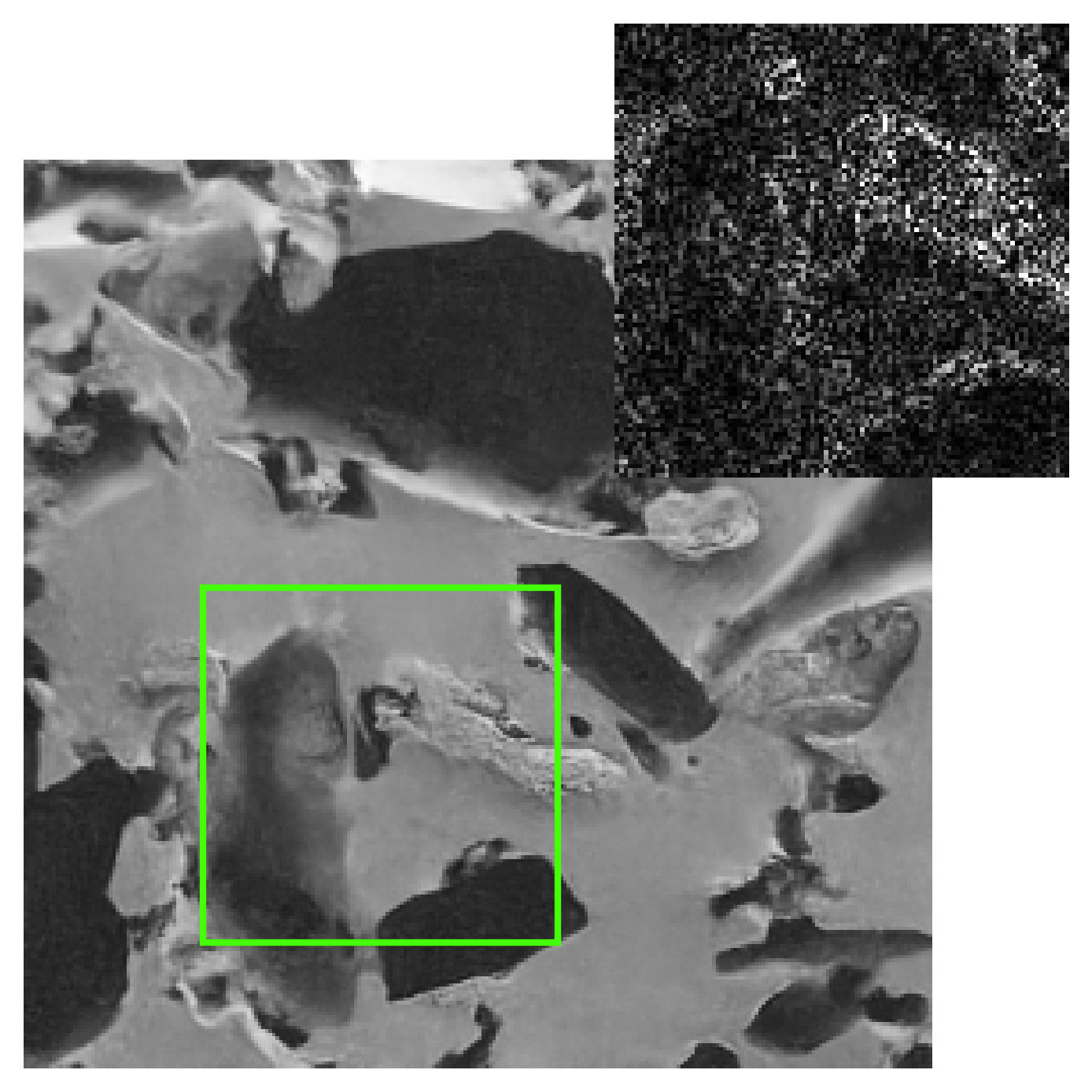}
        \captionsetup{justification=centering}
        \caption*{$\mathrm{RMSE} = 0.240$\\ $\mathrm{SSIM} = 0.788$}
    \end{subfigure} &
    \begin{subfigure}[t]{\restcolwidth}
        \includegraphics[height=\figHeight]{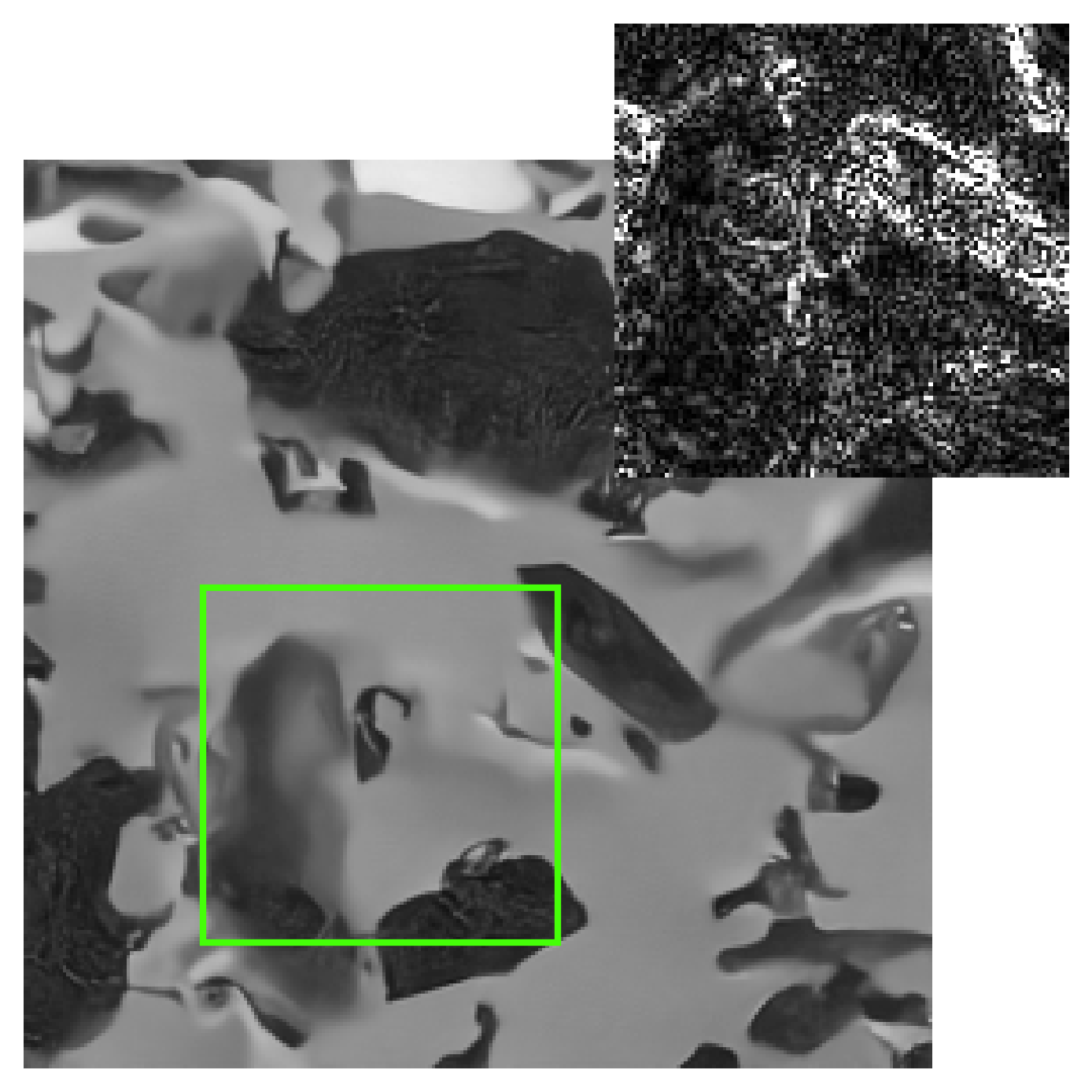}
        \captionsetup{justification=centering}
        \caption*{$\mathrm{RMSE} = 0.368$\\ $\mathrm{SSIM} = 0.633$}
    \end{subfigure} & 
    \begin{subfigure}[t]{\restcolwidth}
        \includegraphics[height=\figHeight]{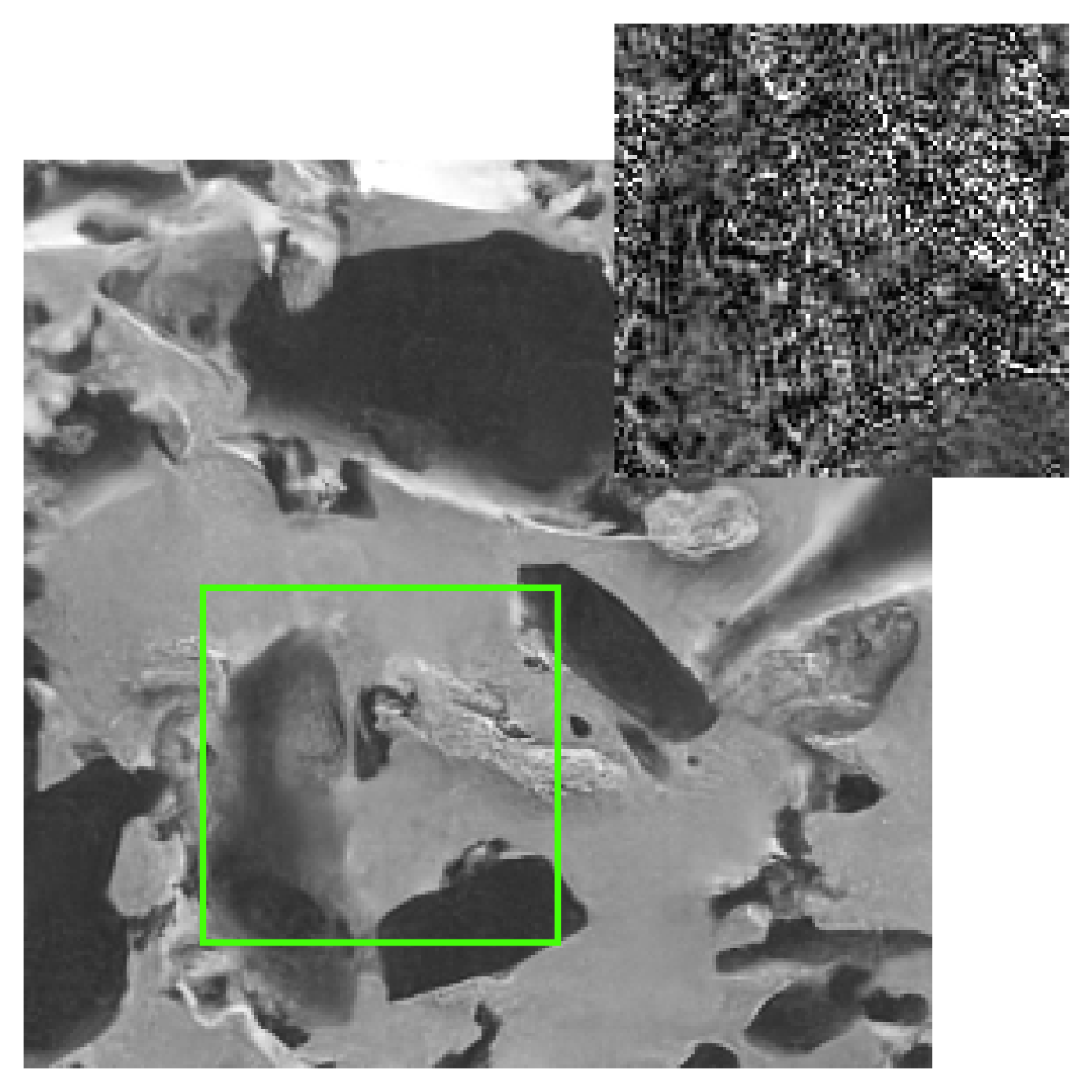}
        \captionsetup{justification=centering}
        \caption*{$\mathrm{RMSE} = 0.384$\\ $\mathrm{SSIM} = 0.784$}
    \end{subfigure} & 
    \begin{subfigure}[t]{\restcolwidth}
        \includegraphics[height=\figHeight]{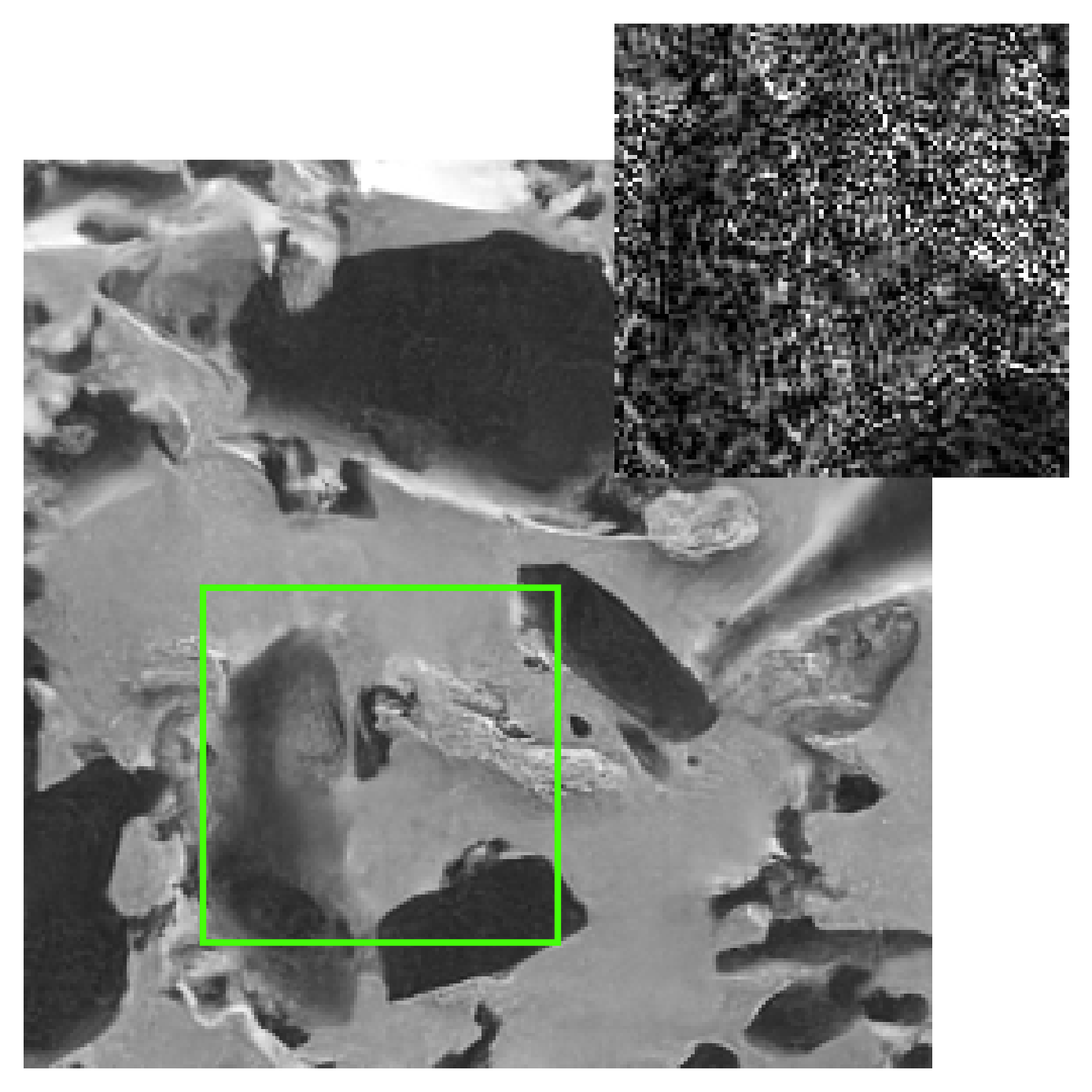}
        \captionsetup{justification=centering}
        \caption*{$\mathrm{RMSE} = 0.346$\\ $\mathrm{SSIM} = 0.780$}
    \end{subfigure} & 
    \begin{subfigure}[t]{\restcolwidth}
        \includegraphics[height=\figHeight]{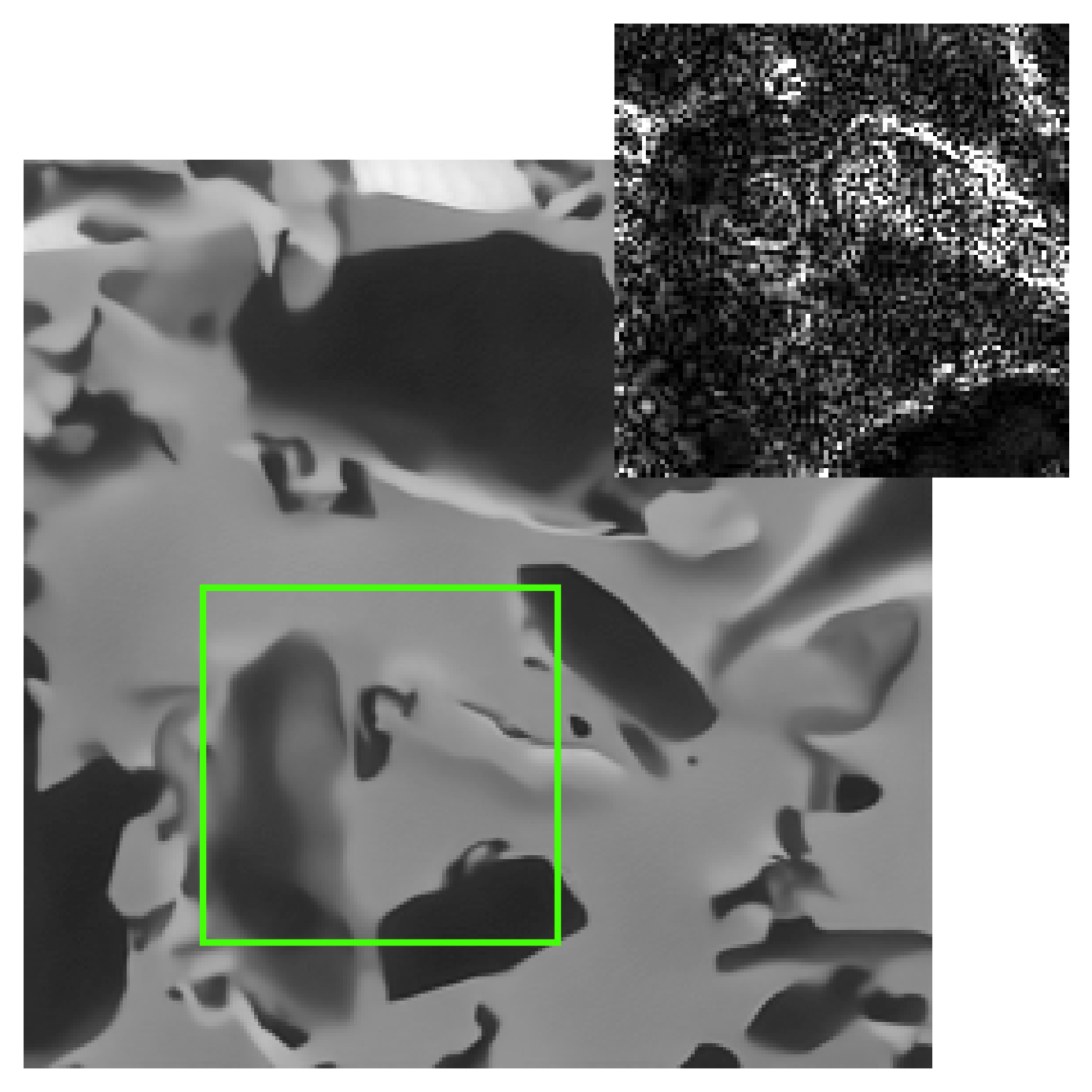}
        \captionsetup{justification=centering}
        \caption*{$\mathrm{RMSE} = 0.324$\\ $\mathrm{SSIM} = 0.693$}
    \end{subfigure}
  \end{tabular}
  \caption{HIM sample with ground truth $\eta \in [2, 8]$,
  total dose $\lambda = 20$, and $n = 200$.
  All images are shown on the same scale as in the ground truth image. Absolute error images for the sub-region in each green square are included for reconstructed micrographs.} 
  \label{fig:brick}
\end{figure*}

\begin{figure*}
  \centering
  \begin{tabular}{@{}c@{\,}c@{\,\,}c@{\,}c@{\,}c@{\,}c@{\,}c@{}}
    & \multicolumn{1}{c|}{{\small \bf Ground truth}}
    & {\small Oracle}
    & {\small Conventional}
    & {\small QM}
    & {\small LQM}
    & {\small TRML}
    \\
    \rotatebox[origin=l]{90}{\quad \small No regularization}
    &
    \multicolumn{1}{l|}{
    \begin{subfigure}[t]{\firstcolwidth}
        \includegraphics[height=1.0\figHeight]{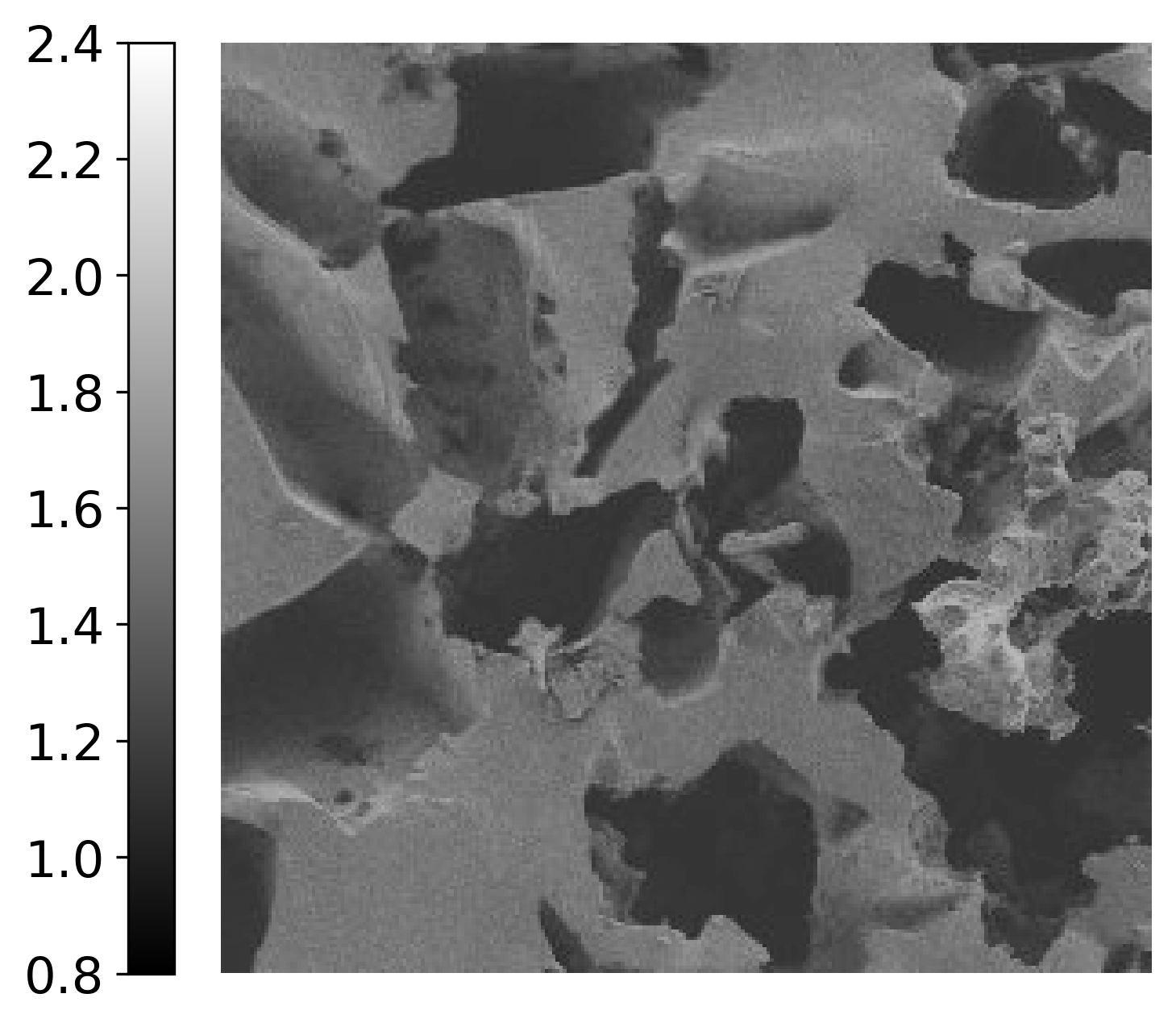}
    \end{subfigure}\,} &
    \begin{subfigure}[t]{\restcolwidth}
        \,\includegraphics[height=1.01\figHeight]{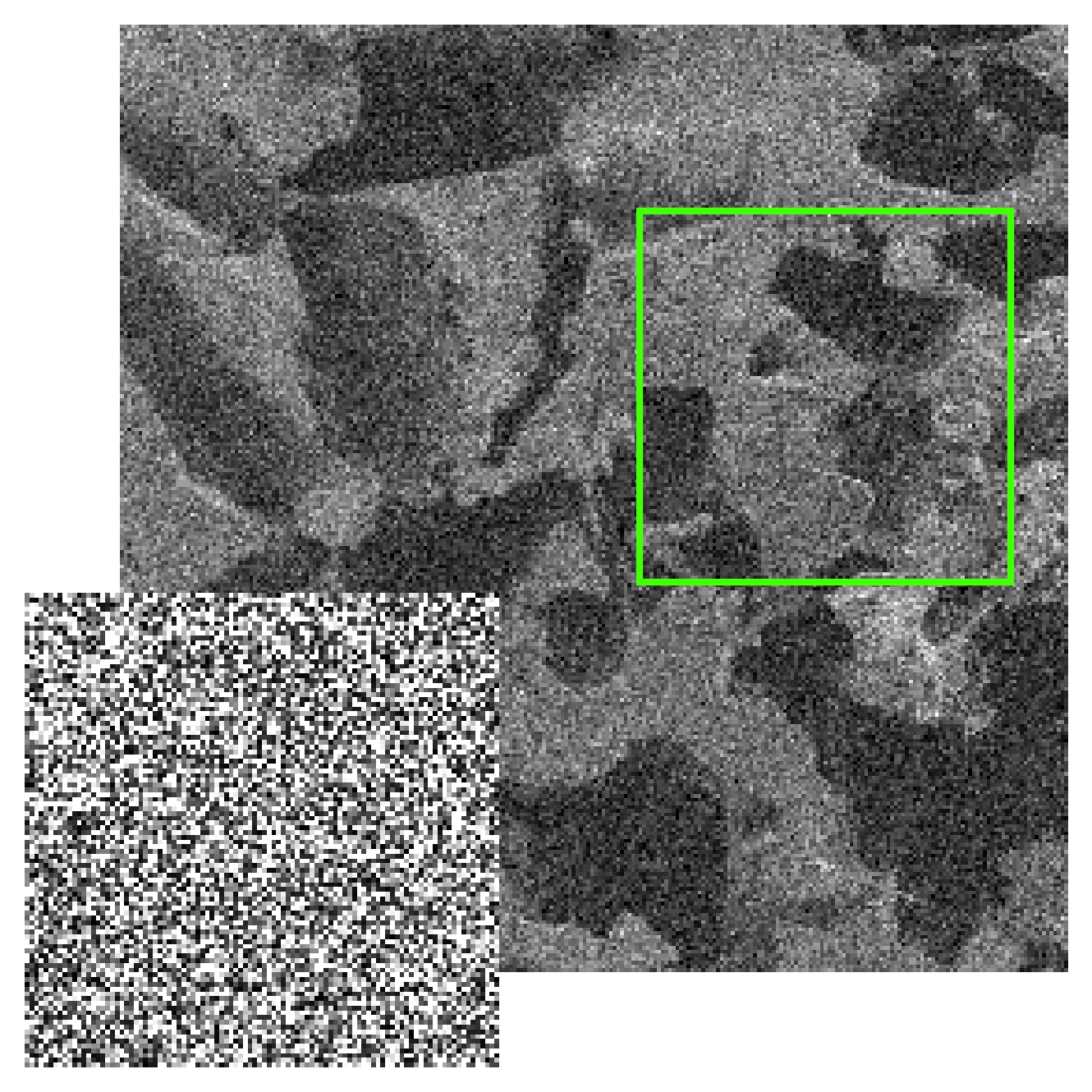}
        \captionsetup{justification=centering}
        \caption*{$\mathrm{RMSE} = 0.169$\\ $\mathrm{SSIM} = 0.273$}
    \end{subfigure} &
    \begin{subfigure}[t]{1.01\restcolwidth}
        \includegraphics[height=\figHeight]{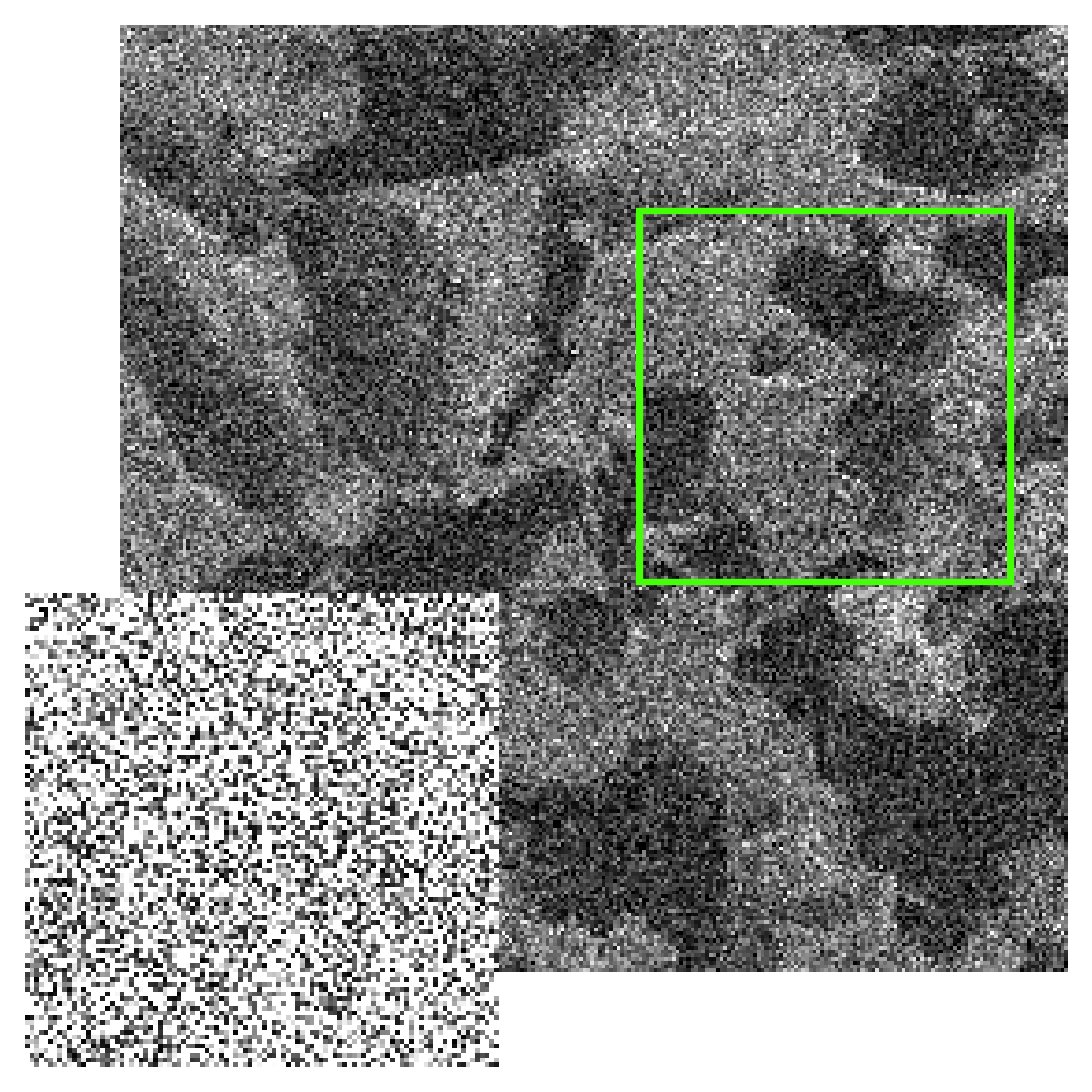}
        \captionsetup{justification=centering}
        \caption*{$\mathrm{RMSE} = 0.259$\\ $\mathrm{SSIM} = 0.161$}
    \end{subfigure} &
    \begin{subfigure}[t]{\restcolwidth}
        \includegraphics[height=\figHeight]{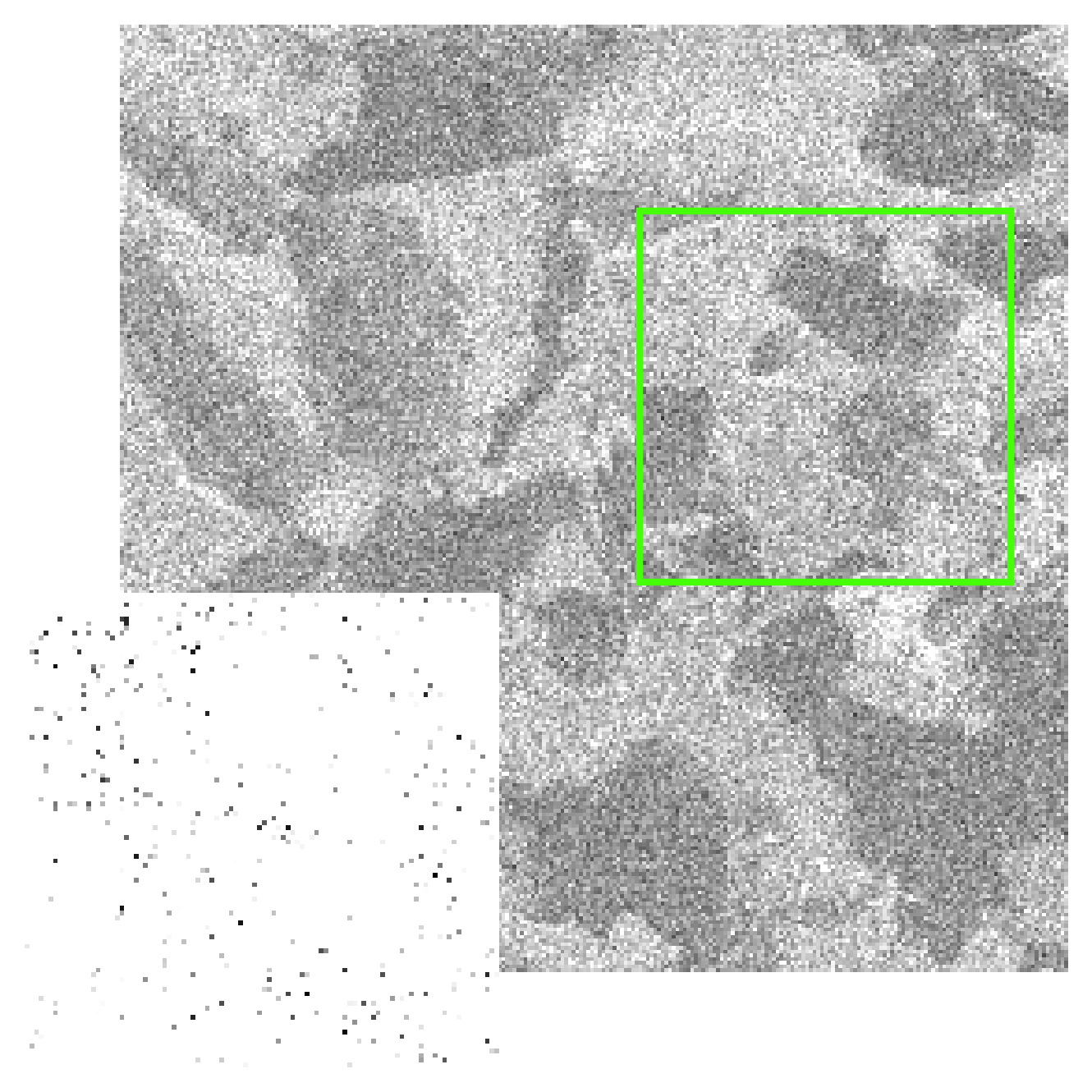}
        \captionsetup{justification=centering}
        \caption*{$\mathrm{RMSE} = 0.565$\\ $\mathrm{SSIM} = 0.204$}
    \end{subfigure} &
    \begin{subfigure}[t]{\restcolwidth}
        \includegraphics[height=\figHeight]{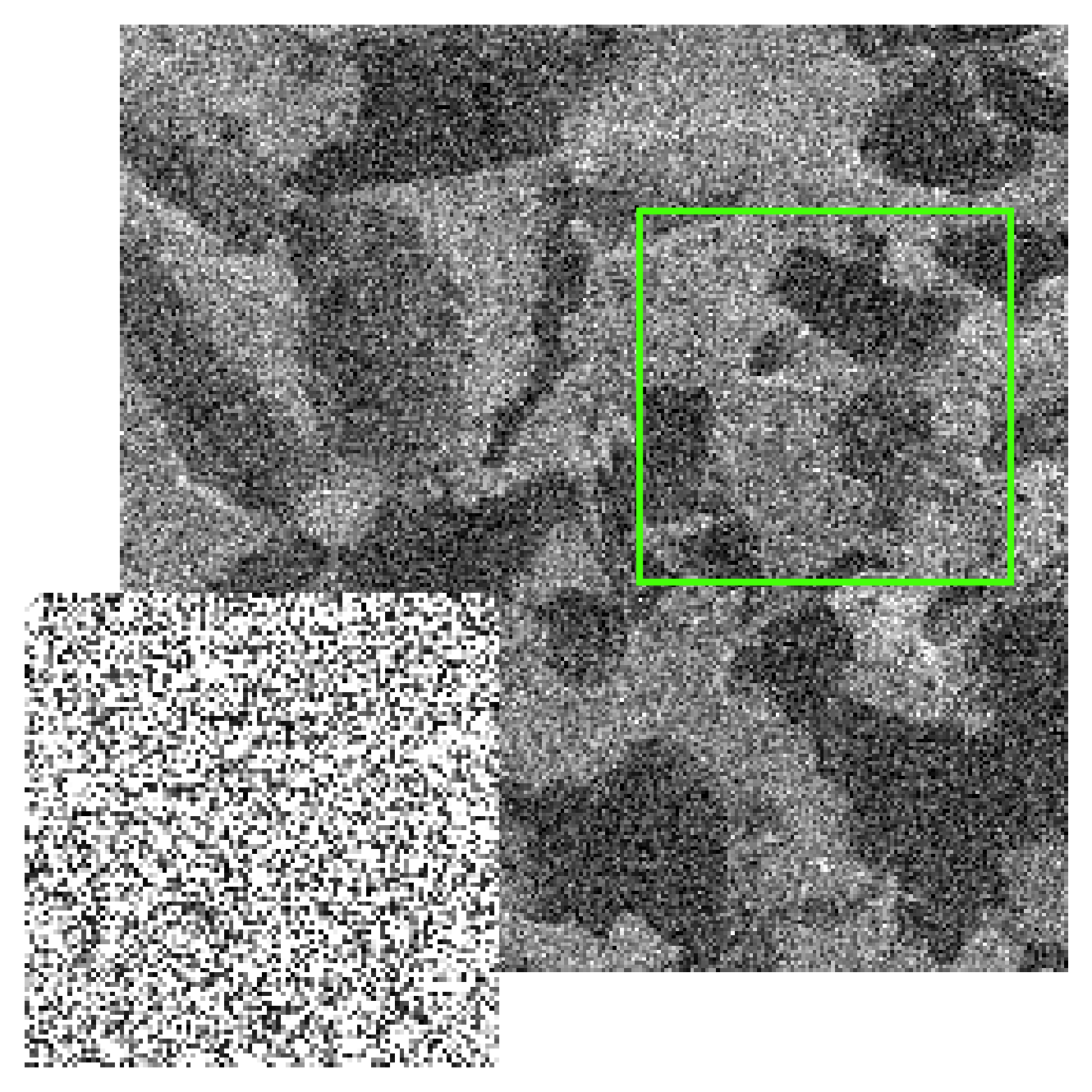}
        \captionsetup{justification=centering}
        \caption*{$\mathrm{RMSE} = 0.266$\\ $\mathrm{SSIM} = 0.175$}
    \end{subfigure} &
    \begin{subfigure}[t]{\restcolwidth}
        \includegraphics[height=\figHeight]{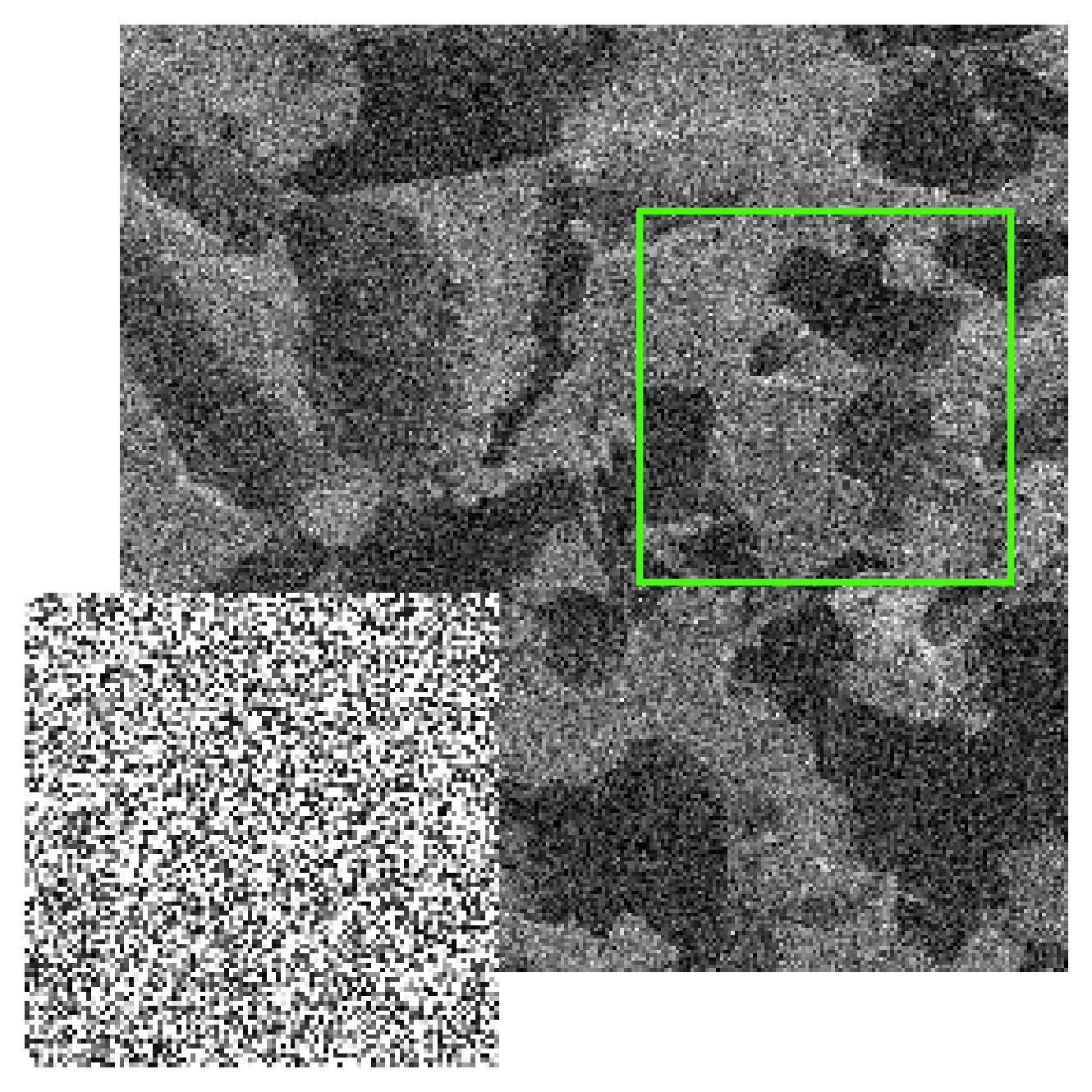}
        \captionsetup{justification=centering}
        \caption*{$\mathrm{RMSE} = 0.214$\\ $\mathrm{SSIM} = 0.206$}
    \end{subfigure}
    \\[5ex]
    \hline
    \\[-1.3ex]
    \multicolumn{3}{l}{{\small \bf PnP estimators}}
    \\
    & {\small Na\"ive}
    & {\small Oracle}
    & {\small Gaussian}
    & {\small QM}
    & {\small LQM}
    & {\small TRML}
    \\
    \rotatebox[origin=l]{90}{\qquad \quad \small TV}
    &
    \begin{subfigure}[t]{\firstcolwidth}
        \includegraphics[height=\figHeight]{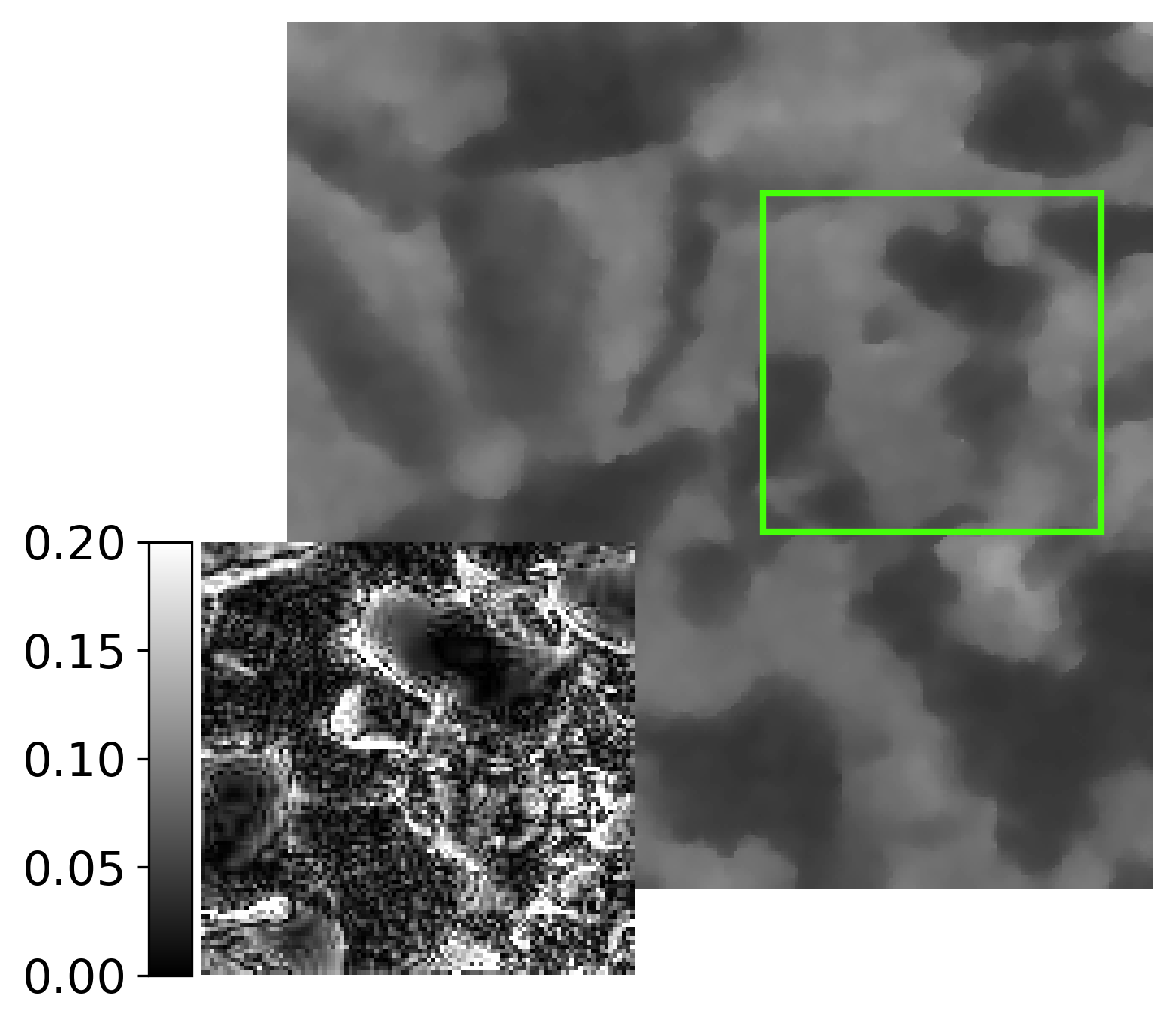}
        \captionsetup{justification=centering}
        \caption*{$\mathrm{RMSE} = 0.077$\\ $\mathrm{SSIM} = 0.580$}
    \end{subfigure}\,\,\,\,\,\, & 
    \begin{subfigure}[t]{\restcolwidth}
        \includegraphics[height=1.01\figHeight]{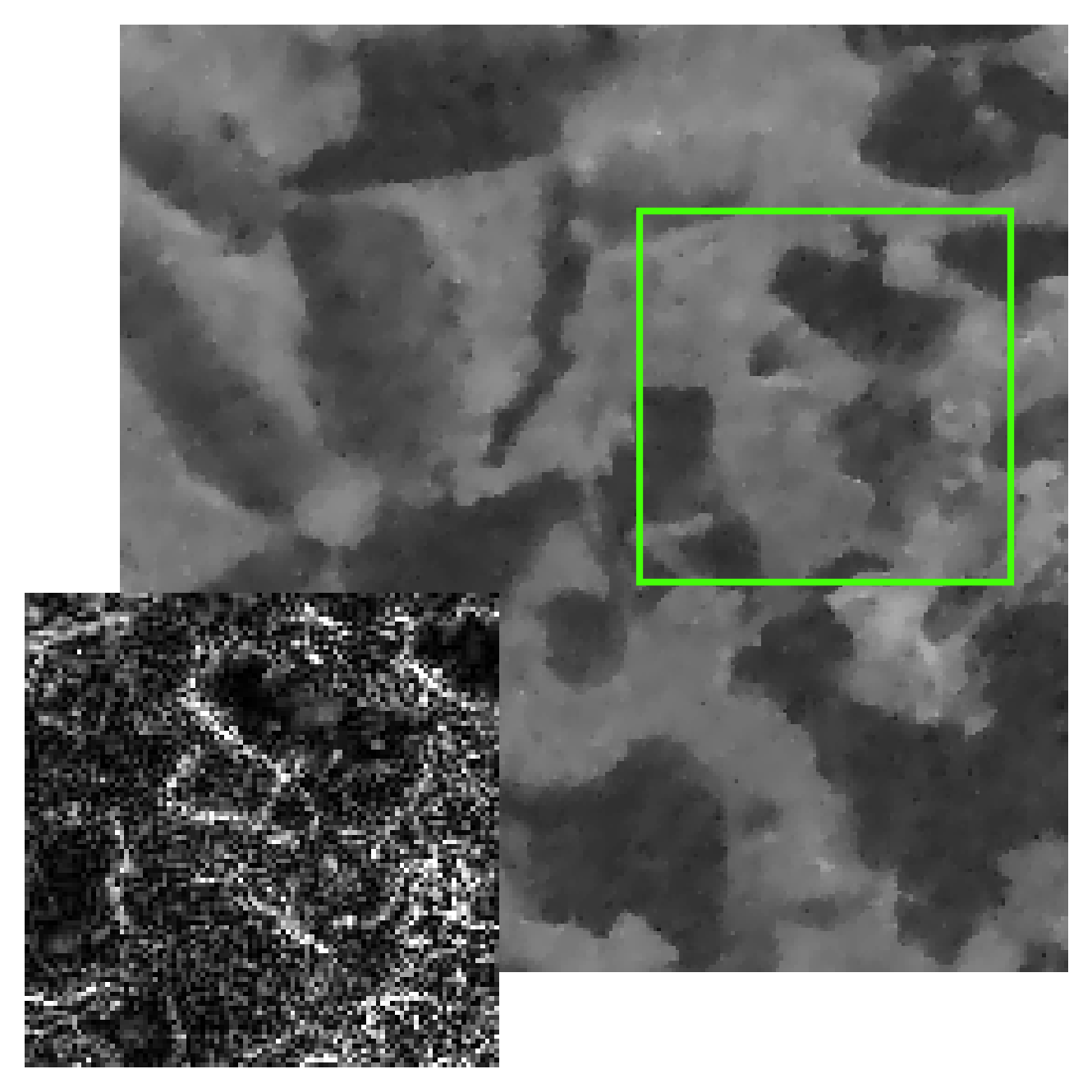}
        \captionsetup{justification=centering}
        \caption*{$\mathrm{RMSE} = 0.057$\\ $\mathrm{SSIM} = 0.672$}
    \end{subfigure} &
    \begin{subfigure}[t]{\restcolwidth}
        \includegraphics[height=\figHeight]{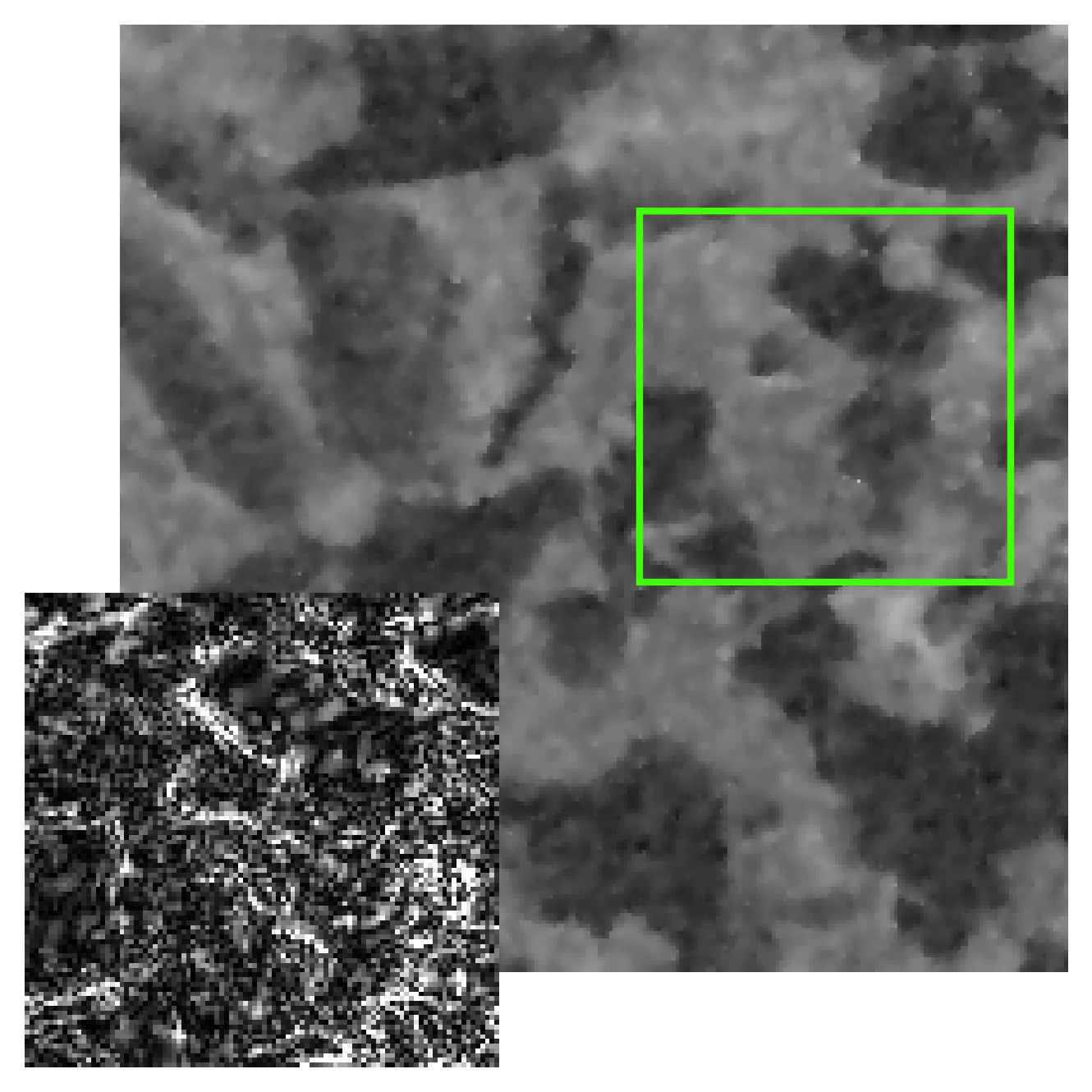}
        \captionsetup{justification=centering}
        \caption*{$\mathrm{RMSE} = 0.069$\\ $\mathrm{SSIM} = 0.565$}
    \end{subfigure} & 
    \begin{subfigure}[t]{\restcolwidth}
        \includegraphics[height=\figHeight]{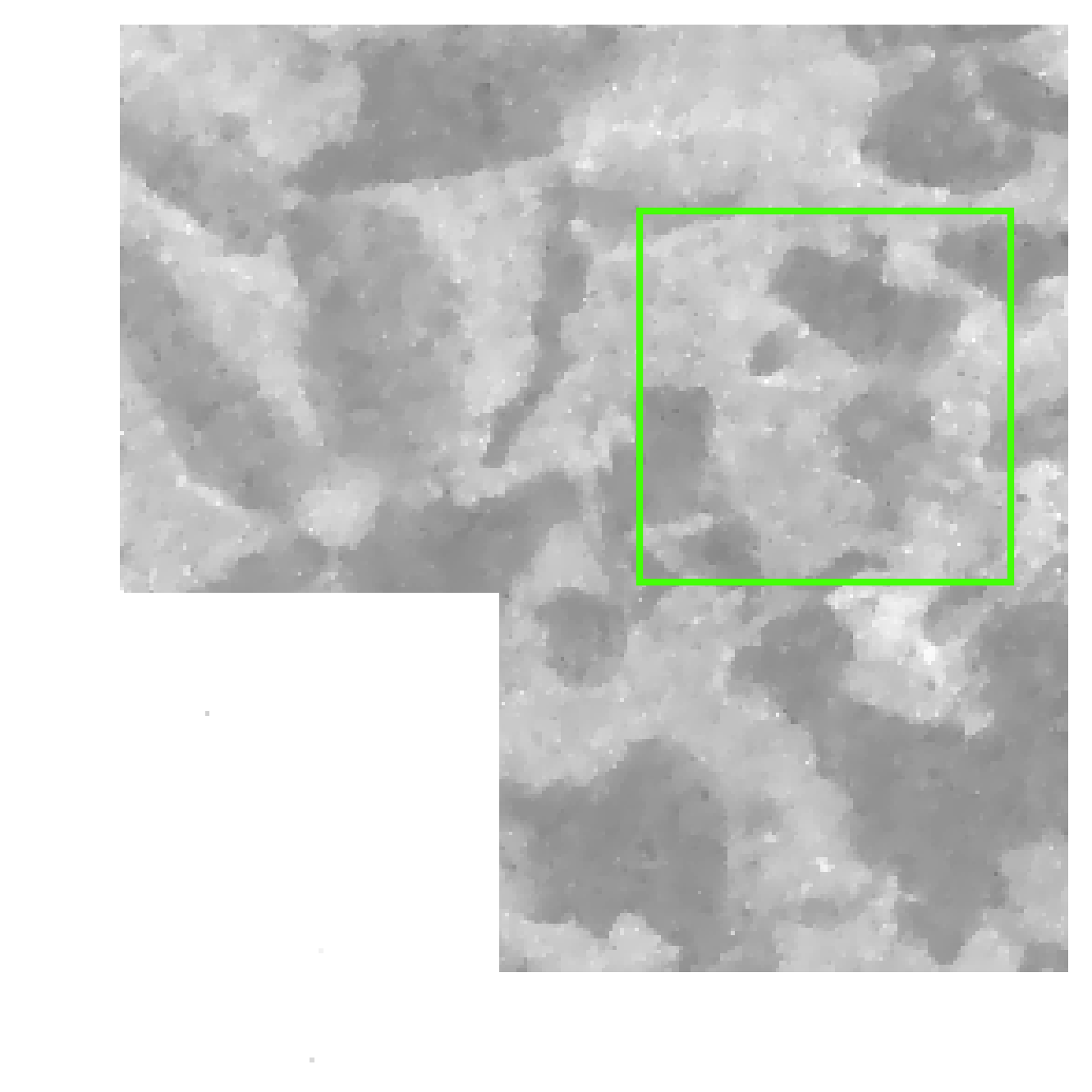}
        \captionsetup{justification=centering}
        \caption*{$\mathrm{RMSE} = 0.538$\\ $\mathrm{SSIM} = 0.562$}
    \end{subfigure} &
    \begin{subfigure}[t]{\restcolwidth}
        \includegraphics[height=\figHeight]{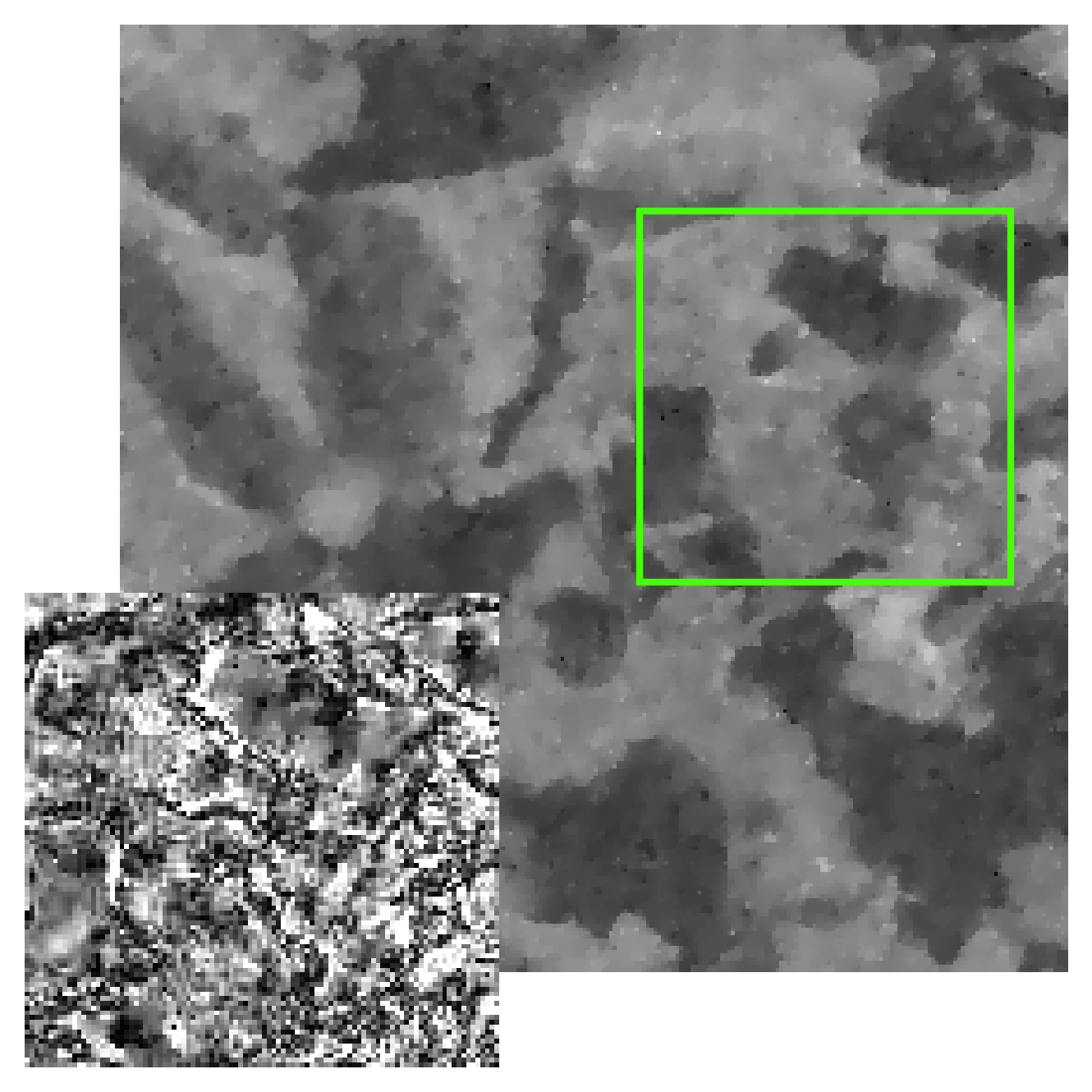}
        \captionsetup{justification=centering}
        \caption*{$\mathrm{RMSE} = 0.115$\\ $\mathrm{SSIM} = 0.596$}
    \end{subfigure} & 
    \begin{subfigure}[t]{\restcolwidth}
        \includegraphics[height=\figHeight]{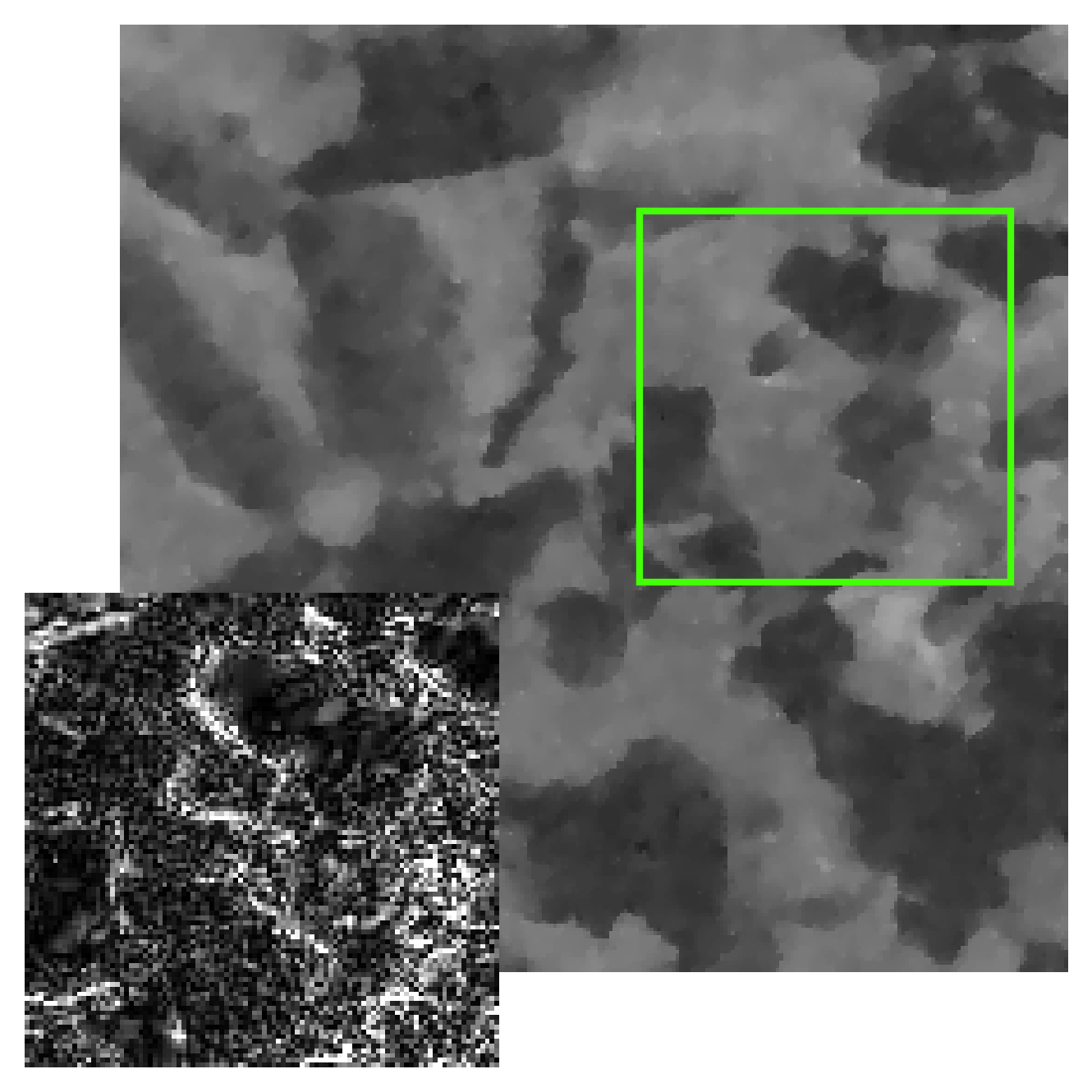}
        \captionsetup{justification=centering}
        \caption*{$\mathrm{RMSE} = 0.062$\\ $\mathrm{SSIM} = 0.642$}
    \end{subfigure}
    \\
    \rotatebox[origin=l]{90}{\qquad \small BM3D}
    &
    \multicolumn{1}{r}{
    \begin{subfigure}[t]{\firstcolwidth}
        \includegraphics[height=\figHeight]{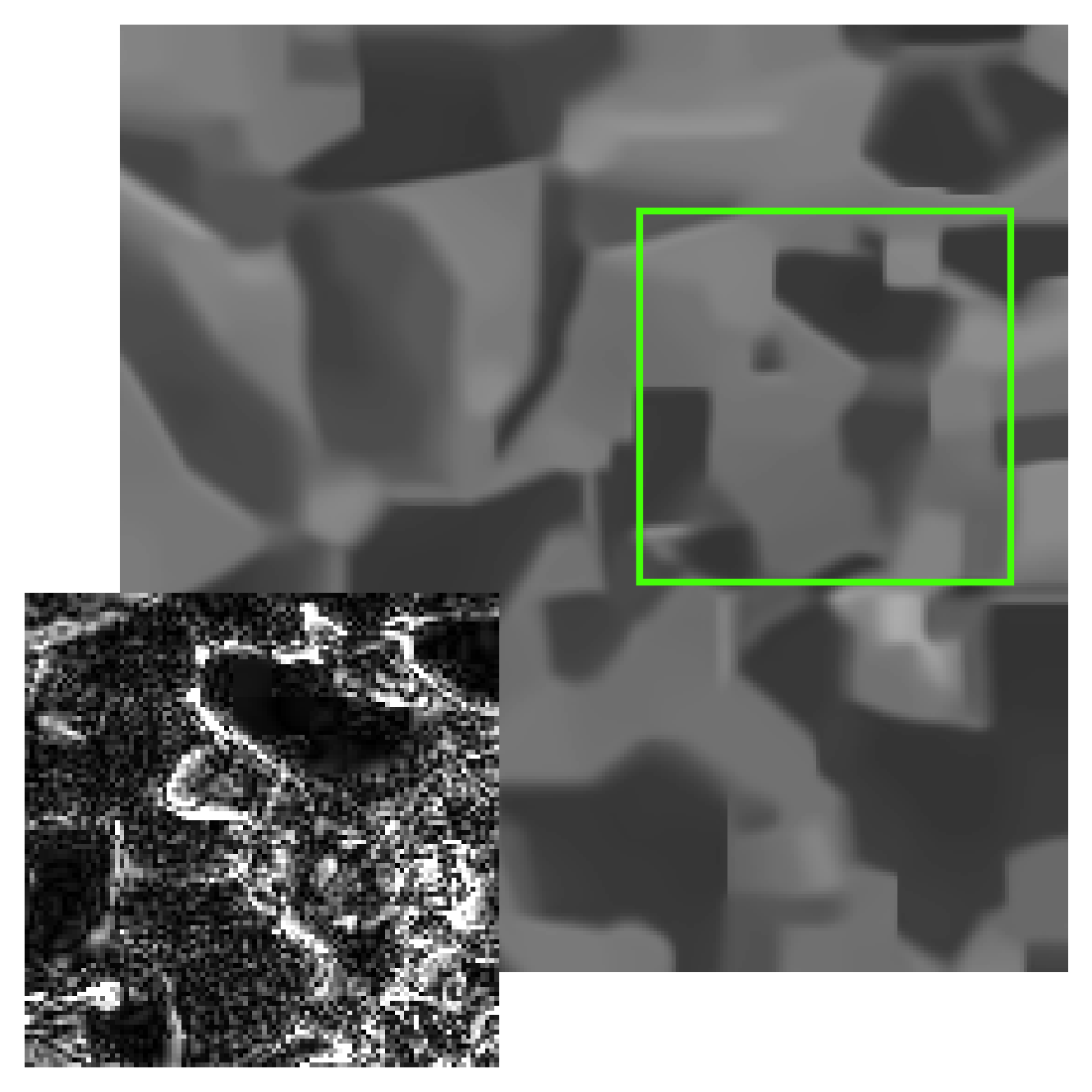}
        \captionsetup{justification=centering}
        \caption*{$\mathrm{RMSE} = 0.072$\\ $\mathrm{SSIM} = 0.602$}
    \end{subfigure}} & 
    \begin{subfigure}[t]{\restcolwidth}
        \includegraphics[height=\figHeight]{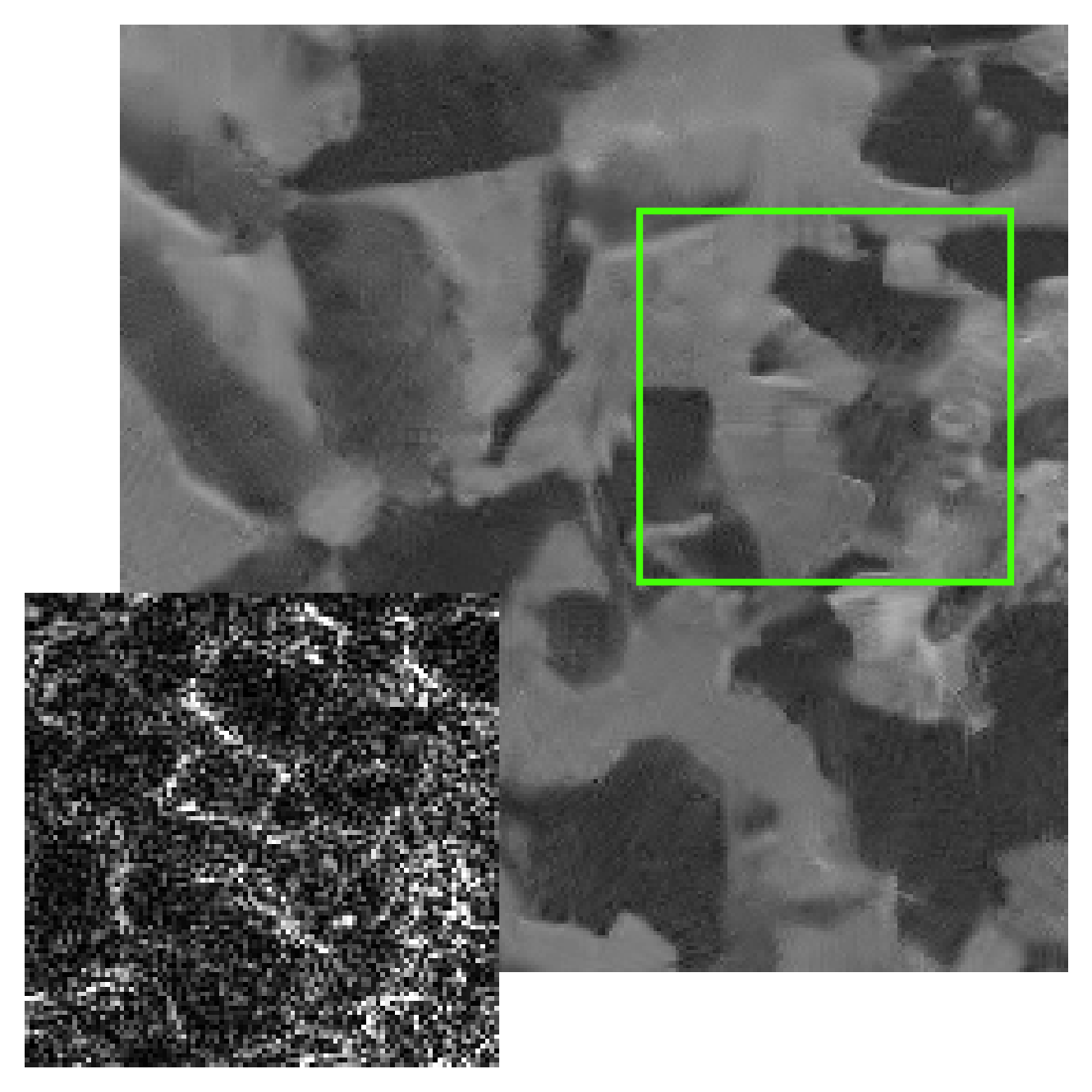}
        \captionsetup{justification=centering}
        \caption*{$\mathrm{RMSE} = 0.058$\\ $\mathrm{SSIM} = 0.641$}
    \end{subfigure} &
    \begin{subfigure}[t]{\restcolwidth}
        \includegraphics[height=\figHeight]{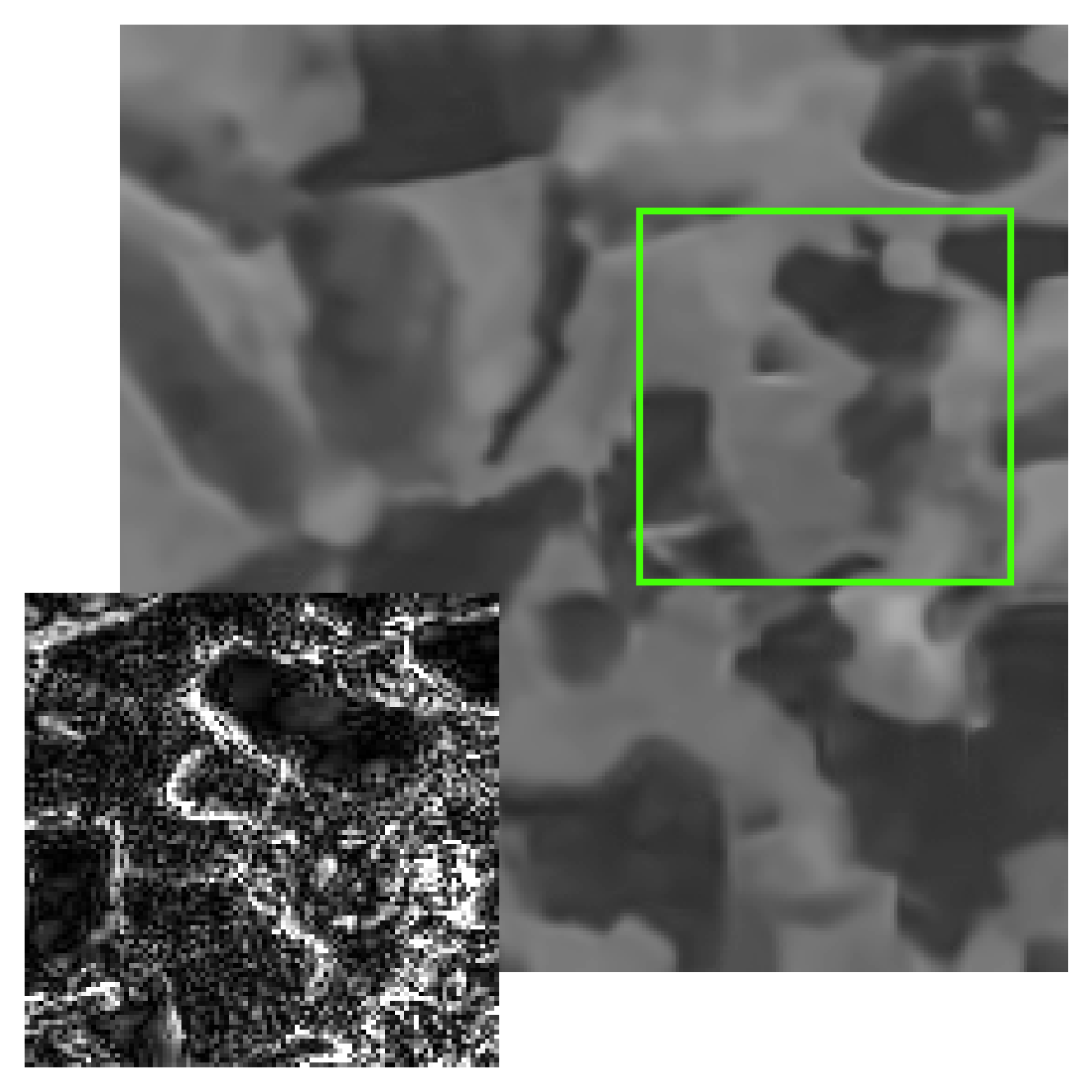}
        \captionsetup{justification=centering}
        \caption*{$\mathrm{RMSE} = 0.067$\\ $\mathrm{SSIM} = 0.625$}
    \end{subfigure} & 
    \begin{subfigure}[t]{\restcolwidth}
        \includegraphics[height=\figHeight]{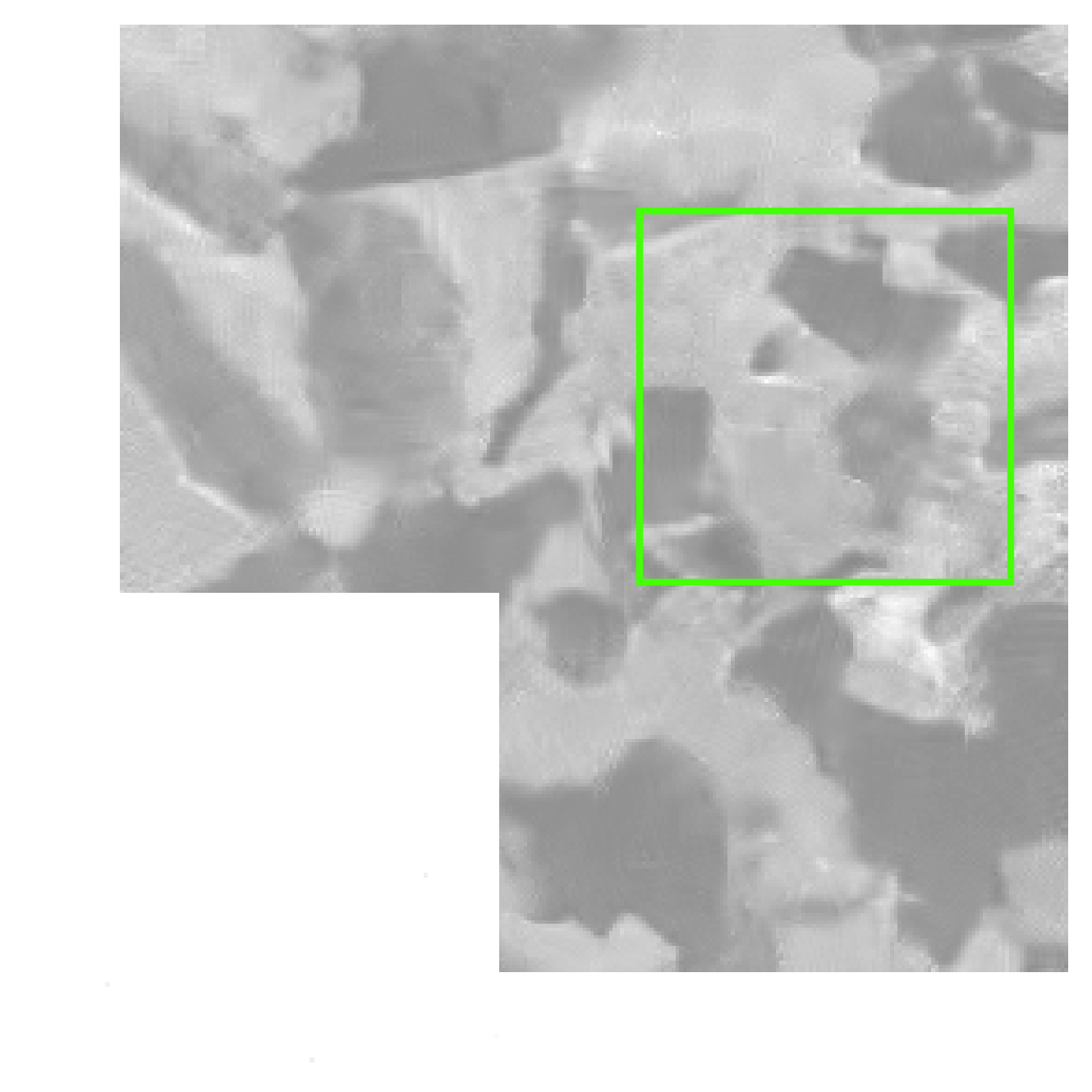}
        \captionsetup{justification=centering}
        \caption*{$\mathrm{RMSE} = 0.536$\\ $\mathrm{SSIM} = 0.597$}
    \end{subfigure} & 
    \begin{subfigure}[t]{\restcolwidth}
        \includegraphics[height=\figHeight]{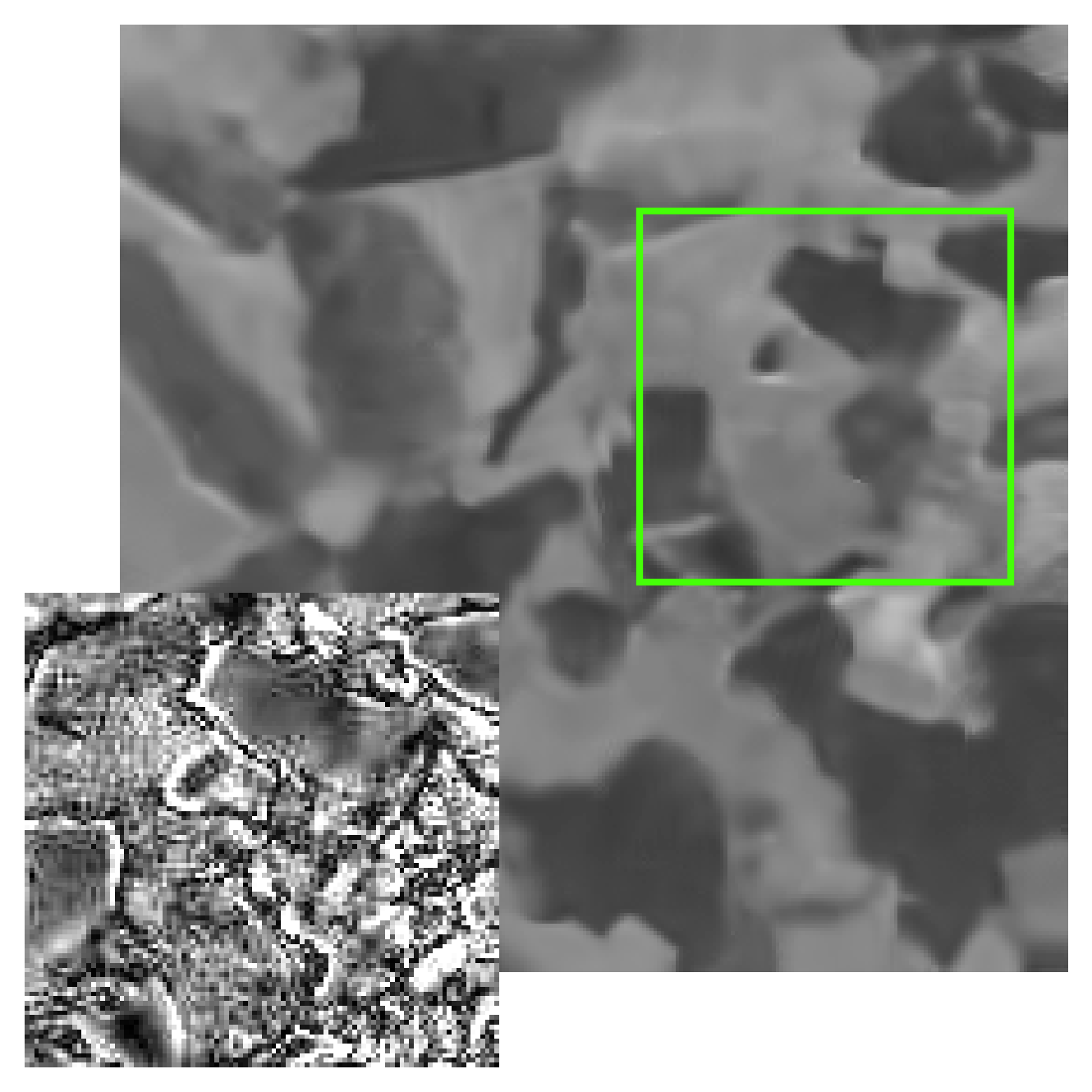}
        \captionsetup{justification=centering}
        \caption*{$\mathrm{RMSE} = 0.115$\\ $\mathrm{SSIM} = 0.647$}
    \end{subfigure} & 
    \begin{subfigure}[t]{\restcolwidth}
        \includegraphics[height=\figHeight]{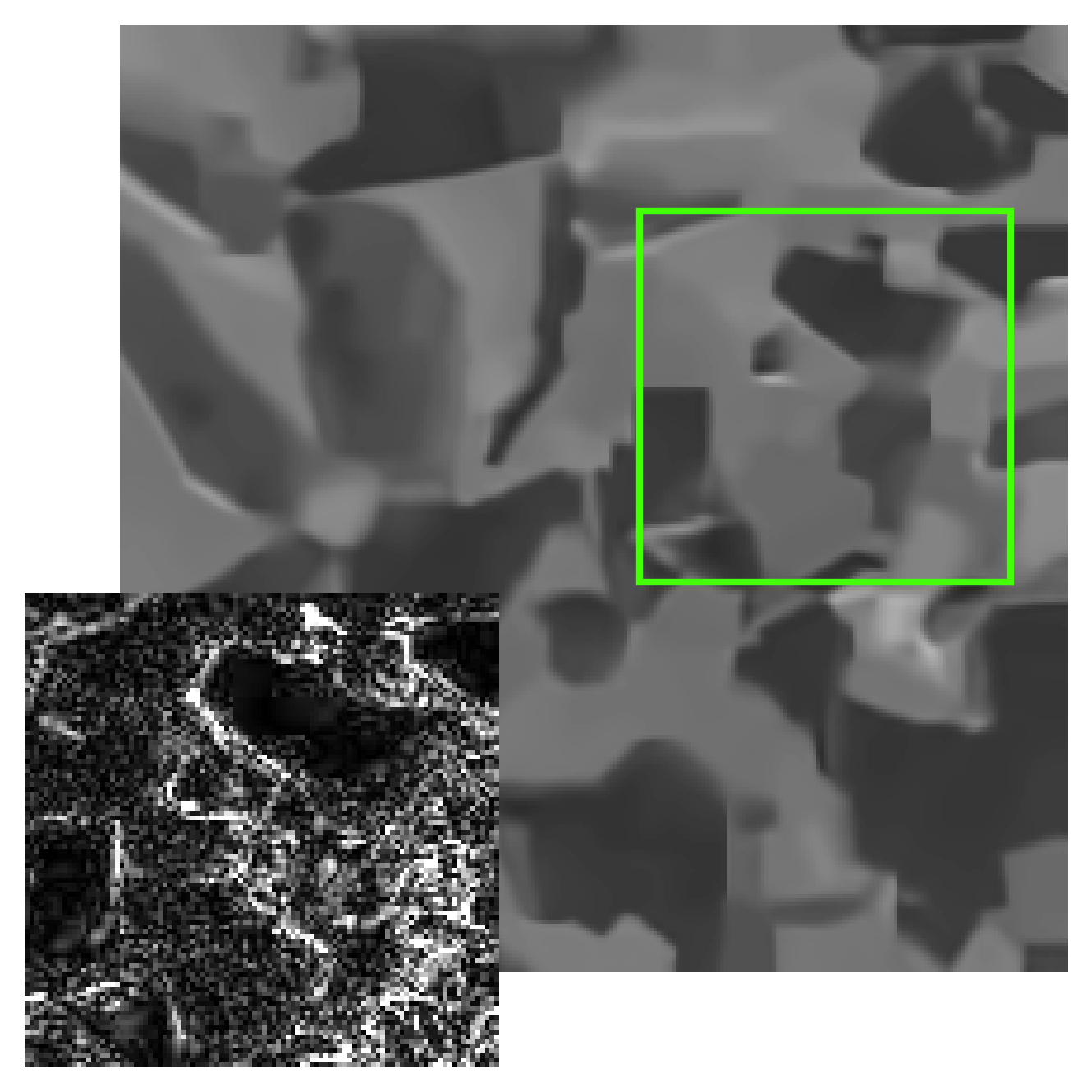}
        \captionsetup{justification=centering}
        \caption*{$\mathrm{RMSE} = 0.062$\\ $\mathrm{SSIM} = 0.655$}
    \end{subfigure}
    \\
    \rotatebox[origin=l]{90}{\qquad \small DnCNN}
    &
    \multicolumn{1}{r}{
    \begin{subfigure}[t]{\firstcolwidth}
        \includegraphics[height=\figHeight]{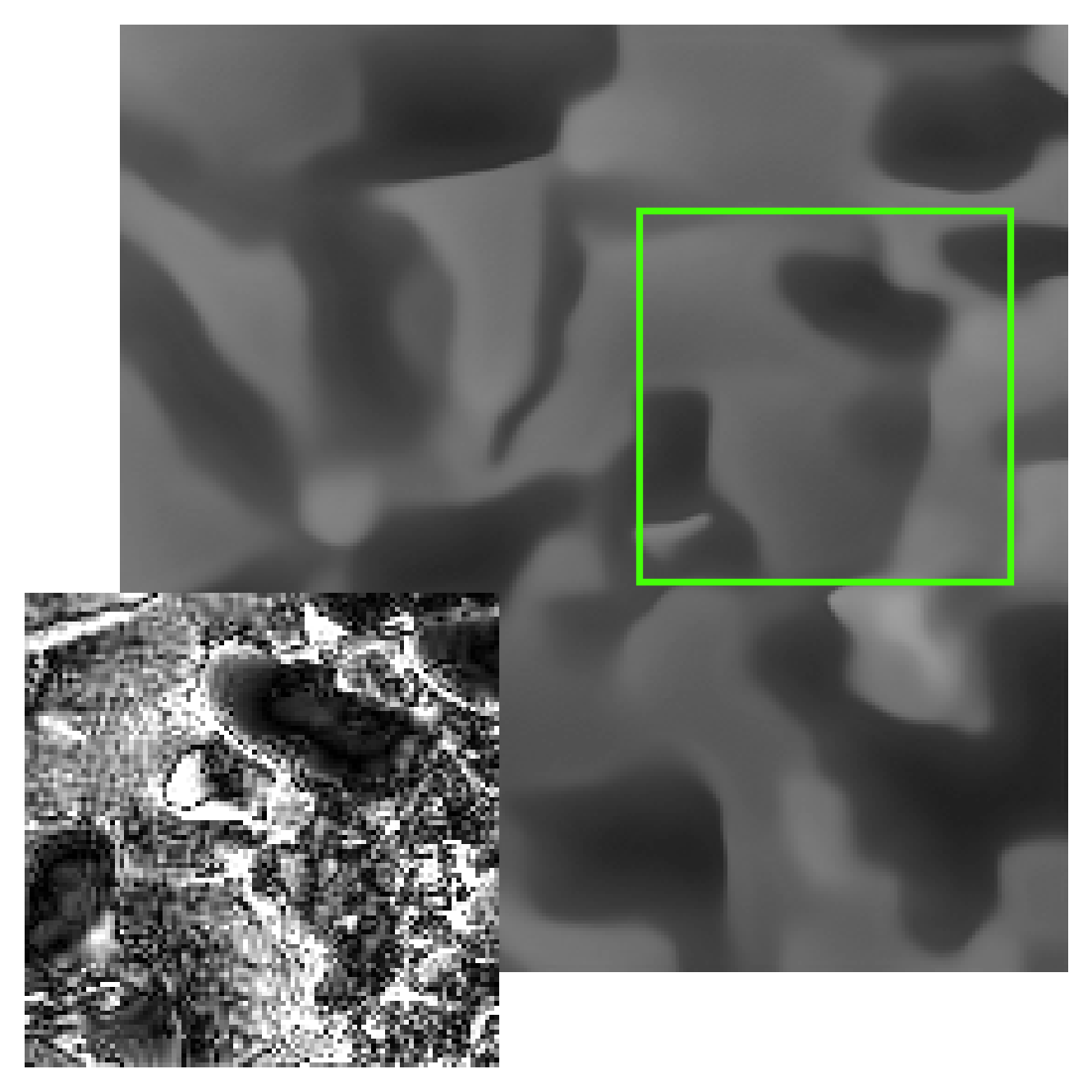}
        \captionsetup{justification=centering}
        \caption*{$\mathrm{RMSE} = 0.113$\\ $\mathrm{SSIM} = 0.498$}
    \end{subfigure}} & 
    \begin{subfigure}[t]{\restcolwidth}
        \includegraphics[height=\figHeight]{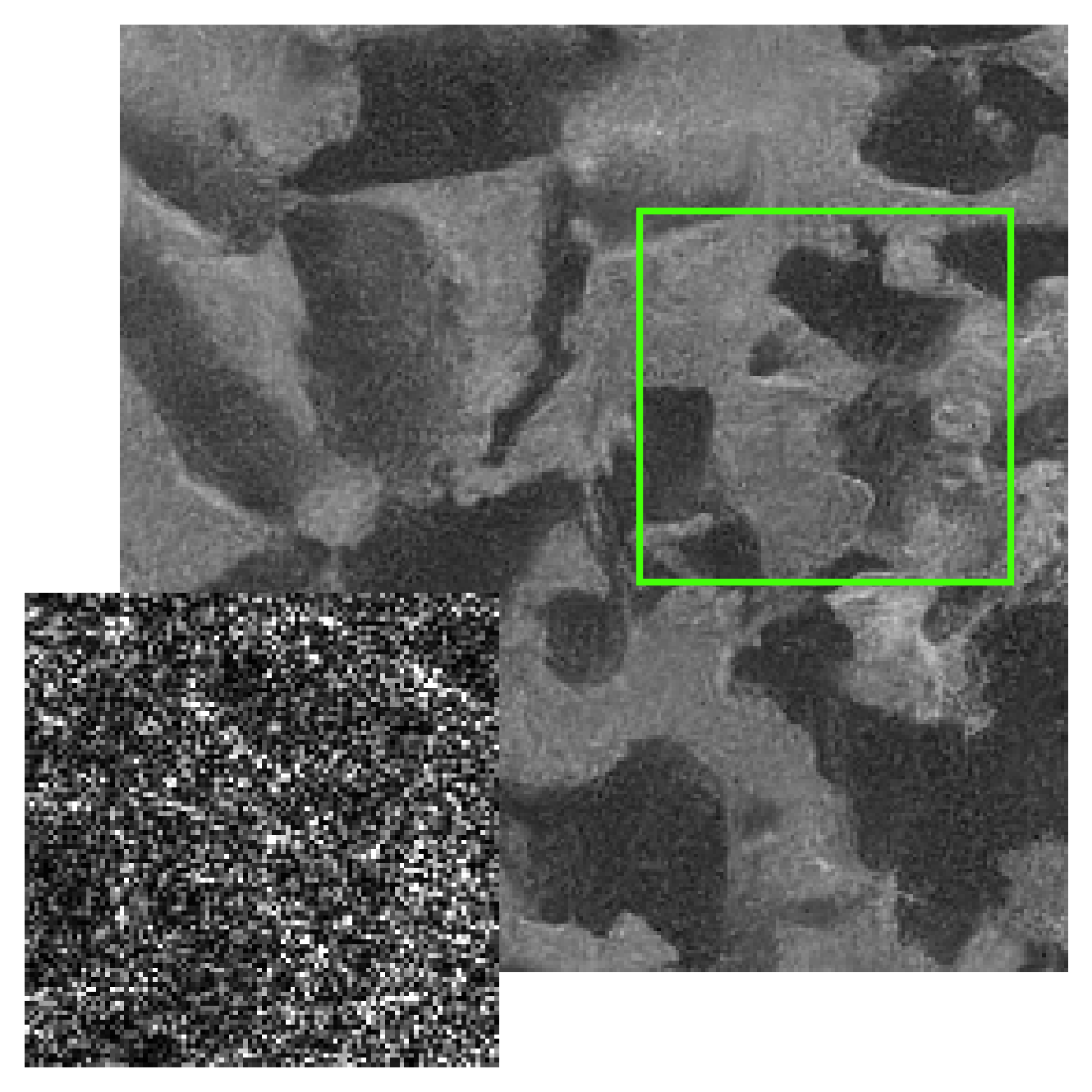}
        \captionsetup{justification=centering}
        \caption*{$\mathrm{RMSE} = 0.071$\\ $\mathrm{SSIM} = 0.536$}
    \end{subfigure} &
    \begin{subfigure}[t]{\restcolwidth}
        \includegraphics[height=\figHeight]{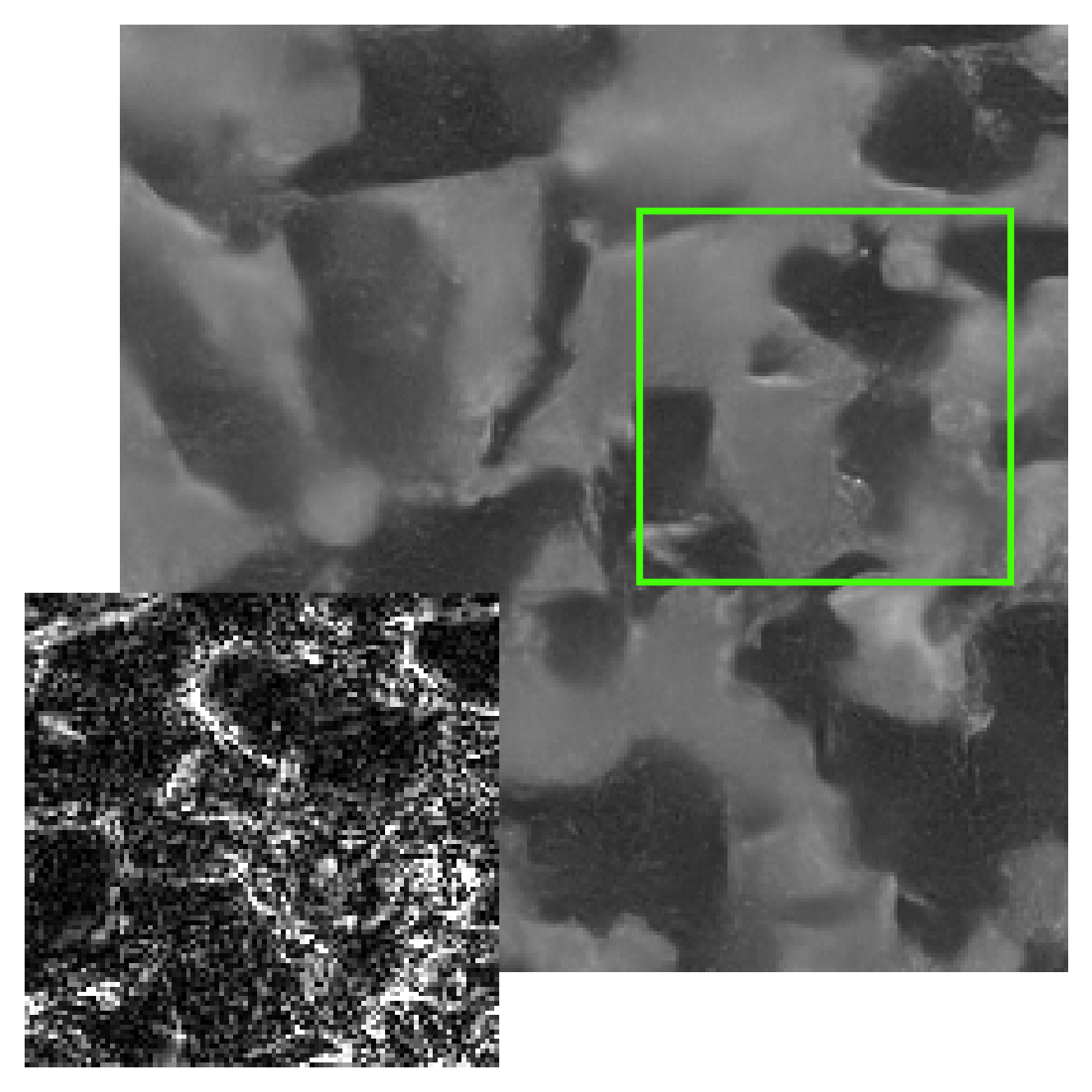}
        \captionsetup{justification=centering}
        \caption*{$\mathrm{RMSE} = 0.066$\\ $\mathrm{SSIM} = 0.600$}
    \end{subfigure} & 
    \begin{subfigure}[t]{\restcolwidth}
        \includegraphics[height=\figHeight]{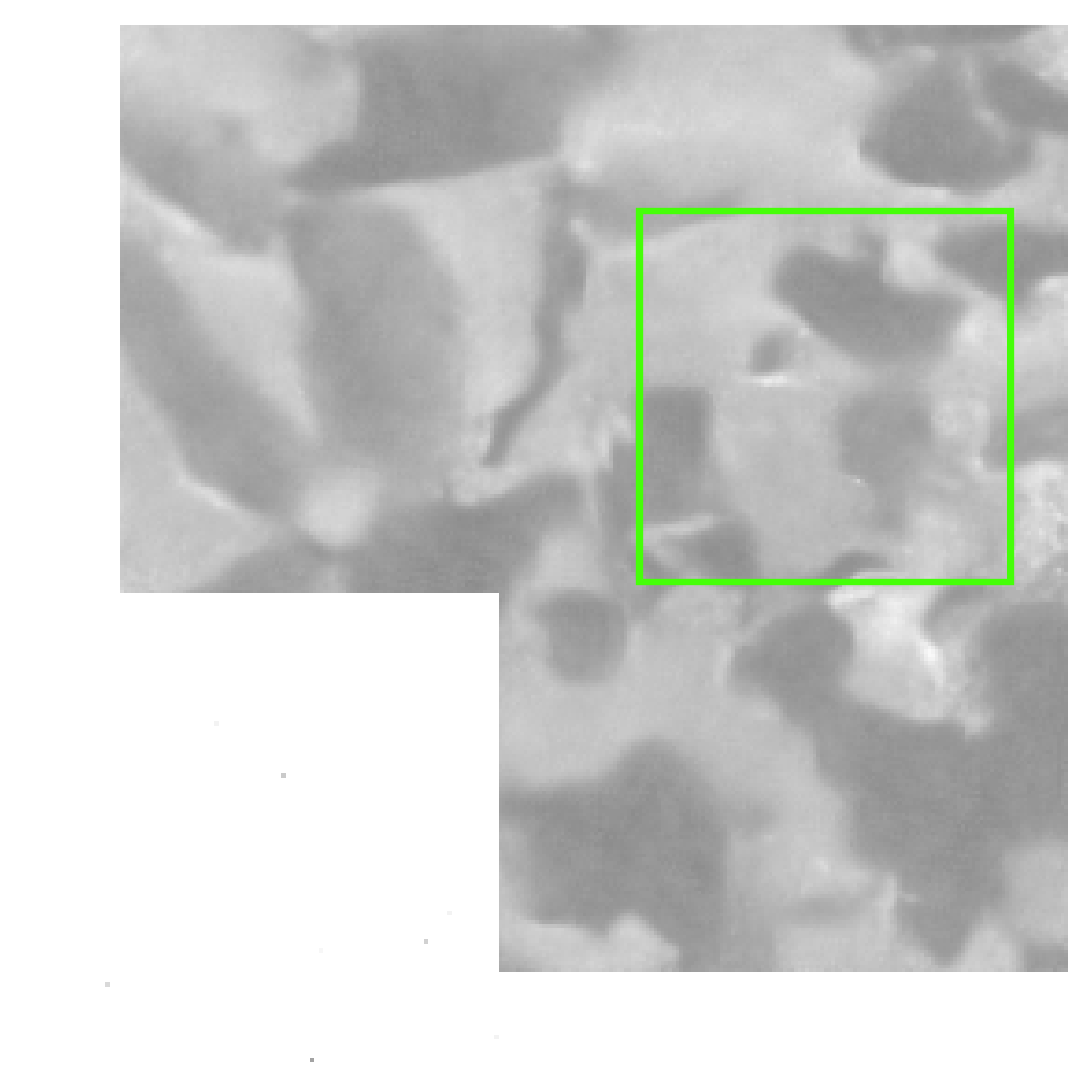}
        \captionsetup{justification=centering}
        \caption*{$\mathrm{RMSE} = 0.540$\\ $\mathrm{SSIM} = 0.542$}
    \end{subfigure} & 
    \begin{subfigure}[t]{\restcolwidth}
        \includegraphics[height=\figHeight]{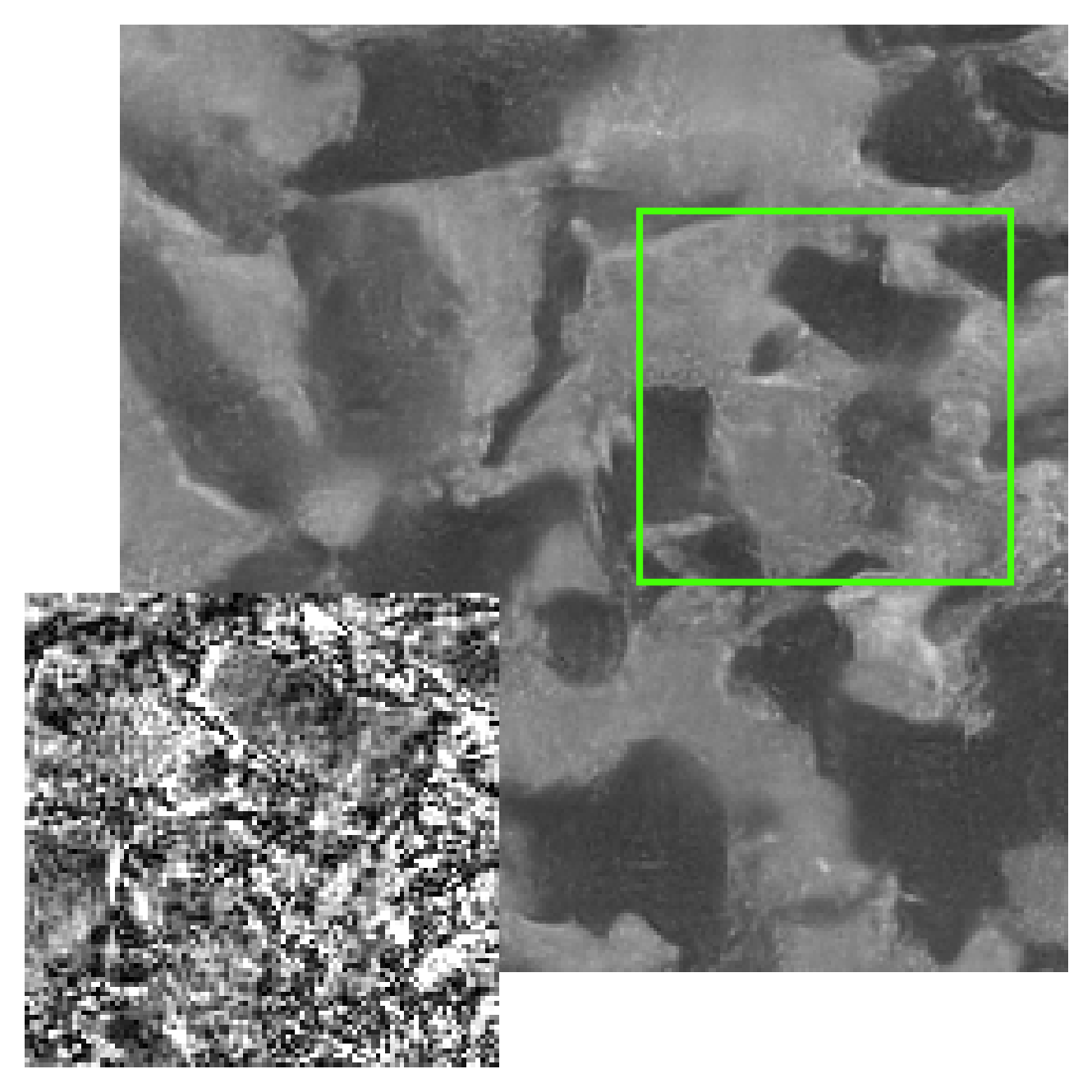}
        \captionsetup{justification=centering}
        \caption*{$\mathrm{RMSE} = 0.114$\\ $\mathrm{SSIM} = 0.605$}
    \end{subfigure} & 
    \begin{subfigure}[t]{\restcolwidth}
        \includegraphics[height=\figHeight]{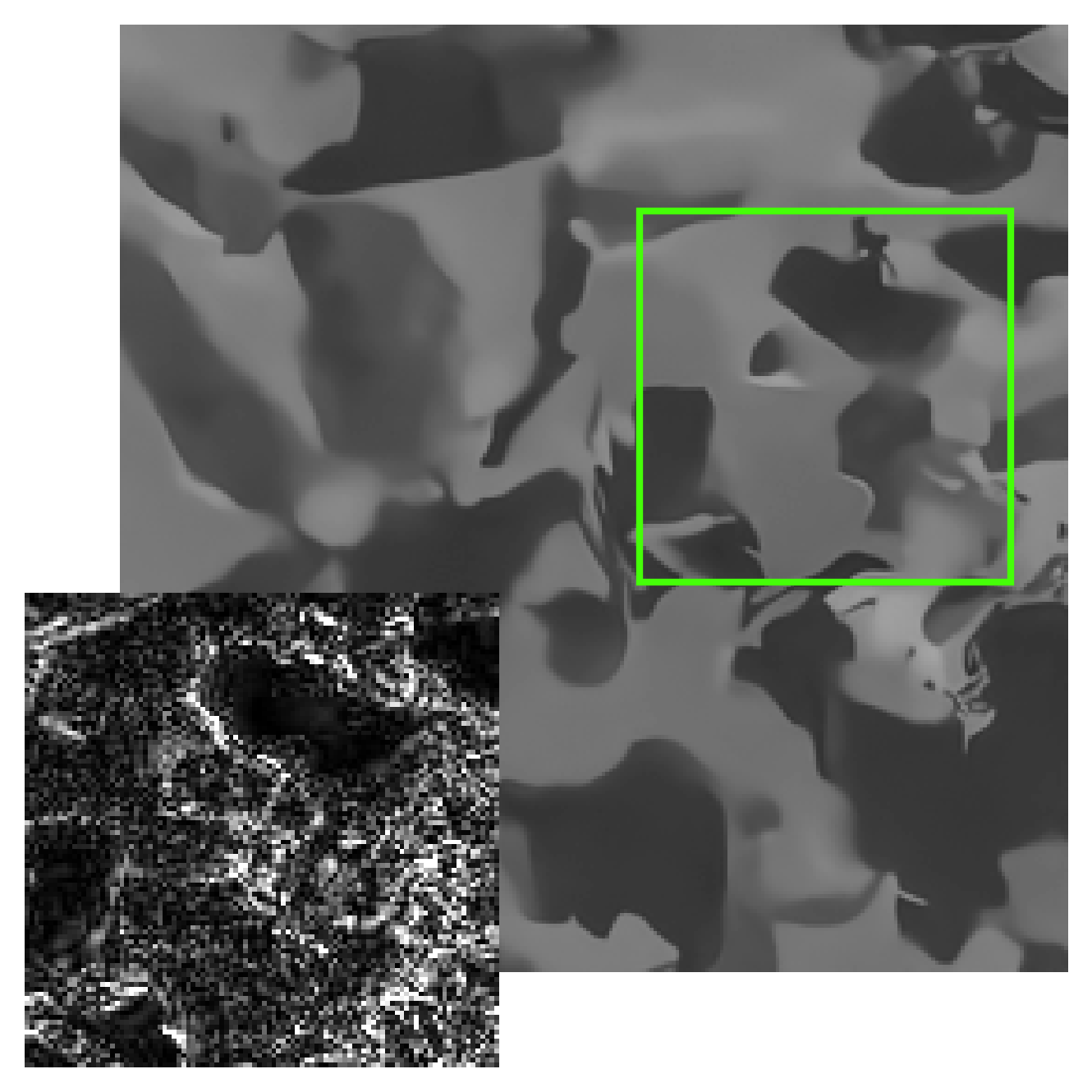}
        \captionsetup{justification=centering}
        \caption*{$\mathrm{RMSE} = 0.059$\\ $\mathrm{SSIM} = 0.674$}
    \end{subfigure}
  \end{tabular}
  \caption{SEM sample with ground truth $\eta \in [1, 2]$, total dose $\lambda = 50$ and $n = 500$. All images are shown on the same scale as in the ground truth image. Absolute error images for the sub-region in each green square are included for reconstructed micrographs. }
  \label{fig:wave}
\end{figure*}

\subsection{Reconstruction Quality Comparisons}
We compare the reconstruction qualities of pixelwise estimations and proposed PnP methods with different data fidelity terms and denoisers.
\Cref{fig:brick,fig:wave} show reconstructed $\eta$
micrographs under HIM and SEM settings, respectively. 
All micrographs are on the same scale shown in the ground truth image. 
The colorbars are chosen so that no more than 2\% of pixels are saturated 
in any given image.
Absolute error images $|\etaHatVec - \etaVec|$ for sub-regions marked by green boxes are displayed for each micrograph as well.
We report RMSE
and 
also the structural similarity index measure (SSIM) to quantify the perceived image quality. 

\paragraph*{Pixelwise estimators (top rows)}
To validate the use of regularization in general,
we show the performance of pixelwise application of each of the five estimation methods (right five columns of the top row in \Cref{fig:brick,fig:wave}).
Without regularization, 
the best TR methods significantly outperform 
the conventional method,
as demonstrated in \cite{PengMBG:21} as well.
The RMSE reduction of the best implementable method of these---TRML---is by a factor of 2.1 for HIM and 3.6 for SEM;
these factors are consistent with theoretical predictions
(see \Cref{ssec:pixelwise-performance} and~\cite{PengMBG:21}).

\paragraph*{Data fidelity terms (comparisons by column)}
The bottom three rows in \Cref{fig:brick,fig:wave} 
present results obtained by PnP methods. 
Moving from the second column
to the rightmost column, each unregularized estimator is improved through regularization by comparing the top row with the bottom three rows in a single column for both HIM and SEM samples.

In comparing the methods using TR data and including regularization,
it is uniformly true that the TRML data fidelity term gives the best performance,
LQM second, and QM the worst.
For SEM emulation, the QM-related PnP results suffer from large bias due to the large bias of the QM estimator when $\eta < 2$~\cite{PengMBG:21}.
Interestingly, the SSIM metric when using the QM data fidelity term is still decent, which is consistent with invariance properties of SSIM~\cite{WangBSS:04}\@.
For HIM emulation, the performance of the Gaussian data fidelity term
is similar to QM and worse than TRML by a factor of 1.4 in RMSE\@.
For SEM emulation, the performance of the Gaussian data fidelity term
is competitive with the best TR method, but still worse than TRML in terms of RMSE\@. 
However, in terms of SSIM, both TRML and LQM data fidelity terms outperform the Gaussian data fidelity term by a noticeable margin.
These empirical results with regularization are consistent with
theoretical results from~\cite{PengMBBG:20,PengMBG:21} that show
increasing utility of TR measurements over conventional measurements as $\eta$ increases.

Along with the combinations of data fidelity terms and denoisers yielding fifteen PnP
methods (bottom three rows and right five columns in \Cref{fig:brick,fig:wave}),
we compute three additional na\"ive estimates
to demonstrate the virtue of accurate modeling of the acquisition process.
The na\"ive estimators assume that $\yconv/\lambda$ 
is Gaussian with mean $\etaVec$ and constant variance
(bottom three rows of the left column in \Cref{fig:brick,fig:wave}).
These are computed by solving
\begin{equation}
\label{eq:gaussian_denoise}
{\etaG} = \argmin_{\etaVec \in \mathbb{R}^d} \, \left\lVert\etaVec-\Frac{\yconv}{\lambda}\right\rVert^2_2 + \beta g(\etaVec)
\end{equation}
for each of the three regularizers
$g$ given explicitly or implicitly in \Cref{ssec:regularization}.
The large reconstruction error from assuming a spatially invariant Gaussian distribution reinforces the significance of having accurate modeling of the acquisition process.

\paragraph*{Denoisers (comparisons by row)}
In the HIM emulation, 
for each of the four PnP ADMM data fidelities
(oracle, Gaussian, QM, LQM),
the DnCNN denoiser and BM3D give nearly equal performance, better than TV\@.
PnP FISTA with TRML data fidelity exhibits the best performance coupled with BM3D denoiser and nearly equal performance with TV denoiser;
these are markedly better in the metrics and recovery of fine details compared to coupling with DnCNN denoiser.

Denoiser comparisons in SEM emulation have different trends.
The impact of denoiser choice on RMSE is nearly negligible.
We observe that BM3D and DnCNN reconstructions provide 
better visual reconstruction results than the TV denoiser as the image with TV denoiser displays patchy artifacts.
Arguably, BM3D denoiser also results in oversmoothing.
This may justify employing a deep network, which can represent more complex structural properties of the images.

\subsection{Accuracy and Runtime Trade-Offs}
\paragraph*{Accuracy summary}
Table~\ref{tab:accuracy} summarizes the average quantitative results from the test images.
The \emph{Regularized} columns present the best metrics over the choices of denoisers and data fidelity terms, excluding the oracle.
The summary highlights the effectiveness of the PnP methods introduced in this paper for both HIM and SEM emulation and both conventional and time-resolved data.
Comparing with pixelwise estimators across the four combinations in the table,
regularization with PnP methods reduces RMSE by factors ranging from 2.24 to 4.11.
Comparing the regularized results using TR data against using conventional data,
the RMSE is reduced by a factor of 1.35 for HIM emulation
and 1.10 for SEM emulation.
This shows that the improvement from regularization is at least partially complementary to the improvement from the use of TR data.
Also, the lesser improvement from TR data for SEM emulation is consistent with earlier results~\cite{PengMBBG:20,PengMBG:21}.\@

\newlength{\vstretch}
\setlength{\vstretch}{0.4ex}
\begin{table}
    \caption{Summary of quantitative accuracy metrics.}
    \centering
    \begin{tabular}{|cl|rrrr|}
        \hline 
        & & & & & \\[-2\vstretch]
        & & \multicolumn{2}{c}{\textbf{Conventional data}}
        & \multicolumn{2}{c|}{\textbf{Time-resolved data}} \\[\vstretch]
                     & & Pixelwise & Regularized & Pixelwise & Regularized \\[\vstretch]
        \hline
        & & & & & \\[-2\vstretch]
        \multirow{2}{*}{\begin{sideways}\textbf{HIM}\end{sideways}}
                 & RMSE & 1.173 & 0.339 & 0.561 & 0.251 \\[\vstretch]
                 & SSIM & 0.225 & 0.643 & 0.448 & 0.741 \\[\vstretch]
        \hline 
        & & & & & \\[-2\vstretch]
        \multirow{2}{*}{\begin{sideways}\textbf{SEM}\end{sideways}}
                 & RMSE & 0.267 & 0.065 & 0.218 & 0.059 \\[\vstretch]
                 & SSIM & 0.147 & 0.563 & 0.190 & 0.626 \\[\vstretch]
        \hline 
    \end{tabular}
    \label{tab:accuracy}
\end{table}

For each of the 4 test images,
we repeat Monte Carlo emulations of HIM and SEM with 10 different random seeds.
For the HIM emulations, 
for estimators with regularization,
the standard deviation of RMSE of $\etaHatVec$ relative to the mean RMSE was 5.1\% on average,
with the highest not exceeding 10\%.
The relative standard deviation of RMSE for the conventional estimator without regularization is only slightly lower at 4.1\%,
suggesting reasonably robust convergence of the PnP methods.
In addition, the performance ranking stays unchanged in each simulation, indicating a fairly robust ranking of methods for the HIM case.
The variation for SEM emulations is higher, suggesting that one should consider
some of the performances in \Cref{fig:wave} virtually indistinguishable.
The greater variation in SEM emulations may be attributable to higher noise level in the data
(the SSIM value for the oracle estimator without regularization
is significantly higher in \Cref{fig:brick} than that in \Cref{fig:wave})
and convergence challenges (see \Cref{sec:convergence}).

\paragraph*{Computation times}
Table~\ref{tab:algo_run_time} shows the average runtimes of Algorithms~\ref{alg:plugNplay} and~\ref{alg:plugNplay_FISTA}
across the test images for different applicable data fidelity terms and denoisers in HIM emulation.

The runtimes for all PnP ADMM methods are similar across different data fidelity terms
while having much more dependence on the choice of denoiser;
using DnCNN is the fastest, TV about 10 times slower, and BM3D about another 60 times slower.
Considering that none of the PnP ADMM methods with BM3D give appreciably lower RMSE or higher SSIM than the corresponding methods with DnCNN,
BM3D is generally unattractive in PnP ADMM.

Using PnP FISTA with the TRML data fidelity term,
which significantly improves reconstruction quality over the PnP ADMM methods in most cases,
the runtimes for TV and DnCNN denoisers increase by approximately 40 and 180 times, respectively, compared to their PnP ADMM counterparts.
On the contrary, the runtime for PnP FISTA using BM3D decreases slightly.
Comparing the runtimes of PnP FISTA using different denoisers, the runtime when using BM3D is only 1.7 times longer than that of TV and 3 times longer than that of DnCNN\@.
Consequently, BM3D becomes an attractive option in PnP FISTA, especially when it also gives better reconstructions, such as in the case of HIM emulation as shown in \Cref{fig:brick}.

\setlength{\vstretch}{0.5ex}
\begin{table}
    \caption{Runtimes in seconds for all combinations of data fidelity term and denoiser. PnP FISTA is used for the TRML data fidelity term, while PnP ADMM is used for all other data fidelity terms.}
    \centering
    \begin{tabular}{|cl|rrrrr|}
        \hline 
& & & & & & \\[-2\vstretch]
& & \multicolumn{5}{c|}{\textbf{Data fidelity}}\\[\vstretch]
& & Gaussian & Oracle & QM & LQM & TRML \\[\vstretch]
        \hline
& & & & & & \\[-2\vstretch]
        \multirow{3}{*}{\begin{sideways}\textbf{Denoiser}\end{sideways}}
& TV & 0.23 & 2.86 & 0.68 & 1.42 & 46.50 \\[\vstretch]
& BM3D & 84.82 & 85.25 & 85.13 & 84.88 & 78.98 \\[\vstretch]
& DnCNN & 0.14 & 0.17 & 0.14 & 0.13 & 26.14 \\[\vstretch]
        \hline 
    \end{tabular}
    \label{tab:algo_run_time}
\end{table}

\subsection{Noise Level and Optimal Regularization Strength}
\label{ssec:optimal_reg_strength}

\begin{figure}
    \flushright
    \begin{subfigure}[t]{0.49\columnwidth}
        \centering
        \includegraphics[width=\columnwidth]{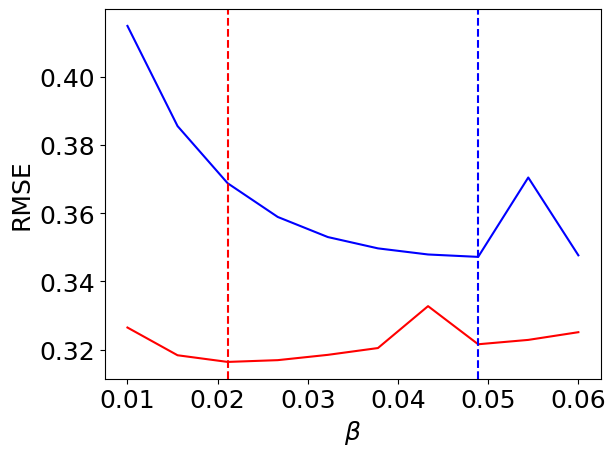}
        \captionsetup{oneside, margin={0.6cm,0cm}}
        \caption{TV denoiser}
    \end{subfigure}
    \begin{subfigure}[t]{0.49\columnwidth}
        \centering
        \includegraphics[width=\columnwidth]{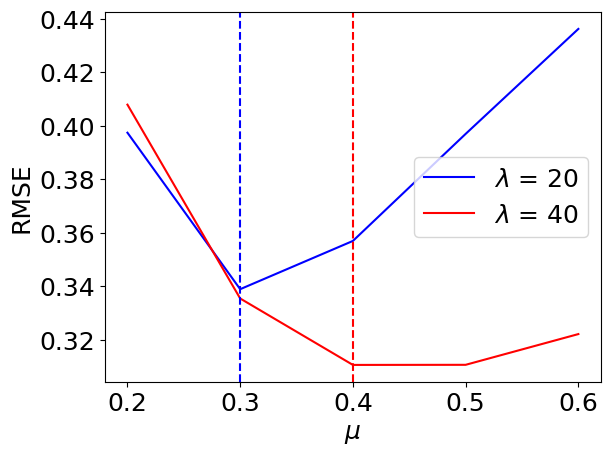}
        \captionsetup{oneside, margin={0.6cm,0cm}}
        \caption{DnCNN denoiser}
    \end{subfigure}
    \caption{Relationship between RMSE of the reconstruction and the regularization parameters, $\beta$ for TV and $\mu$ for DnCNN, at $\lambda = 20$ and $40$. The vertical lines indicate the regularization parameter value when the RMSE is minimum for the corresponding $\lambda$.}
    \label{fig:lambdaTrend}
\end{figure}

Reconstructions from different pixelwise estimators have different expected MSE depending on $\lambda$ and $\eta$ as explained in \Cref{ssec:estimator_no_reg,ssec:pixelwise-performance}.
However, it is unclear how this effective noise level determines the optimal choice of the regularization strength $\beta$ in \eqref{eqn:problem}.
In our experiments, we tune the regularization parameters $\sigma$ and $\mu$, both of which imply $\beta$, on a hold-out validation image.
Still, the question of how $\lambda$ influences the optimal choice of $\beta$ remains.

To study the relationship between $\lambda$ and the optimal regularization strength, we simulated two measurements of the hold-out validation image: one with $\lambda = 20$ and $n = 200$ and the other with $\lambda = 40$ and $n = 400$.%
\footnote{Scaling $n$ with $\lambda$ maintains the advantage of TR measurement and the validity of approximations used to justify the LQM estimator~\cite{PengMBG:21}.}
Then, we compare the RMSE of reconstructions from PnP ADMM with the $\flqm$ data fidelity term using different choices of $\beta$.
We scale $\flqm$ by $1 / n$ to make the comparison fair, since $\flqm$ scales approximately linearly with $n$ under a fixed expected number of incident ions per subacquisition. The ADMM inversion step then becomes
\begin{subequations}
\label{eq:f-lqm-scaled}
\begin{equation}
  \etaVec_k^{(t + 1)} = -\frac{1}{2} \varphi + \frac{1}{2}\sqrt{ \varphi^2 + \frac{4\yconv_k}{\rho n}}
\end{equation}
where
\begin{equation}
  \varphi = \frac{\lVec_k}{\rho n (1 - e^{-\etaVecPrev_k})} - \vVec_k + \uVec_k.
\end{equation}
\end{subequations}

We fix $\rho = 0.005$ and $\alpha = 5 \times 10^{-4}$, and we use the TV denoiser and DnCNN with varying $\beta$ and $\mu$.
Note that $\beta$ cannot be specified exactly when using DnCNN, because the relationship between $\mu$ and the Gaussian noise standard deviation $\sigma$ relies on assumptions about the denoiser that are not true in practice \cite{xu2020boosting}. 

The RMSE of the reconstructions are shown in \Cref{fig:lambdaTrend}.
As expected, the higher the dose $\lambda$, the greater the information in the measurements and thus the smaller the error.
Consequently, as $\lambda$ increases, we observe that the optimal regularization strength decreases, i.e., $\beta$ decreases and $\mu$ increases.

\subsection{Empirical Convergence}
\label{sec:convergence}

The empirical fixed point convergence behaviors of our PnP methods agree with the theory in Section~\ref{sec:convergence}.
As shown in Table~\ref{tab:conv}, the PnP methods with a convex data fidelity term---$\foracle$, $\fqm$, or $\flqm$---and a denoiser with contractive residue---TV denoiser or DnCNN---always converge to some fixed point.
The combination of BM3D with any data fidelity term always fails to converge because BM3D has expansive residues~\cite{ryu19pnp}.
While there is no convergence guarantee for PnP methods with the nonconvex data fidelity terms $\fconv$ and $\ftrml$, they sometimes converge.
One explanation is that the iterates converge in a locally convex region of the data fidelity term. We observe similar convergence behaviors in SEM emulations.
However, with SEM data, fixed point convergence when using $\ftrml$ is more difficult.
This may be partially understood through examination of $\partial^2 \ftrml(\etaVec) / \partial \etaVec_k^2$.
The derivative of the leading term of \eqref{eq:trml_gradient} is nonpositive.
When $\etaVec_k$ is small (as in SEM),
$n - \lVec_k$ is likely to be relatively large,
making the magnitude of the derivative of the leading term of \eqref{eq:trml_gradient} relatively large;
empirically, this term is causing a lack of local convexity of $\ftrml(\etaVec)$.

\begin{table}
    \caption{Fractions of 40 independent HIM emulations for which each combination of data fidelity term and denoiser results in convergence to a fixed point.}
    \centering
    \begin{tabular}{|cl|rrrrr|}
        \hline 
& & & & & & \\[-2\vstretch]
& & \multicolumn{5}{c|}{\textbf{Data fidelity}}\\[\vstretch]
& & Gaussian & Oracle & QM & LQM & TRML \\[\vstretch]
        \hline
& & & & & & \\[-2\vstretch]
        \multirow{3}{*}{\begin{sideways}\textbf{Denoiser}\end{sideways}}
& TV & 0.70 & 1 & 1 & 1 & 1 \\[\vstretch]
& BM3D & 0 & 0 & 0 & 0 & 0 \\[\vstretch]
& DnCNN & 0.55 & 1 & 1 & 1 & 0.025 \\[\vstretch]
        \hline 
    \end{tabular}
    \label{tab:conv}
\end{table}

\begin{figure}
    \centering
    \includegraphics[width=0.95\columnwidth]{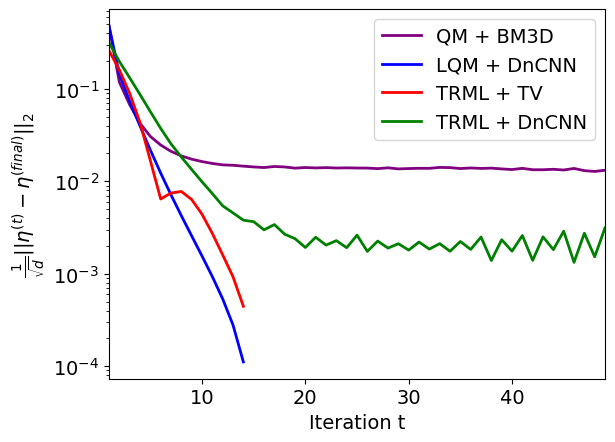}
    \caption{Root mean-square distance between the reconstruction at each iteration $\etaVec^{(t)}$ and the final reconstruction $\etaVec^{(\text{final})}$ for selected combinations of data fidelity terms and denoisers from an HIM emulation.}
    \label{fig:conv}
\end{figure}

Typical convergence behaviors of some combinations of data fidelity terms and denosiers are shown in \Cref{fig:conv}.
For PnP methods that converge to fixed points, the final iterate $\etaVec^{(\text{final})}$ is approximately the fixed point.
For PnP methods that do not converge to fixed points, the root mean-square distance between the iterates $\etaVec^{(t)}$ and $\etaVec^{(\text{final})}$ either plateaus or fluctuates.

\section{Conclusion}
In this paper,
we develop data fidelity terms to make model-based reconstruction efficiently applicable to the distinctive measurement model for particle beam microscopy.
We use these data fidelity terms in plug-and-play methods for estimation of the mean secondary electron yield $\etaVec$.
Because of the Neyman Type~A likelihood for PBM data,
using the negative log-likelihood directly would be computationally intractable.
We introduce approximations with different computational complexities and accuracies
applicable to conventional or time-resolved measurements, 
and we compare their efficacies when combined with three different denoisers.
In synthetic experiments emulating helium ion microscopy and scanning electron microscopy,
we demonstrate that our approaches outperform pixelwise (non-regularized) methods
substantially in RMSE, SSIM, and qualitative appearance;
RMSE reduction is by a factor of 2.24 to 4.11\@.
We provide the first systematic demonstration that improvements due to regularization and to time-resolved measurement can be complementary.

\section*{Acknowledgment}
The authors thank Akshay Agarwal for insightful discussions.
The authors also acknowledge the Boston University Research Computing Services group for providing computational resource support which has contributed to the results reported in this paper.

% Good for checking accuracy of .bib files: \bibliographystyle{plain}
\bibliographystyle{IEEEtran}
% \bibliography{bibfile,bibl}
\bibliography{bibfile,bibl}

\end{document}